\documentclass[twocolumn,amssymb,showpacs, amsmath,superscriptaddress,pre]{revtex4-1}

\usepackage{graphicx}
\usepackage{dcolumn}
\usepackage{bm}

\begin{document}

\title[Two-band superconductivity]{A review of two-band superconductivity: materials and effects on the thermodynamic and reversible mixed-state properties.}

\author{M. Zehetmayer}

\affiliation{Vienna University of Technology, Atominstitut, 1020 Vienna, Austria}

\begin{abstract}
Two-band superconductivity has become an important topic over the last ten years. Extensive experimental and theoretical studies started with MgB$_2$ and are now focused on the iron-based and other new superconductors. In this review, I describe how important thermodynamic, reversible mixed-state, and other superconducting properties are changed by two-band and, for comparison, by other effects such as anisotropy in a single-band material or an energy gap structure different from the conventional s-wave symmetry. The work consists of three main parts, in which I review (i) theoretical models and what they predict for experimentally accessible properties in the two-band and other scenarios, (ii) experimental methods applied for investigating superconducting properties and the results obtained in potential two-band materials, and (iii) materials, for which two- or multi-band superconductivity has been suggested. It is shown that two-band effects appear in most of the analyzed properties and that they can be quite significant but usually fade away as interband interactions increase. Anisotropy often leads to similar modifications in single-band superconductors, which is why the distinction of two-band and anisotropy effects is usually difficult, particularly when the temperature dependence of the quantities is examined. In contrast, the field dependent effects are more often different and thus more often allow a reliable distinction between the models. 
\end{abstract}

\maketitle

\tableofcontents

\section{Introduction}
In this article, my guideline was to work out how two-band superconductivity shows up in
experiment. To be more precise, my intention was to provide a literature
overview on a multitude of effects by which  conjectural two-band materials
have been identified. Furthermore, it was also my intention to compare these
effects with those expected from other models, such as the anisotropic
single-band  and the d-wave scenario and thus to find out which effects are
unique to and hence unambiguously mark out two-band superconductivity. 

Two-band superconductivity has been a prominent issue since the discovery of MgB$_2$ in 2001 \cite{Nag01a}, though it was sometimes considered for describing the unconventional behavior of materials before. In MgB$_2$ numerous experiments revealed unconventional results such as a positive curvature of the upper critical field near the transition temperature, a shoulder in the specific heat at intermediate temperatures, and different anisotropies for different quantities. Those significant deviations from the expected (standard) single-band behavior initiated profound theoretical investigations of the two-band scenario. Accordingly, MgB$_2$ has formed today's understanding of two-band superconductivity, and many effects have mainly been investigated for parameters close to those of MgB$_2$. Today the focus has shifted to the iron-based superconductors. Though there is little doubt that these materials are two- or even multi-band superconductors, their unconventional behavior is partly quite different from that of MgB$_2$, which may be due to quite different Fermi surfaces and gap structures. Theoretical work on the iron-based materials is yet largely missing but is anticipated to provide further impact on our knowledge of multi-band effects. Beside these two major players, other materials, partly known much longer than MgB$_2$, have been found to resemble MgB$_2$ in some properties and have thus been supposed to be two-band materials. 

Unconventional behavior does not necessarily mean two-band behavior but can also emerge from strong anisotropy in single-band materials, from gap-symmetries different from s-wave, from a second superconducting phase, and from other effects. It is therefore a major task of this paper to point out that features originating from two-band effects are often similar to those from other effects, which often makes distinguishing the scenarios difficult. This holds, for instance, for quantities that mainly depend on the variation of the gap values or the Fermi velocities, as such a variation is available in both, the anisotropic single-band and the two-band scenario. Nevertheless, differences are expected to show up in some properties, particularly when they are examined as a function of magnetic field, since in two-band materials superconductivity is often suppressed in one of the bands at sufficiently high magnetic fields.

Apart from the introduction (i) and the final summary (v), the review contains three main sections: (ii) multi-band models, where some theoretically predicted results calculated by applying different models are summarized, (iii) experiments, where two-band effects observed in some selected experiments are reviewed, and (iv) materials, where most of the presumed two-band superconductors are listed.

\section{Multi-band models \label{Sec:Models}}

This section is devoted to the theoretical description of two-band
superconductivity by applying Ginzburg Landau theory, the separable model, BCS,
and Eliashberg theory. We are mainly interested in the predictions of how
experimentally accessible magnetic and thermodynamic superconducting properties
are influenced by the presence of a second band.  Properties of the anisotropic
single-band and the d-wave model are partly discussed for comparison. The
two-band model is assumed to consist of two basically independent Fermi
bands, which differ in their properties. In particular, the coupling strength
and thus the energy gaps should be distinctly different for the two bands.
Moreover, different Fermi velocities, anisotropies, gap symmetries, impurity
scattering rates, densities of states, etc. are possible. The two bands are
connected by interband coupling and impurity scattering. Because of the
interaction, the system should have only one transition temperature ($T_{\rm
c}$), one upper critical field ($B_{\rm c2}$), etc., and the values of the
common characteristic properties should differ from both values of the single
(unconnected) bands. The formal extension to a multi-band model with more than
two bands and connections between all of them is a straightforward process. In
experiment, such conditions may be met by materials, whose Fermi surface is
crossed by two or more unconnected bands on which the energy gaps are
distinctly different. In contrast, the anisotropic single-band model is assumed
to consist of one band on which the energy gap and the Fermi velocity vary
rather continuously between a maximum and a minimum value. This may fit to materials, in which the superconducting parts are connected.

\subsection{Ginzburg Landau theory}

Let us start with the two-band Ginzburg Landau model \cite{Til64a}  which
provides insight into superconducting parameters directly accessible to
experiment. Deriving the free energy Ginzburg Landau functional from a two-band
BCS model, Zhitomirsky and Dao \cite{Zhi04a} found
\begin{equation}
 	\label{eqn:GL}
	F_{\rm{GL}} = \int {\rm{d}}^3r\, \left( f_1 + f_2 + f_{12} +
\frac{\vec{B}^2}{2 \mu_0}\right)
\end{equation}
with
\begin{eqnarray}
 	\label{eqn:GL_fi1}
	f_j &=& \alpha_j |\Psi_j|^2 + \frac{1}{2}\beta_j |\Psi_j|^4 + K_j
\left|\left(
\vec{\nabla}
+ {\rm i} \frac{2 \pi}{\Phi_0} \vec{A}\right) \Psi_j\right|^2  \\
	 \label{eqn:GL_fi2}
  f_{12} &=& \sigma (\Psi_1^\star \Psi_2 + \Psi_2^\star \Psi_1)
\end{eqnarray}
with $j = 1,2$ the index of the individual bands, $\vec{B}$ the magnetic
induction, and $\vec{A}$ the corresponding vector potential; $\mu_0$ ($= 4 \pi
\times 10^{-7}$\,T\,m\,A$^{-1}$) denotes the vacuum permeability and $\Phi_0$
($\simeq 2.07 \times 10^{-15}$\,V\,s) the magnetic flux quantum. $\Psi_j$ is the
order parameter of the superconducting state of each band. Finally, $\alpha_j$,
$\beta_j$, and $K_j$ are numerical coefficients expressed by BCS quantities. It was shown that $\alpha_j$ deviates from the single-band definition by a constant that depends on the inter- and intraband coupling strengths (cf. also \cite{Kog11a}). Possible dependencies of the quantities on the spatial coordinates and temperature are omitted in the equations. 

Expression (\ref{eqn:GL_fi2}) represents a Josephson-like interband interaction
with coupling parameter $\sigma$. This term ensures minimal coupling and is
responsible for a common single transition temperature of the system, usually
larger than the transition temperatures of both individual bands (i.e. the
transition temperatures of the two bands when coupling is not present), and a single value
of the critical fields. Several groups added additional coupling terms
\cite{Ask06a,Ask02a,Ask02b} or extended the Ginzburg Landau functional
\cite{Sha11a, Kom11a} to higher orders of $(1 - T/T_{\rm c})$ to extend the
validity of the model to lower temperatures. In \cite{Sil12a}, the applicability
of the two-band Ginzburg Landau model with the minimal coupling term was studied
by comparing the results with those from the Eilenberger equations. The authors
found reliable agreement from the transition temperature to quite low
temperatures, particularly in case of weak interband coupling and/or when
adjustable parameters (i.e. for $\alpha_j$, $\beta_j$, and $K_j$) are used
instead of the microscopically derived values. The issue of the validity of the
two-band Ginzburg Landau theory for MgB$_2$ was also addressed in \cite{Dao05a}, where an interval from about 30\,K to $T_{\rm c}$ ($\sim 39$\,K)
was derived.

If interband coupling is weak in comparison to the intraband properties, traces
of the individual bands are likely to be observable in various thermodynamic or
magnetic properties. In such a case, experiment may allow us to uncover two-band superconductivity. Different effects of two-band superconductivity on
such experimentally accessible properties have been calculated within the
Ginzburg Landau model and were shown to resemble some of the unconventional
experimental results. Note, however, that not only two-band superconductivity
leads to unconventional behavior, as will be shown in the next subsection.

In reference~\cite{Eis05a}, the two-band Ginzburg Landau equations were
expressed in terms of more conventional quantities such as the condensation
energies of each band and the magnetic penetration depths. Fitting this model to
experimental data of MgB$_2$ allowed the authors to determine various Ginzburg
Landau properties of the individual bands and of the total system. It was shown
that the superconducting properties of the band with the smaller gap are
suppressed by a rather low magnetic field, which is the reason for significant
deviations of the in-field behavior of various properties from the single-band
behavior. The field above which superconductivity is significantly suppressed in one of the bands is often assumed the upper critical field of this band, but due to interband coupling, traces of superconductivity should be available in this band up to the global upper critical field.  Further properties derived from Ginzburg Landau theory are the upper
and lower critical fields, the characteristic lengths, and the corresponding
anisotropies as well as thermodynamic quantities \cite{Zhi04a,
Ask06a,Udo06a,KAr10a,Dao05a,Min07a}. For example, it was demonstrated that a
positive curvature of the upper critical field near the transition temperature
may result from two-band superconductivity \cite{Zhi04a,Ask06a,Udo06a}. This
would also result in a temperature dependent anisotropy, which could be quite
different from that of the magnetic penetration depth \cite{Ask06a}.
Furthermore, a second band was found to reduce the jump of the specific heat
\cite{Ask06a} at the transition temperature, and aspects of point defect
scattering were investigated \cite{Min07a}.

In summary, the two-band Ginzburg Landau theory is a typical example of a
simple two-band model, namely the sum of two independent single-band
and one or more coupling terms. The coupling term is responsible for a single
common transition temperature, usually different from the two single-band
values, and similar effects on other properties. 

\subsection{\label{sec:sepmodel} Separable model}

In this section, I will discuss the separable model, which introduces anisotropy
in a very simple way. Most interestingly, it helps to clarify whether
modifications caused by a second band can be distinguished from those by
anisotropy. We will see that two-band and anisotropy effects can be described
by an equivalent set of equations within this model, which indicates that
discriminating the two models by the temperature dependence of various
experimental results may be difficult in most cases.

The separable model was originally introduced into BCS theory by Markovitz and
Kadanoff
\cite{Mar63a}, who defined the anisotropic pairing potential via 
\begin{equation}
 	\label{eqn:SEP:Vk}
	V_{\vec{k} \vec{k}'} = (1 + a_{\vec{k}}) V (1 + a_{\vec{k}'}),
\end{equation}
where $V$ denotes the isotropic, i.e. averaged, value of the
BCS coupling potential; $\vec{k}$ and $\vec{k}'$ are the momentum vectors of the
electrons (or quasiparticles) before and after a scattering event, and
$a_{\vec{k}}$
specifies the anisotropy.
We assume
\begin{equation}
 	\label{eqn:SEP:akFS}
	\langle a_{\vec{k}} \rangle = 0,
\end{equation}
so that the Fermi surface average of this function vanishes. The concept
was later adapted for Eliashberg theory \cite{Dam81a} by defining the
anisotropic Eliashberg spectral function - $\alpha^2 F(\omega)_{\vec{k}
\vec{k}'}$ - in exactly the same way as the BCS potential, and hence the
anisotropic electron - phonon interaction function is given by
\begin{equation}
 	\label{eqn:SEP:CS}
	\lambda_{\vec{k} \vec{k}'}(\omega) = (1 + a_{\vec{k}}) \lambda(\omega)
(1 + a_{\vec{k}'}),
\end{equation}
with $\lambda(\omega)$ the Fermi surface average of $\lambda_{\vec{k}
\vec{k}'}(\omega)$. Setting $\omega = 0$ yields the coupling strength
$\lambda(0) = \lambda$.

Dividing the Fermi surface into two sheets, on which $a_{\vec{k}}$ is constant,
e.g. $a_1$ on sheet 1 and $a_2$ on sheet 2, with $a_1 = -a_2$ due to equation
(\ref{eqn:SEP:akFS}), leads to the simplest form of the model.
Additionally, we can assume different electronic densities of states (DOS) with
relative weights $n_i$ ($n_1 + n_2 = 1$) on each sheet, and hence
\begin{equation}
 	\label{eqn:SEP:ai}
	a_i = (-1)^{i+1}\sqrt{\langle a_{\vec{k}}^2 \rangle n_j / n_i},
\end{equation}
where $i, j = 1, 2$ or $2, 1$. The anisotropy is thus fully determined by a
single parameter, namely $\langle a_{\vec{k}}^2 \rangle$.

The anisotropic electron - phonon coupling function can be described by a series
of so-called Fermi surface harmonics \cite{All76a}. Aborting this expansion
after the first term, i.e. at zeroth order, leads to the separable model.
Accordingly, the separable model may not provide a correct quantitative
description, except for very small anisotropies, but it is plausible to assume
that the most significant features of the system are reliably described.

Some properties, as for example the upper critical field, additionally depend 
on the Fermi velocity and its anisotropy, which is taken into account by
\begin{equation}
 	\label{eqn:SEP:FV}
	v_{{\rm F}, \vec{k}} = (1 + b_{\vec{k}}) v_{\rm F},
\end{equation}
where $b_{\vec{k}}$, the anisotropy function, is defined in the same way as
$a_{\vec{k}}$, and $v_{\rm F}$ is the average of the Fermi velocity. To analyze
different field orientations, the average and the corresponding anisotropy
should be taken from the relevant plane, e.g. that perpendicular to the applied
field in case of calculating the upper critical field. Again, it is most simple
to assume $b_{\vec{k}}$ to be constant on the two sheets, $b_1$ and $b_2$, and
to be defined in the same way as $a_i$ in relation (\ref{eqn:SEP:ai}) by using a
single anisotropy parameter $\langle b_{\vec{k}}^2\rangle$; but it should be noticed that the signs of $a_i$ and $b_i$ are opposite in most superconductors (i.e., a large
Fermi velocity implies a small coupling strength at the same part of the Fermi
surface and vice versa).

Now I come to the point that explains our interest in the separable model,
namely its possible interpretation as a special case of the (isotropic) two-band
model \cite{Ent76a,Lan92a,Pro87a,Shu98a}. In the simple two-band model, the
Fermi surface is divided into two spherical surfaces usually having different radii. Each
band is determined by an independent set of parameters, including a constant
value of the coupling strength $\lambda_i$ ($i = 1, 2$) and of the Fermi
velocity $v_{{\rm F},i}$ ($i = 1, 2$). Additionally, interband coupling
parameters $\lambda_{ij}$ ($i \neq j$) are to be defined. Accordingly, the same
set of parameters as in the separable model of the anisotropic single-band
superconductor is used, and the same set of equations has to be solved. If the
two-band quantities are fixed by the relations introduced for the anisotropic
single-band model, namely equations (\ref{eqn:SEP:CS}) and
(\ref{eqn:SEP:FV}), both models will lead to the same results. 

\begin{figure}
    \centering
    \includegraphics[clip, width = 8.5cm]{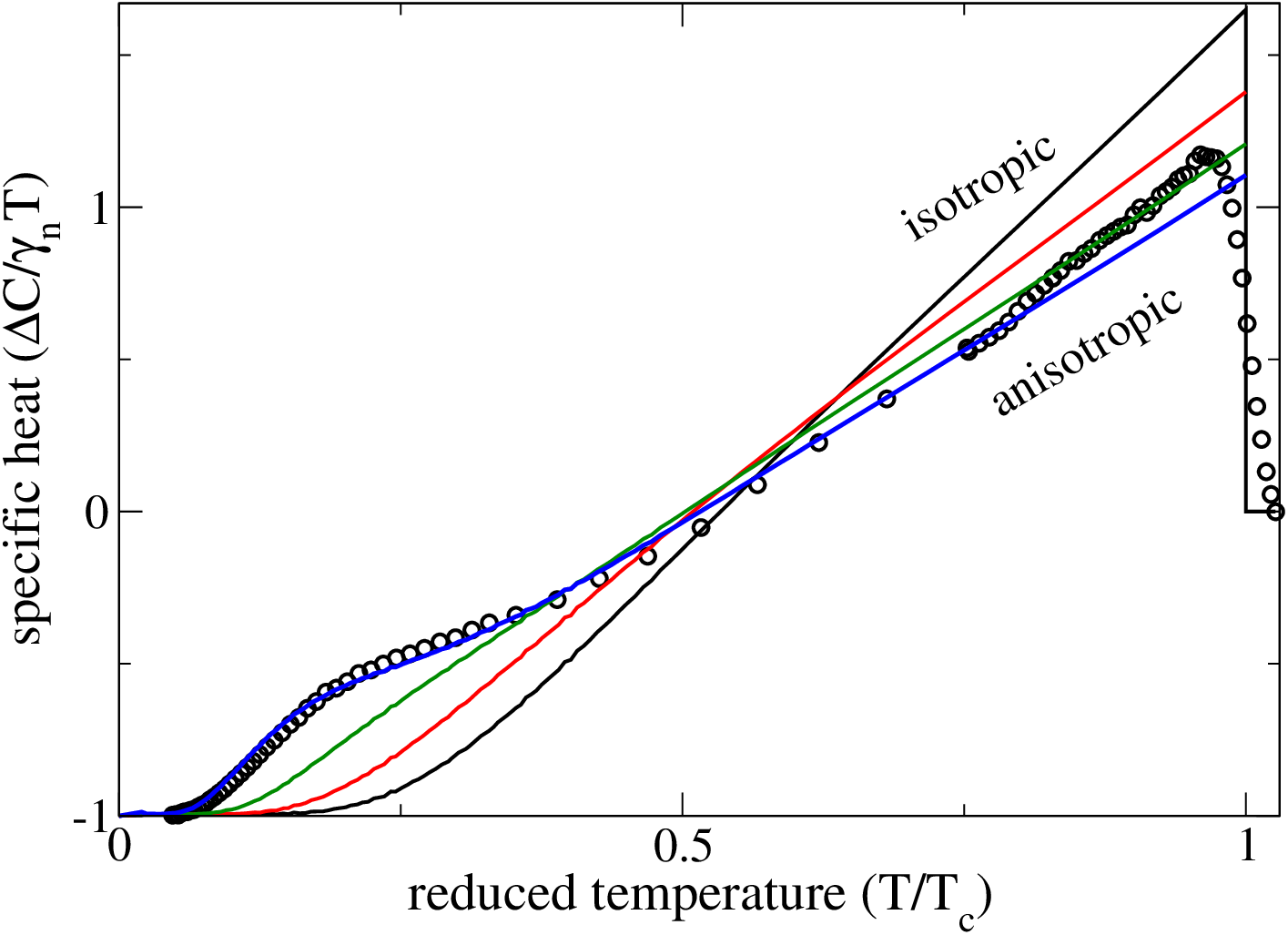}
   \caption{\label{SepModel1} The specific heat (difference) of MgB$_2$,
normalized by $\gamma_{\rm n} T$, from experiment (symbols \cite{Bou01a}; data provided by F. Bouquet) and
from separable model calculations for different anisotropies (solid lines
\cite{Zeh03a}). The lines illustrate how increasing the anisotropy of the
(electron-phonon) interaction function from $\langle a_{\vec{k}}^2 \rangle$ = 0
(isotropic case) to $\langle a_{\vec{k}}^2 \rangle$ = 0.1, 0,2, and 0.3 changes
the specific heat. In the single-band case, the anisotropy refers
to different values in one band, in the two-band case, to the
different coupling strengths of the two different spherical bands. The
averaged coupling strength, $\lambda$, is 0.61 in all cases, the density of
states weights are 0.5, and the interband coupling strengths are equal for
the single-band and the two-band model. At the transition temperature, the
specific heat jump decreases from 1.65 to 1.38, 1.21, and 1.11; at low
temperatures the curves start to grow faster when we increase the anisotropy. A
low-temperature shoulder emerges only for the highest anisotropy.}
\end{figure}

\begin{figure}
    \centering
    \includegraphics[clip, width = 8.5cm]{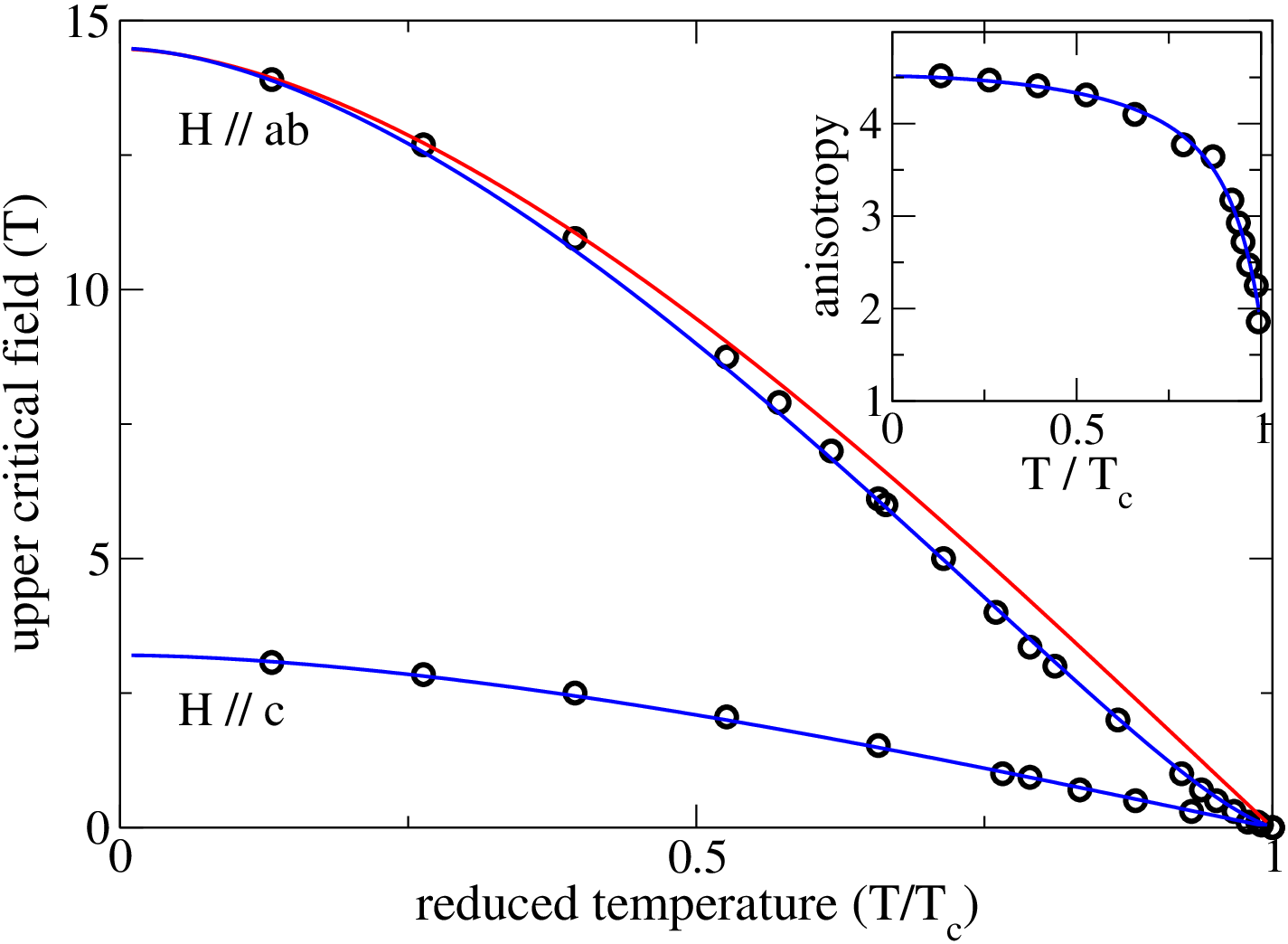}
   \caption{\label{SepModel2} The upper critical field of MgB$_2$ (symbols)
compared with separable model calculations for different Fermi velocity
anisotropies (solid lines \cite{Zeh03a}). For $H \parallel ab$, the upper line
refers to an isotropic Fermi velocity and the lower line, which matches the
experimental data, to an anisotropic one ($\langle b_{\vec{k}}^2 \rangle$ =
0.4). For $H \parallel c$ the line was calculated for almost isotropic behavior
($\langle b_{\vec{k}}^2 \rangle$ = 0.03). In the single-band case, the
anisotropy refers to different values in one band and in the two-band case, to
the different Fermi velocities of the two bands. For all lines, the relative density of states weights are 0.5 and the coupling function
anisotropy given by $\langle a_{\vec{k}}^2 \rangle$ = 0.3 (cf. figure
\ref{SepModel1}). The experimentally observed upward curvature at high
temperatures for fields parallel to $ab$ emerges and becomes more prominent with
increasing Fermi velocity anisotropy. As a result, the upper critical field
anisotropy, shown in the inset, becomes temperature dependent.}
\end{figure}

Note that the two-band model provides an additional parameter, the interband
coupling strength, which is fixed in the single-band model by equation
(\ref{eqn:SEP:CS}) but can be varied freely in the two-band model.
Nevertheless, the separable model was found to match experimental data not
only of anisotropic single-band but also of two-band superconductors quite well
\cite{Web91a,Man01a,Zeh03a}. This was, for instance, demonstrated for the
transition temperature and the thermodynamic critical field of the two-band
superconductor MgB$_2$ in \cite{Zeh03a}. Moreover, very good
agreement was also achieved for the temperature dependence of the
electronic specific heat, including the reproduction of
the specific heat jump at the transition temperature,
which is much smaller than anticipated from BCS theory, and the
unconventional low-temperature behavior with its shoulder at
about 7\,K (see figure \ref{SepModel1}). Using the second anisotropy parameter
$\langle b_{\vec{k}}^2\rangle$ and Fermi velocities from literature allowed
calculating the upper critical fields for the two main crystallographic axes,
resulting in the well known upward curvature for one of the field directions and
the corresponding temperature dependence of the anisotropy in excellent
agreement with experiment (see figure \ref{SepModel2}).

The separable model is of particular use for analyzing some basic effects of
anisotropy \cite{Zeh03a,Man01a}. In case of the two-band model, the anisotropy
refers to the difference of the two constant values on the two spherical Fermi
bands, and in case of the single-band model, to the variation within one band.
Enhancing the anisotropy of the coupling function increases the transition
temperature, for instance by a factor of roughly 2 in MgB$_2$ \cite{Zeh03a}. As
for the specific heat (cf. figure \ref{SepModel1}), the jump at the transition
temperature decreases when the anisotropy grows, whereas it increases when the
mean coupling strength grows. A larger anisotropy weakens
the coupling strength on parts of the Fermi surface, which makes the electronic
specific heat grow faster with temperature at low temperatures, while a stronger mean coupling, which results in larger energy gaps, makes it grow more slowly. A shoulder or kink at low
temperatures, which reflects the weak coupling parts, emerges only for a rather
high anisotropy. The situation may be somewhat different in the two-band model
due to the possibility of varying  the interband coupling strength
independently, and hence the shoulder may emerge at a smaller or larger anisotropy.
Furthermore, anisotropy reduces the thermodynamic critical field and slightly
changes its temperature dependence, which is nevertheless still close to the
parabolic behavior. In contrast to the specific heat and the critical field,
which are thermodynamic properties, the upper critical field (cf. figure
\ref{SepModel2}) depends on the field orientation via the Fermi velocity
anisotropy. Both, the coupling and the velocity anisotropy enhance the upper
critical field, but they affect its temperature dependence in a rather contrary way. The
Fermi velocity anisotropy perpendicular to the field is responsible for the
upward curvature at high temperatures. This effect becomes more prominent
and extends to lower temperatures as the velocity anisotropy grows but is
moderated as the coupling function anisotropy  grows.

I conclude this section by discussing the effects of non-magnetic point
impurities. In the anisotropic single-band model, such impurities reduce the
transition temperature and at the same time increase the slope of the upper
critical field at the transition temperature (which will enhance the upper
critical field at low temperatures if the suppression of the transition
temperature is not too large). For the two-band model, we may define different
impurity scattering rates for intra- and interband scattering. In the isotropic
two-band model, intraband scattering increases the upper critical field slope
but does not affect the transition temperature, as known from the Anderson
theorem; only interband impurity scattering reduces the transition temperature.
Accordingly, the upper critical field should be modifiable almost independently from
the transition temperature, which might explain the observation of quite
different upper critical field values in MgB$_2$ samples with equal transition
temperature \cite{Eis07a}.

In conclusion, though the separable model is a very simple model for the
anisotropy, significant deviations from the isotropic (BCS) behavior can nicely be
described, and the results often match the experimental data very well. This suggests that
many of the unusual effects, found for instance in MgB$_2$, are rather general
for strongly anisotropic superconductors, and do not heavily rely on the
particular shape and properties of the Fermi surface but rather on the mean
values of the anisotropies. Because the two-band and the anisotropic single-band
model can be defined by the same set of parameters, anisotropy and two-band
effects are expected to modify the temperature dependence of many
superconducting properties in a similar way, making it difficult to distinguish between
the models by experiment. As for the specific heat, increasing  the anisotropy
of the coupling function reduces its jump at the transition temperature, leads
to a faster grow at low temperatures, and eventually generates a shoulder-like
behavior at low temperatures. The curvature of the upper critical field becomes
positive at high temperatures as a consequence of Fermi velocity anisotropy in
this model. 

\subsection{\label{sec:BCS} BCS theory and quasi-classical equations}

Two-band superconductivity was first formulated within BCS theory \cite{Suh59a,
Mos59a}
using the following Hamiltonian
\begin{equation}
 	\label{eqn:BCS:1}
	\hat{\mathcal{H}} = \hat{\mathcal{H}}_1 + \hat{\mathcal{H}}_2 +
\hat{\mathcal{H}}_{12} +
\hat{\mathcal{H}}_{21}
\end{equation}
with
\begin{eqnarray}
 	\label{eqn:BCS:2}
	 \hat{\mathcal{H}}_i &=& \sum_{\vec{k}, \sigma}  \epsilon_{\vec{k}i}
\hat{c}^\dagger_{\vec{k}\sigma i} \hat{c}_{\vec{k}\sigma i}  - V_i
\sum_{\vec{k},
\vec{k'}} \hat{c}^\dagger_{\vec{k}\uparrow i}
\hat{c}^\dagger_{-\vec{k}\downarrow i} 
\hat{c}_{\vec{k}'\uparrow i} \hat{c}_{-\vec{k}'\downarrow i}  \\
	 \label{eqn:BCS:3}
 \hat{\mathcal{H}}_{ij} &=&  - V_{ij} \sum_{\vec{k},
\vec{k'}} \hat{c}^\dagger_{\vec{k}\uparrow i}
\hat{c}^\dagger_{-\vec{k}\downarrow i} 
\hat{c}_{\vec{k}'\uparrow j} \hat{c}_{-\vec{k}'\downarrow j}. 
\end{eqnarray}
Here, $\epsilon_{\vec{k} i}$ denotes the renormalized normal-state energy with
respect to the chemical potential of a single particle with momentum vector
$\vec{k}$ in band $i$ ($i = 1, 2$); $\hat{c}^\dagger_{\vec{k}\sigma i}$ and
$\hat{c}_{\vec{k}\sigma i}$ are the corresponding creation and annihilation
operators; $\sigma$ denotes the spin direction, which can be up ($\uparrow$) or
down ($\downarrow$), and $V_i$ the effective coupling potential of the BCS
theory. Definition (\ref{eqn:BCS:3}), where $i,
j$ = 1,2 or 2,1 and $V_{12}$ and $V_{21}$ are the interband coupling
potentials, introduces the interband effects.

\begin{figure}
    \centering
    \includegraphics[clip, width = 8.5cm]{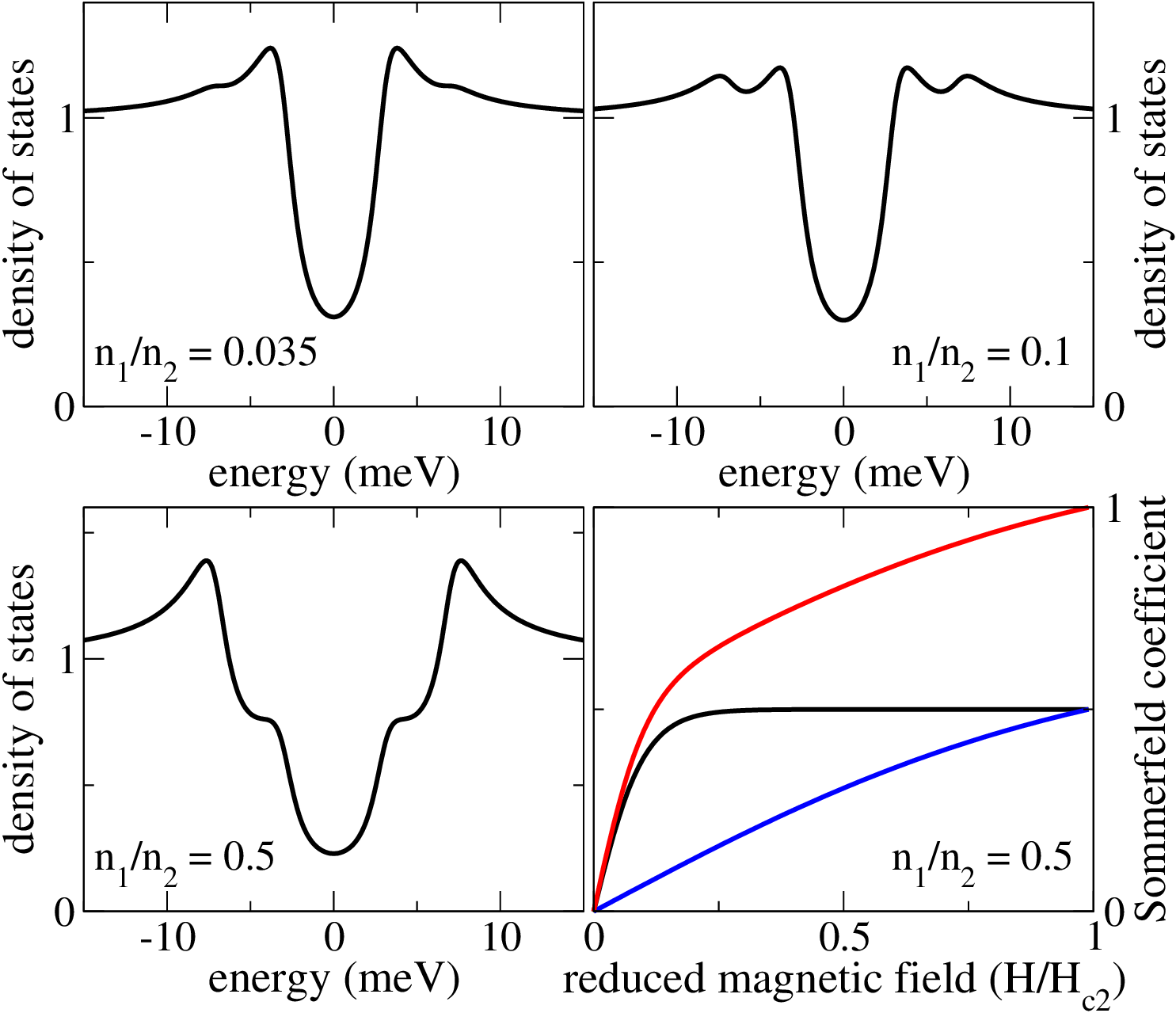}
   \caption{\label{Fig:BCS} Schematic representation of the quasiparticle density of
states around the Fermi level as a function of energy (upper and left panels)
and the Sommerfeld coefficient as a function of magnetic field (bottom right
panel) in two-band superconductors. The quasiparticle spectra show $n_1 \rho_1$
$+$ $n_2 \rho_2$, where $\rho_i$ are single-band BCS spectra and $n_i$ the
corresponding weights. For all three spectra, the energy gaps are $\Delta_1$ =
7\,meV, $\Delta_2$ = 3\,meV, and the broadening parameters $\Upsilon_1$ =
$\Upsilon_2$ = 1\,meV (cf. equation~\ref{eqn:gap:2}). The three panels
illustrate how the outcome of a tunneling spectroscopy measurement changes with
the weights of the two bands, namely from a single peak plus a shoulder (on both
sides of the Fermi level), which may barely be visible, at a higher energy when
only a small part of the tunneling current comes from the band with the larger
gap (top left panel, $n_1 / n_2$ = 0.035), to a two-peak structure when we
increase the contribution from the large band (top right panel, $n_1 / n_2$ =
0.1), to a structure in which the larger gap dominates and the lower one appears
only as a shoulder (bottom left panel, $n_1 / n_2$ = 0.5). The Sommerfeld
coefficient, shown in the bottom right panel, can be acquired from specific heat
measurements and is proportional to the volume averaged density of states in the
mixed-state of a superconductor. The lower line indicates the field dependence
of the Sommerfeld coefficient of the band with the larger upper critical field,
which should be roughly linear in case of an isotropic s-wave material. The
middle line refers to the second band, in which superconductivity is assumed to
be suppressed and the coefficient thus becomes almost constant at higher
fields. Accordingly, the overall Sommerfeld coefficient, shown by the upper
line, is governed by a large slope at low and a smaller slope at high magnetic
fields.}
\end{figure}

Applying the standard techniques from single-band BCS theory allows evaluating
various properties of the two-band system, such as the density of states,
presented in figure~\ref{Fig:BCS}, the temperature dependence of the energy
gaps, and the transition temperature, which was shown to be higher than the
values of each single-band \cite{Bus03a}. Fitting two-band BCS theory to
experimental specific heat data worked well for MgB$_2$
\cite{Kri03a,Zhi04a,Kon04a,Mis05a}, revealing that the jump at the transition
temperature is smaller than in the single-band case. Furthermore, the effects on
the  superfluid density and the magnetic penetration depth were reported to agree
with the experimental data quite well. \cite{Bus07a,Bus10a} 

Aspects of anisotropy in single-band s-wave superconductors were construed in
Refs. \cite{Haa01a,Pos02a} and compared with experimental data on MgB$_2$. It
was shown that the anisotropy of the gap function results in a two-peak
structure in the energy dependence of the tunneling conductance, similarly as anticipated in a two-band system. Also thermodynamic properties, such as the specific heat and the
critical magnetic field, could be brought into reliable agreement with
experimental data on MgB$_2$. Consequently, the modifications due to
anisotropy are again similar to those derived from two-band models.

More complex situations including finite magnetic fields and non-spherical Fermi
surfaces can be studied by using the quasi-classical approximation
of the Green's function, i.e. by solving the corresponding Eilenberger
\cite{Eil68a}
or - in case of strong disorder (dirty limit) - the Usadel \cite{Usa70a}
equations. The Eilenberger equations are derived from Gorkov's Green's function
approach to the BCS theory and simplify calculations at finite  magnetic fields.

Several groups applied the approach to multi-band superconductivity. They
showed that the thermodynamic properties \cite{Kus04a,Mug05a,Kog09a} were modified in a
similar way as already mentioned for the previously discussed models. Interband
scattering by non-magnetic impurities was found to reduce the transition
temperature much faster when the interband coupling strength is repulsive
\cite{Kog09a} (i.e. $\lambda_{12} < 0$, implying that the gaps have opposite
sign on the different sheets; the $s_{\pm}$ scenario) than in the ordinary case
($\lambda_{12} > 0$).

The energy gap and the electronic density of states near or at the Fermi level
can be calculated as a function of applied magnetic field and compared with
experimental data from tunneling spectroscopy, which reveals the energy gap and
the local density of states at a certain point of the sample surface, or with
the Sommerfeld coefficient ($\gamma$), which is proportional to the volume
average of the density of states and can be determined from specific heat
measurements. Assuming the volume averaged density of states to be proportional to the
total volume of the normal-conducting vortex cores leads to a linear field
dependence of the Sommerfeld coefficient in an isotropic s-wave superconductor,
which is however changed to a concave function in real materials due to the field
dependence of the coherence length and the overlap of vortex cores at high
magnetic fields \cite{Ich99a,Nak04a,Son06a}. Above the upper critical field, the
Sommerfeld coefficient matches the normal-conducting value which is usually
constant. Similarly, $\gamma_{\rm n} \propto B^{0.5}$ is expected for an
isotropic d-wave material \cite{Vol93a,Ich99a}.

Two-band effects were mainly calculated for MgB$_2$ by considering one isotropic
band ($\pi$), having a small gap and a low upper critical field, and one
anisotropic band ($\sigma$), having a large gap and a high upper critical field
\cite{Mug05a,Ich04a,Nak04a}. The slope of the Sommerfeld coefficient was shown
to be rather steep at low fields due to the dominating influence of the
$\pi$-band, the contribution of which becomes saturated, i.e. constant, above
its band upper critical field value, and thus to become flatter at higher fields
when the $\sigma$-band dominates, which is illustrated in figure~\ref{Fig:BCS}.
Interband coupling effects make the overall curve of the systems smoother,
which may roughly (but not very well) be characterized by $\gamma(B) \propto
B^{\alpha}$ with $\alpha < 1$. The density of states of such a system would be
rather isotropic at low fields (when dominated by the isotropic $\pi$-band) but
strongly anisotropic at high fields (due to the anisotropic $\sigma$-band). 

It was further reported that even when the density of states of the smaller
gap-band becomes saturated above a particular magnetic field the corresponding
energy gap can be finite over the whole field range, i.e. both gaps close at the
same field due to interband coupling \cite{Mug05a,Ich04a}. As for the energy
dependence of the local density of states, it was indicated that the peak
associated with the smaller gap may vanish at higher applied fields due to
different band topologies \cite{Gra04a}. Similar results were obtained for the
dirty limit by solving the Usadel equations \cite{Kos03a}.

It is again interesting to compare the two-band with anisotropic single-band
results \cite{Ich04a}. The exponent $\alpha$ in $\gamma(B) \propto
B^{\alpha}$, which should be roughly one for isotropic superconductors, was
shown to become smaller with increasing anisotropy.
Hence the anisotropy effects are again similar to the two-band effects. Indeed,
reliable agreement between the experimental data from the two-band
superconductor NbSe$_2$ and calculated results from the anisotropic single-band
model was demonstrated. Additional minor modifications are expected from
impurity scattering \cite{Tew03a}.

In summary, the quasi-classical approximation makes calculations at finite
magnetic fields possible. The field dependence of the Sommerfeld coefficient
should be close to linear, though slightly curved, in case of isotropic s-wave
and close to a square root behavior in case of d-wave symmetry. For a two-band
superconductor in which one band is suppressed at relatively low magnetic
fields, we expect a steep gradient at low and a flatter course at high fields
and the kink, indicating the upper critical field of the weaker band, to be
smeared out by interband coupling. A similar behavior is expected in anisotropic
single-band superconductors, where the isotropic linear dependence was
demonstrated to become more concave with increasing anisotropy. Concerning the
energy gaps, both should close at the same (upper critical) field even when the
band upper critical fields are different. The gap structure of the density of
states near the Fermi energy may change considerably with magnetic field.

\subsection{Eliashberg theory}

In the final subsection, I will summarize results from the Migdal Eliashberg
theory, which takes strong coupling effects into account. Even if the
mean coupling strength of a two-band superconductor is weak, one of the bands
could be in the strong coupling regime, making strong coupling effects relevant.
Some publications, which will be reviewed below, have provided deep insight into the influence of the anisotropy, the interband coupling strengths, and the impurity scattering rates, etc. on the thermodynamic and magnetic properties of two-band materials.

Results from the Eliashberg equations have often been found to agree with
experimental data better than those from other models and to describe the
superconducting properties of many conventional (i.e. electron - phonon
mediated) superconductors even quantitatively precisely \cite{Car90a}.
Basically, the Green's function of the system has to be calculated by
taking into account the self-energies related to the electron-phonon
interaction, which is characterized by the Eliashberg spectral function
$\alpha^2 F(\omega)_{\vec{k} \vec{k}'}$, the electron - electron Coulomb
interaction, characterized by the Coulomb pseudopotential $\mu^\star_{\vec{k}
\vec{k}'}$, and the electron - impurity interaction, usually in first Born's
approximation, specified by the mean scattering time \cite{Sca69a,
All82a,Mar08a}. The model allows us to determine all thermodynamic properties
that can be derived from the free energy, and other properties, such as the
transition temperature, the energy gap, the density of states, etc. Except for the
upper critical fields, which additionally depend on the Fermi velocity
$\vec{v}_{{\rm F}, \vec{k}}$, magnetic field effects are usually inaccessible.

Two-band effects are naturally included when we solve the complete anisotropic form of the equations. To do so, the local values of the input parameters, at least of $\lambda_{\vec{k} \vec{k}'}(\omega)$, which can be determined from $\alpha^2 F(\omega)_{\vec{k} \vec{k}'}$, have to be known on the whole Fermi surface. To avoid that, the Fermi surface topology is often simplified. As already mentioned, the Fermi surface anisotropy may be approximated by the separable model or by the corresponding two-band model with two spherical bands, specified by two densities of states, four (frequency dependent) coupling functions, four pseudopotentials, and the impurity scattering rates. Moreover, cylindrically, elliptically, and similarly shaped Fermi bands have been used.

Choi et al. \cite{Cho02a,Cho02b} solved the fully anisotropic Eliashberg
equations for the two-band superconducting material MgB$_2$. The $\vec{k}$,
$\vec{k}'$ and phonon energy $\omega$ dependence of $\alpha^2 F(\omega)_{\vec{k}
\vec{k}'}$ was acquired from ab-initio methods, while the Coulomb
pseudopotential $\mu^\star$, supposed to have only minor influence on the
superconducting properties and hence taken isotropic, was chosen by adjusting the
results to the correct value of the transition temperature. Using these parameters,
they were able to calculate the energy gap distribution, the specific heat
behavior, and the isotope effect in excellent quantitative agreement with
experiment. It was concluded that the anisotropy and the two-band nature are
responsible for the rather high transition temperature (which was approximately
twice the value of the corresponding isotropic system) and that the low energy
excitations, which come along with the smaller values within the wide range of
the energy gap distribution, lead to the low-temperature shoulder of the specific
heat.

Several other groups investigated two-band properties - mainly of MgB$_2$ - by
Eliashberg theory but applied simpler models, such as mentioned above, for the
anisotropy. Nevertheless, they were able to obtain similarly good
agreement with experimental data as with the fully anisotropic theory. This again indicates that prominent two-band effects such as the
low-temperature anomaly in the specific heat do not depend on details of the
Eliashberg spectral function \cite{Cho06a} but are common for anisotropic or
two-band superconductors that have a wide distribution of energy gap values. For
instance, taking the same mean value of the coupling strength
($\lambda$) and the peak of the Eliashberg spectral function, simplified by
an Einstein-like spectrum, at the same position as in the fully
anisotropic model of Choi et al. \cite{Cho02a,Cho02b} allowed reproducing
various experimental results on MgB$_2$ by the separable model of the
Eliashberg equations very well \cite{Zeh03a}. 

\begin{figure}
    \centering
    \includegraphics[clip, width = 8.5cm]{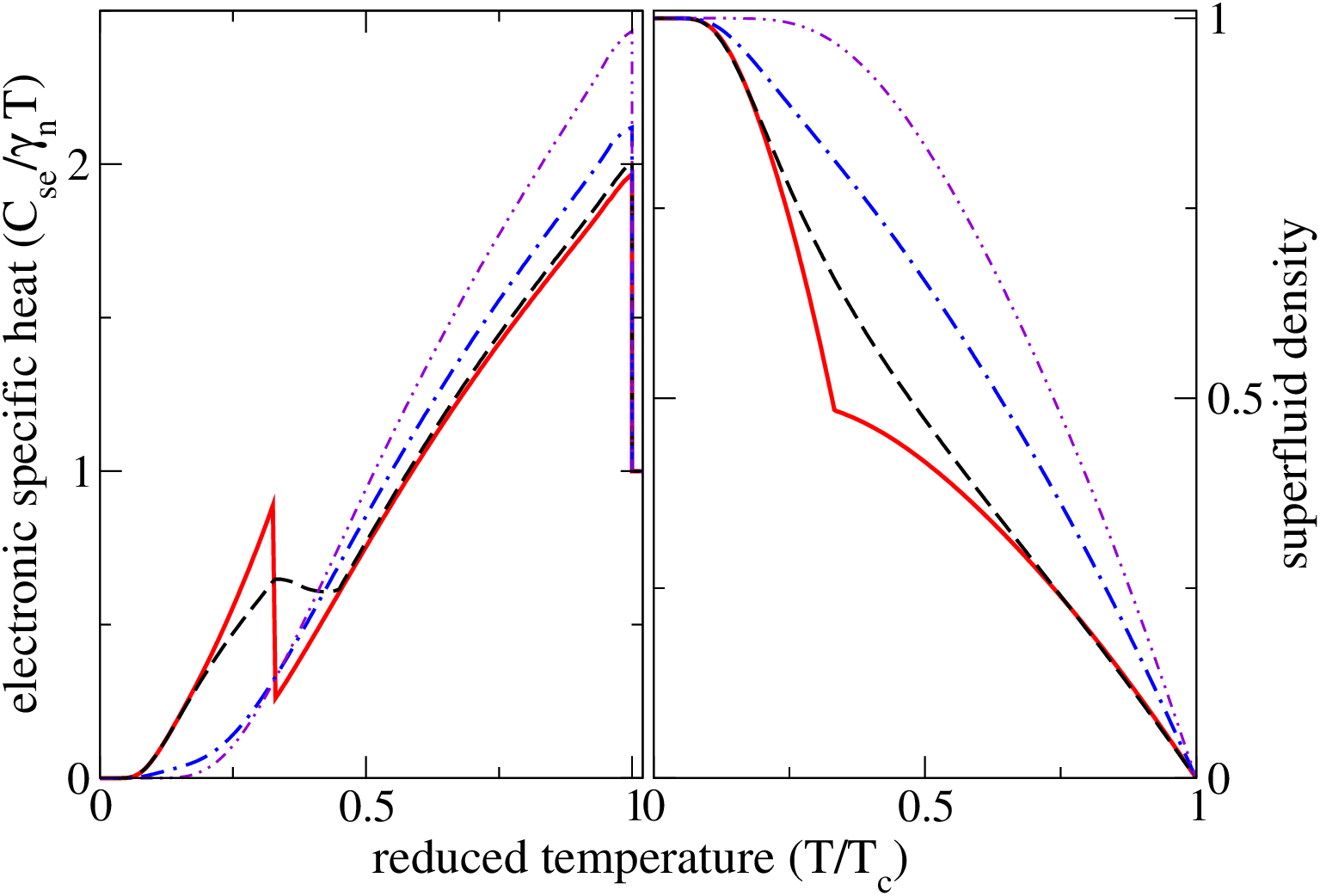}
   \caption{\label{Fig:EliashA} The electronic specific heat normalized by its
normal-conducting value (left panel) and the superfluid density (right panel) as
a function of reduced temperature. Both panels schematically illustrate the
effect of increasing the interband coupling parameter in a two-band
superconductor. For detailed numerical results see
\cite{Nic05b,Dol05a,Gol02a,Moc02a}. The solid lines show results for negligible
interband coupling, and thus equal the sum of two independent single-band
materials having different properties, which leads to a jump in the specific
heat and a sharp kink in the superfluid density where the band with the lower
transition temperature becomes normal-conducting. The effect of very small
interband-coupling is illustrated by the dashed lines and mainly consists in
washing out the sharp low-temperature transition of the uncoupled case. As
interband coupling increases further, the low-temperature anomalies, caused by
the second band, fade away, and the curves are gradually shaped towards a
conventional-like form, as shown by the dashed-dotted lines. For comparison, BCS
results are included (dashed-dotted-dotted). Note that increasing the
interband impurity scattering rate instead of the coupling constant results in
almost the same effects.}
\end{figure}
 
Diverse other effects of a second band have been investigated. For instance, the
superfluid density (which is proportional to $1 / \Lambda_x^2$, with $\Lambda_x$
the magnetic penetration depth for current flow in $x$ direction) of an
isotropic s-wave superconductor is almost constant at low temperatures, but
starts to decrease rapidly above a certain temperature. Similarly as for the
specific heat, low energy excitations, associated with the smaller gap values,
reduce the threshold temperature and may significantly change the overall
behavior from a concave towards a more linear or even convex behavior over a
large temperature range \cite{Nic05b,Gol02a,Moc02a}, see
figure~\ref{Fig:EliashA}. In case of small interband effects, a kink indicating
the weaker band may be observed at low temperatures. Strong impurity scattering
in one band, particularly if the smaller-gap-band is in the dirty limit, can
almost recover the conventional BCS behavior \cite{Gol02a}.

\begin{figure}
    \centering
    \includegraphics[clip, width = 8.5cm]{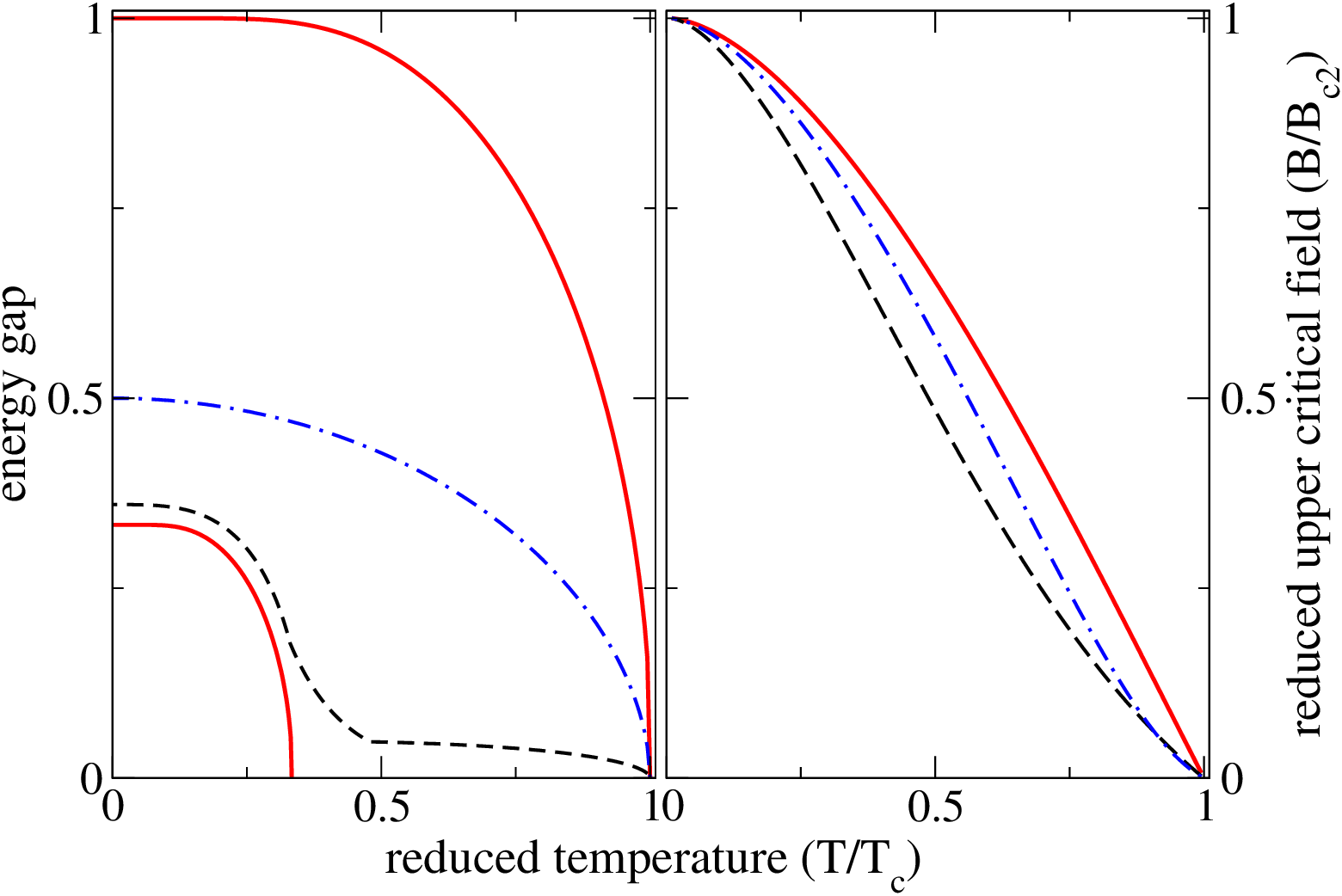}
   \caption{\label{Fig:EliashB} The energy gaps (left panel) and the reduced
upper critical field (right panel) as a function of reduced temperature in a
two-band superconductor. The left panel schematically shows the effect of the
interband coupling strength on the superconducting gaps. The solid lines refer
to the uncoupled case, and hence show two BCS gaps with different absolute values
and transition temperatures. Applying weak interband coupling washes out the
transition of the smaller gap, illustrated by the dashed line, so that both gaps
close at the same temperature. As the coupling is increased, the kink in the
smaller gap disappears gradually, and the shape of the smaller gap becomes
more conventional (dashed-dotted line). The effects on the larger gap are not so
significant and hence not shown in the panel. Applying interband impurity
scattering instead of the coupling modifies the gaps in a similar way. For
detailed numerical results see e.g. \cite{Nic05b}. The right panel indicates how
the Fermi velocity anisotropy affects the upper critical field shape. The solid
line shows the behavior in case of an isotropic Fermi velocity. The curvature of
the upper critical field, which is negative in the isotropic case, becomes
positive near the transition temperature when the velocity in one of the bands
is changed (dashed dotted line). The effect becomes more pronounced and extends
to lower temperatures as the Fermi velocity difference of the two bands
increases (dashed line). For detailed numerical results see e.g.
\cite{Man05a,Zeh03a}.}
\end{figure}

Calculations of the upper critical field reproduced the upward curvature found
in many experiments. Applying the separable model \cite{Zeh03a} showed that the
temperature at which the curvature of the upper critical field switches from
negative to positive shifts to lower values when we increase the anisotropy of
the Fermi velocity perpendicular to the applied field but to higher values when
we increase the anisotropy of the coupling function. A two-band Eliashberg model
based on MgB$_2$ with one spherical weak coupling ($\pi$) and one elliptical
stronger coupling ($\sigma$) band was studied in \cite{Man05a}. The kink
indicating the upper critical field of the $\pi$-band in the absence of
interband interaction was demonstrated to be smeared out by quite small
interband coupling parameters. The upward curvature of the upper critical field
near the transition temperature was predicted to become more pronounced when the
anisotropy of the Fermi velocities (here, the ratio of the values of the two bands) or the
off-diagonal (interband) couplings increase (cf. figure~\ref{Fig:EliashB}).
Similar effects were calculated for intraband impurity scattering. For
instance, it was reported that $B_{c2}(T)$ may show an upward curvature in
MgB$_2$ even for $H \parallel c$, for which the relevant Fermi velocities are
quite similar and thus the upward curvature is usually not observed, if the
$\sigma$-band is pushed into the dirty limit and the $\pi$-band remains clean.
All these effects usually render the upper critical field anisotropy temperature
dependent.

Further systematic investigations on diverse properties such as the energy gaps,
the specific heat, the superfluid density, and the thermodynamic critical field
deviation function, which measures how strongly the thermodynamic critical field
deviates from a parabolic temperature behavior, were reported in
\cite{Nic05b}. Using parameters roughly based on the properties known for
MgB$_2$, the authors showed that not only anisotropy but also strong coupling
could have a significant influence, so that BCS and Eliashberg results may diverge
significantly in some cases. Furthermore, anisotropy and strong coupling were
reported to lead to opposite corrections in most cases; for instance, strong
coupling increases the specific heat jump at the transition temperature, whereas
anisotropy associated with the ratio of the intraband coupling strengths,
$\lambda_{11}$ and $\lambda_{22}$, decreases it; stronger interband coupling,
however, increases the jump again. Moreover, studying the influence of the
interband coupling strength and of the interband impurity scattering rate
revealed that even a weak interband interaction significantly washes out the
transition of the smaller gap band. Effects of interband impurity scattering are
often similar to that of interband coupling. Concerning the specific heat jump, a small
impurity scattering rate may reduce the value when the transition temperature
decreases strongly, but a high intraband impurity scattering rate should always
raise the jump-height (cf. also \cite{Dol05a} for further impurity effects).
Finally, interband coupling can induce superconductivity in a band with
repulsive intraband coupling.

In conclusion, the Eliashberg equations give access to strong coupling effects
and result in a quantitatively correct description of superconducting properties
in many cases. Though any anisotropy can be implemented, simple Fermi surfaces
with spherical or elliptical shapes have proved successful in describing diverse
two-band effects. For instance, the superfluid density was found to drop faster
at low temperatures and to change from a concave to a more linear or even
convex temperature dependence with increasing intraband coupling anisotropy. The
effects on the specific heat and the upper critical field correspond to those
reported in the 'separable model' section (\ref{sec:sepmodel}). In addition, stronger 
interband coupling enhances the specific heat jump and the kink representing the
transition of the weaker band in the temperature dependence of diverse
properties is washed out. Quite similar effects were predicted for interband
impurity scattering.

\section{Experiments \label{Sec:Exp}}

The following section is considered the main part of this review. I will list
and discuss several experimental techniques that have commonly been
used for identifying two-band materials and will show the results. The selection
of the methods is rather
arbitrary, and omitted methods are not meant to be less important. Except for
the thermal conductivity and the energy gaps, I concentrate on thermal and
mixed-state properties that are somehow linked to the reversible magnetic
properties of the superconductor, namely the specific heat and the corresponding
Sommerfeld coefficient, the superfluid density, the upper critical fields, the
torque, the reversible magnetization, the anisotropy, and the field dependence
of the characteristic lengths. For each subsection, I will give a brief
introduction on the property and then review some results obtained on different
samples,
including not only two-band materials.

\subsection{Specific heat \label{sec:SH}}

Measurements of the specific heat  in MgB$_2$ have uncovered striking deviations
from the so-called standard curve (i.e. that obtained from BCS theory), which
were soon attributed to two-band superconductivity. Those remarkable features
are the jump-height at the transition, the unconventional low-temperature
behavior and a kink or shoulder in between. Later, similar effects have been
found in other materials.

The difference in the specific heat of the normal-conducting and the
superconducting state follows from the difference in the free energy density
$\Delta F$ via $\Delta C = -T \partial^2 \Delta F / \partial T^2$. The normal-conducting part $C_{\rm n}$ includes the electron $C_{\rm ne}$ and the
phonon $C_{\rm np}$ contribution, from which the linear electron part,
\begin{equation}
 	\label{eqn:SpecHeat:1}
	C_{\rm ne} = \gamma_{\rm n} T,
\end{equation}
with the Sommerfeld constant
\begin{equation}
 	\label{eqn:SpecHeat:2}
	\gamma_{\rm n} = \frac{2}{3} \pi^2 k_{\rm B}^2 z_0 (1 + \lambda),
\end{equation}
dominates at low temperatures. In equation~(\ref{eqn:SpecHeat:2}), $k_{\rm B}$
denotes the Boltzmann constant, $z_0$ the total electronic density of states
at the Fermi energy, and $\lambda$ the coupling strength. The phonon
contribution is normally modeled by $b T^3 + c T^5$, with field independent
constants $b$ and $c$, or just by $b T^3$, taken from the low-temperature limit
of Debye's theory, and is assumed to be equal in the normal- and superconducting
state; $\Delta C$ is thus the difference in the electronic parts. At very low
temperatures, the Schottky anomaly may become noticeable. 

An easily accessible characteristic value of a superconducting material is the
jump of $\Delta C$ at the transition temperature normalized by its
normal-conducting value $\gamma_{\rm n} T_{\rm c}$, i.e.
\begin{equation}
 	\label{eqn:SpecHeat:3}
	\frac{\Delta C(T_{\rm c})}{\gamma_{\rm n} T_{\rm c}} = k.
\end{equation}
Weak coupling BCS theory results in a material-independent universal value of $k
= 1.43$, which was indeed measured in aluminum, a material meeting the BCS
preconditions fairly well due to its weak coupling strength ($\lambda \simeq 0.43$ in Al) and
a near spherical Fermi surface. Both, experiment and
(Eliashberg) theory \cite{Car90a} demonstrated that this value should grow with
the coupling strength - for instance to about 2.8 in lead ($\lambda \simeq
1.55$). Although the coupling strength of MgB$_2$ is approximately 0.6, the
jump-magnitude was found to be much smaller than the weak coupling threshold, namely,
as illustrated in figure~\ref{SepModel1}, at about $1 \pm 0.2$
\cite{Bou01a,Yan01a,Wan01a}. Similar results have been reported for some other
materials. Theory was shown to predict that a second band should reduce that
value as the disparity of the (two) intraband coupling strengths increases; the same should happen when the anisotropy of a single-band material increases \cite{Nic05b,Zeh03a}.

At very low temperatures, BCS theory predicts 
\begin{equation}
 	\label{eqn:SpecHeat:4}
	C_{\rm se}(T) = A T^{-3/2} e^{-\Delta(0) / k_{\rm B} T}
\end{equation}
for the electronic specific heat of a superconductor \cite{Fet03a}, with $A$ a
temperature independent constant. Equation (\ref{eqn:SpecHeat:4}) is governed by
the exponential function at low temperatures, which reflects the probability of
destroying a Cooper-pair by the thermal energy $k_{\rm B} T$. Accordingly, with
increasing temperature the specific heat remains almost constant up to a
threshold-temperature, and then starts to grow rather quickly. The threshold value
lies
at about $0.2 T_{\rm c}$ in the BCS scenario, set by the ratio $\Delta(0) /
k_{\rm B} T_{\rm c}$ which is equal to 1.77. As that ratio becomes smaller,
$C_{\rm se}$ grows faster at low temperatures. In case of different
gap values, the threshold-temperature is determined by the smaller gaps, and
we thus expect similar low-temperature modifications for the anisotropic and the
two-band scenario. The above is only valid for (fully gapped) s-wave
superconductors, while in d-wave systems the gap nodes allow a significant
occupation of the excitation spectrum at any finite temperature, which makes 
$C_{\rm se}$ increase strongly even at very low temperatures. More precisely, 
a quadratic (at zero magnetic field) or linear temperature dependence was
predicted and confirmed by experiment \cite{Vol93a,Che98a,Bai06a,Wan11a}.

\begin{figure}
    \centering
    \includegraphics[clip, width = 8.5cm]{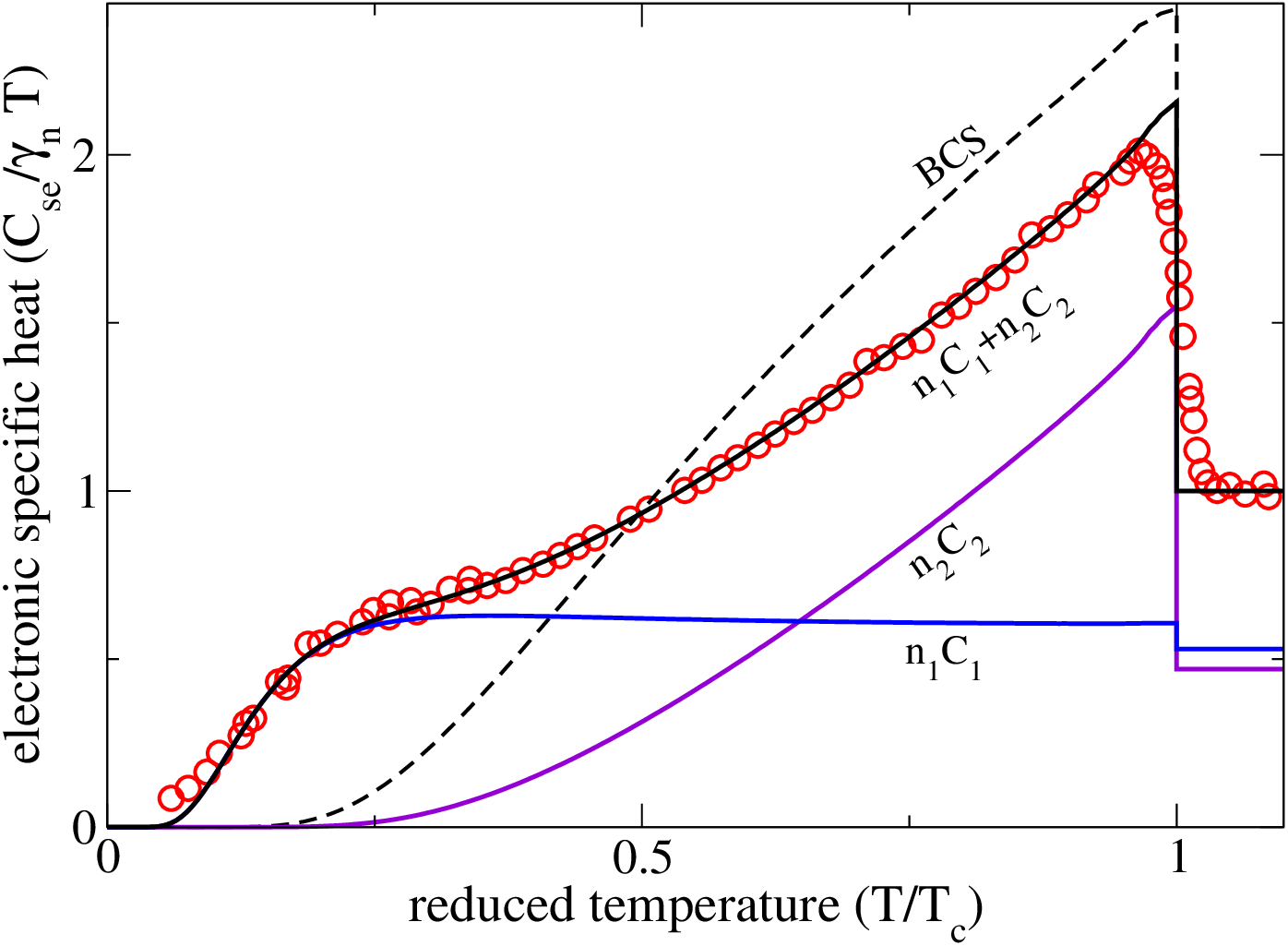}
   \caption{\label{Fig:Spec} The normalized electronic specific heat of
Lu$_2$Fe$_3$Si$_5$ (open circles) and a two-band $\alpha$-model fit
\cite{Nak08a}. The solid lines illustrate the $\alpha$-model, for which BCS
theory, though with a different gap value, is assumed. The lowest curve just
below the transition temperature ($n_1 C_1$) was calculated for $2\Delta(0) /
k_{\rm B} T_{\rm c}$ = 1.1 and a relative weight ($n_1$) of 0.53, the
second lowest  curve below the transition temperature ($n_2 C_2$) for $2\Delta(0) /
k_{\rm B} T_{\rm c}$ = 4.4 and a relative weight ($n_2$) of 0.47. The line
fitting the experimental data well is then the sum of the two others ($n_1 C_1$
+ $n_2 C_2$). Notice that the low-temperature is all but identical to
the small-gap band curve ($n_1 C_1$) while the high-temperature behavior is
determined by the second band ($n_2 C_2$). For comparison, the actual BCS
curve, for which $2\Delta(0) / k_{\rm B} T_{\rm c}$ = 3.54, is presented by the
dashed line.}
\end{figure}

We conclude that the low-temperature behavior of the specific heat of two-band
superconductors is dominated by the band with the smaller gap, while the
high-temperature behavior rather by that with the larger gap (cf.
figure~\ref{Fig:Spec}). In between, a crossover takes place which is smoothed by
interband coupling and impurity scattering; nevertheless, a kink, a shoulder, or
even a small peak may still be present, as for instance observed in MgB$_2$ at
about 7 - 8\,K.

It is obvious that describing the total temperature dependence of those
unconventional specific heat curves by conventional BCS theory would fail. A
common approach to fitting the specific heat and other thermodynamic properties
of superconductors is provided by the so-called $\alpha$-model \cite{Pad73a}, in
which the temperature dependence of the energy gap, $\Delta(T)$, follows BCS
theory, but the absolute value at $T = 0$\,K, i.e. $\Delta(0)$ (or the ratio
$\Delta(0) / k_{\rm B} T_{\rm c}$), is adjusted to fit experiment. We can often
nicely describe two-band properties as the sum of two $\alpha$-model curves by
adjusting the parameters $\Delta_1(0)$ and $\Delta_2(0)$ as well as the
corresponding relative weights of the bands. Although that model is not at all a
true two-band model, for interband interactions are ignored, it has been
successfully used for describing the specific heat in a lot of samples like that
shown in figure~\ref{Fig:Spec}. Moreover, using parameters of MgB$_2$, Dolgov et
al. \cite{Dol05a} showed that the energy gaps resulting from this procedure were
fairly close to the gaps obtained by Eliashberg calculations. Other results, as
for instance the density of states ratio of the two bands, deviated more
considerably from the Eliashberg results. In conclusion, the $\alpha$-model
seems to be useful for roughly estimating the gap values of a two-band
superconductor \cite{Bou01b,Put06a,Wal06a,Kac10a} but may fail in proving a
material to be a two-band superconductor or not.

As already mentioned, MgB$_2$ single crystals display all three characteristic
features of a two-band (or strongly anisotropic single-band) specific heat
curve, namely a rather fast increase at low temperatures, a small jump at the
transition temperature, and a shoulder in between \cite{Bou01b,Wal06a}. The
effect of disorder on MgB$_2$ was studied by Putti et al. \cite{Put06a}. As the
amount of impurities introduced by neutron irradiation increased, the shoulder
at
intermediate temperatures was observed to fade away and the low-temperature
specific heat grew more slowly, only the jump at the transition temperature was
hardly affected. Eventually, the heavily irradiated samples even matched the 
single-band $\alpha$-model. These results confirm the theoretical
prediction that impurity scattering masks two-band effects.

The temperature dependence of the specific heat in the ternary-iron silicide
Lu$_2$Fe$_3$Si$_5$ \cite{Nak08a} was reported to be quite similar to that of
MgB$_2$ (see figure~\ref{Fig:Spec} for Lu$_2$Fe$_3$Si$_5$ and compare with
figure~\ref{SepModel1} for MgB$_2$). The measurements showed a specific heat
jump of roughly 1.05 at the transition temperature, a very pronounced shoulder
at about 0.2 $T_{\rm c}$, and a fast, though still exponential rise of the
curve at low temperatures.

Interband coupling and impurity scattering are supposed to suppress the shoulder
at intermediate temperatures, and the absence of this anomaly does therefore not
refute the two-band hypothesis. For instance, results on the two-band
superconductor NbSe$_2$ reported in \cite{Hua07a} did not display a shoulder
and, moreover, revealed a higher jump $\Delta C(T_{\rm c}) / \gamma_{\rm n}
T_{\rm c} = 2.12$ than in BCS theory. Nevertheless, both the anisotropic gap and
the two-band model agreed well with the specific heat data. Similar results were
reported for NbS$_2$ \cite{Kac10a}, though only the two-band fits revealed gap
values in agreement with scanning tunneling data, while the single-band fits did
not.

The majority of the specific heat measurements on the new iron-based
superconductors has been aimed at unraveling the properties of the order
parameter \cite{Har10a,Gof11a,Mu11a,Lin11a,Wei10a,Sto11a,Pra11a,Pop10a}. Curves
of slightly overdoped Ba(Fe$_{1-x}$Co$_x$)$_2$As$_2$ (Ba122) were reported to grow
exponentially at low temperature \cite{Har10a}, indicating fully gapped bands,
and to be nicely described by the two-band $\alpha$-model with different gaps,
although the jump at the transition was found to be close to BCS theory and no
significant shoulder was observed. A residual normal-state-like specific heat at
0\,K was considered a possible indication of $s_\pm$ symmetry, i.e. of opposite
phases in the order parameters of the two bands. Working on the same kind of
materials, Gofryk et al. \cite{Gof11a} claimed that their optimally-doped samples
could be well described by fully-gapped two-band models, while their over- and
underdoped Ba122 samples should have nodes due to a
power-law-like rise of the specific heat at low
temperatures (cf. also \cite{Mu11a}). The
situation ought to be similarly complex in other iron-based materials. For
instance, the specific heat of FeSe \cite{Lin11a} nicely matched d-wave
and diverse s-wave models. Only the low-temperature data showed that the
two-band behavior with one isotropic and one extended s-wave order parameter agreed
slightly better than the other expressions. However, it is not clear to
which extent the $\alpha$-model is capable of covering the specifics of the
different models in view of ignoring interband effects. Two-band models were
also adjusted to data of LiFeAs \cite{Wei10a,Sto11a}, mainly to decide whether
the gap has nodes on one of the bands or not.
 
Huang et al.\cite{Hua06a} analyzed the specific heat of an YNi$_2$B$_2$C single
crystal. They reported best agreement for the two-band $\alpha$-model but
still reliable matching with other models based on order parameters with nodes. 

In conclusion, the temperature dependence of the specific heat of suspected
two-band superconductors often fits two-band and anisotropic single-band models
equally well. In some cases, it is even difficult to distinguish between s-wave
(when the gaps are very small on parts of the Fermi surface) and d-wave
symmetry. In those problematic cases, a high number of low-temperature
data may allow us to distinguish the fit-quality of the different models, but the usefulness of the simplified models might be questionable, because interband interaction is usually ignored. It may be considered a support for the two-band scenario when the evaluated energy gaps match those determined by other methods such as scanning tunneling spectroscopy.

\subsection{Sommerfeld coefficient}

The volume-averaged quasiparticle density of states (DOS) at
the Fermi surface of a superconductor grows with magnetic field and can be
acquired from the linear part of the specific heat at very low temperatures via
the Sommerfeld constant or coefficient ($\gamma$), as defined in equation
(\ref{eqn:SpecHeat:2}). The field dependence of the Sommerfeld coefficient,
$\gamma (H)$, crucially depends on the Fermi surface and might help to
distinguish different scenarios.

As the magnetic field increases, the number of vortices and the corresponding
total volume of the normal-conducting vortex-cores of the mixed-state grow  and
hence enhance the quasiparticle density of states. We expect the density of
states of a standard s-wave single-band superconductor to be proportional to the
number of vortices and thus to grow linearly with field. In two-band s-wave
materials,
the rise of the density of states should be linear at low fields but should
become flatter at high fields, when superconductivity is suppressed and the
density of states thus saturated in one of the bands. In other words, two linear
parts are anticipated, from which the low-field region should be steeper. The
cross-over indicates the upper critical field of the first band, while the
saturation of $\gamma (H)$ at high fields marks the overall upper critical
field. Finally, d-wave superconductors were predicted to show a $\sqrt{H}$
behavior \cite{Vol93a}, and thus the different models should be easily
distinguishable.

\begin{figure}
    \centering
    \includegraphics[clip, width = 8.5cm]{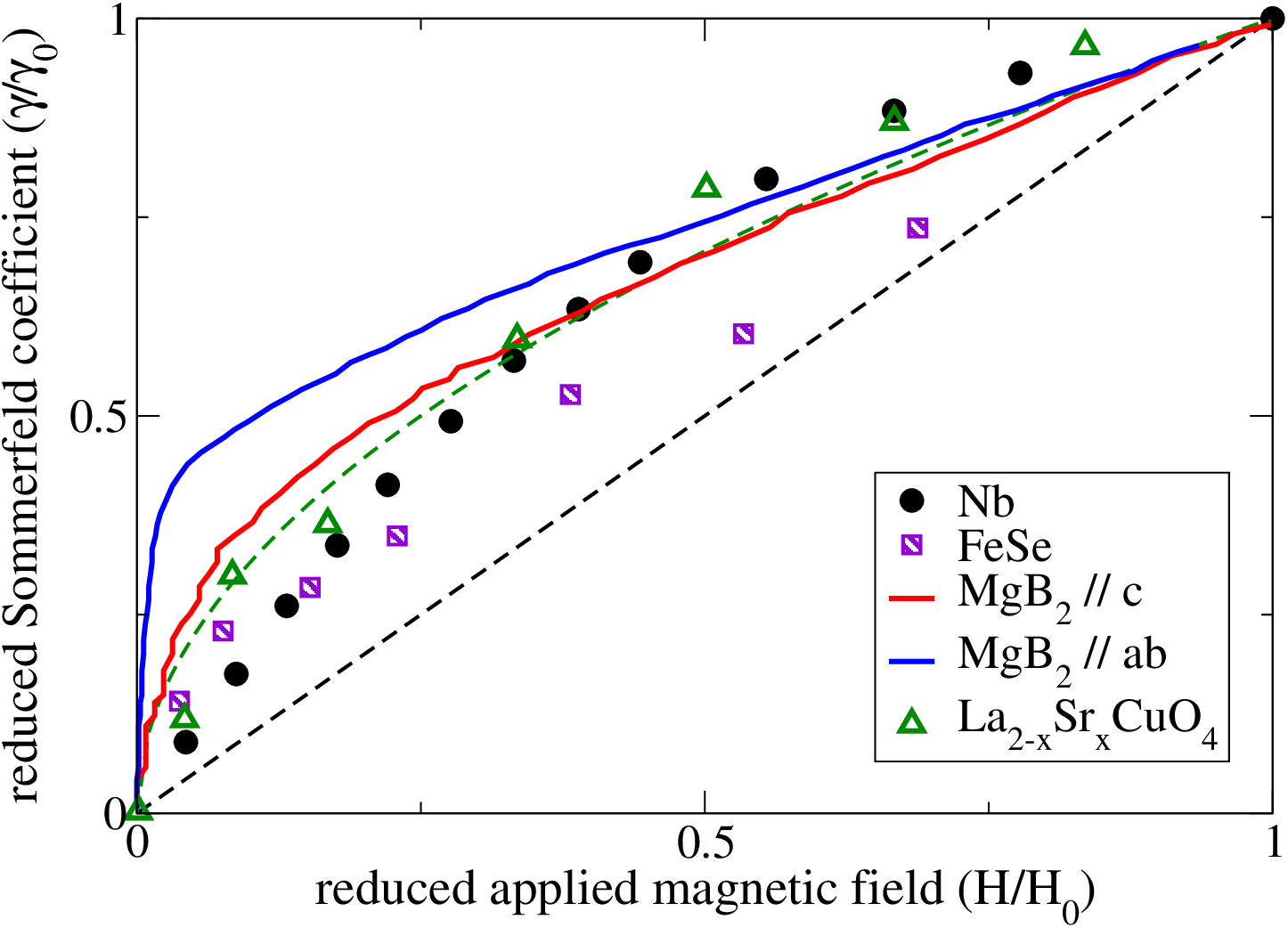}
   \caption{\label{Fig:Sommerfeld} The Sommerfeld coefficient as a
function of applied field in Nb (circles), FeSe (squares), MgB$_2$
(solid lines), and La$_{2-x}$Sr$_x$CuO$_4$ ($x \simeq$ 0.07; triangles)
\cite{Son06a,Lin11a,Pri07a,Wen04a}. The lower solid line for MgB$_2$
refers to a magnetic field parallel to the uniaxial sample axis ($H \parallel
c$) and the upper one to a field perpendicular to this axis ($H \parallel
ab$). Barring La$_{2-x}$Sr$_x$CuO$_4$, for which $H_0$ = 12\,T, the fields are
normalized by the upper critical fields and the Sommerfeld coefficients by
their normal-conducting values. The linear dashed line shows the often expected
behavior of a single-band material, but notice the considerable deviation for
Nb. The second dashed line depicts a $\sqrt{H}$ behavior, expected for d-wave
samples, which indeed matches the cuprate superconductor
La$_{2-x}$Sr$_x$CuO$_4$ but is also close to MgB$_2$ for $H \parallel
c$. In principle, it seems difficult to identify a qualitative aspect by which
the two-band (FeSe and MgB$_2$), the d-wave (La$_{2-x}$Sr$_x$CuO$_4$), and
even the weak-anisotropic single-band (Nb) behavior can be distinguished from
each
other in the figure.}
\end{figure}

As usual, reality is not that simple, because the field dependence of $\gamma
(H)$ is influenced by further effects, such as the overlap of vortex cores,
which becomes more prominent with smaller vortex - vortex distances, and the
shrinkage of the vortex core size with increasing field \cite{Kog05a}. These
effects tend to reduce the slope of $\gamma (H)$, as confirmed by experiments on
the rather isotropic conventional superconductor Nb \cite{Son06a}, shown in
figure~\ref{Fig:Sommerfeld}, and by calculations already mentioned in section \ref{sec:BCS}, that showed the linear dependence
to change to a roughly $H^\alpha$ behavior, with $\alpha < 1$; an even stronger
curvature was predicted for anisotropic materials. As the two-band curves are
influenced by the same effects and additionally by interband interactions, it
becomes again difficult to distinguish anisotropic single-band from two-band or
both from d-wave behavior, which is illustrated in figure~\ref{Fig:Sommerfeld}.

The Sommerfeld coefficient is usually acquired from the low-temperature behavior of the specific heat in magnetic fields. Ignoring the superconducting electronic part at those low temperatures, we can acquire the Sommerfeld coefficient by extrapolating the same expression as used for the normal-conducting region in the previous subsection to 0\,K. In some cases, the superconducting electronic part was taken into account by an exponential part corresponding to equation (\ref{eqn:SpecHeat:4}). 

Several groups have determined $\gamma(H)$ of MgB$_2$
\cite{Bou01a,Bou02a,Pri07a,Kle06a}. In \cite{Yan01a} $\gamma(H) \propto
H^{0.23}$ was found. Having evaluated a lot of data points, the authors of
Refs.~\cite{Bou02a,Pri07a} observed two rather linear regions separated by a
significant drop of the slope at about 0.5\,T, i.e. close to the lower upper
critical field (cf. figure~\ref{Fig:Sommerfeld}). Although Klein et al.
\cite{Kle06a} recorded a large number of data points as well, a similar kink did
not appear in their data, instead the whole field dependence nicely matched a
$\sqrt{H}$ behavior.

In NbSe$_2$ \cite{Hua07a} and NbS$_2$ \cite{Kac10a}, both, the anisotropic
single-band and the two-band model were found to match $\gamma(H)$ reasonably
well. In the case of NbSe$_2$, the two-band fits to
$\gamma(H)$ and to the specific heat resulted in similar values for the gaps and
the relative density of states, while the anisotropic single-band fits did not,
which might indicate that the single-band hypothesis is not appropriate for this
material. Moreover, the experimental data indicated a possible kink in
$\gamma(H)$, which would only be covered by the two-band scenario. For NbS$_2$
\cite{Kac10a}, a very high density of data points revealed a slight inconsistency
with the anisotropic single-band fit at low fields, while the two-band model
matched better.

Studying the Sommerfeld coefficient of iron-based Ba(Fe$_{1-x}$CO$_x$)$_2$As$_2$
superconductors, Gofryk et al. \cite{Gof11a} discovered concave curves close to
the typical d-wave behavior for under- and overdoped samples but an almost
linear behavior, as anticipated for isotropic s-wave materials, for
the optimally
doped sample. The exact classification of the dependence, however, suffered from
uncertainties in the upper critical fields, for these values are very high and
normally not directly measurable. Data on the iron-based FeSe
samples \cite{Lin11a} revealed a rather strong increase of the Sommerfeld
coefficient at very low fields and a much flatter, almost linear behavior at
higher fields, hence indicating the two-band s-wave scenario.

In YNi$_2$B$_2$C, $\gamma(H) \propto H^{0.47}$, which is actually close to the
d-wave prediction, was reported \cite{Hua06a} and claimed to agree with a
two-band model (as was found for the temperature dependence of the specific heat
in this case). A field dependence of $\gamma(H) \propto H^{0.5}$ was confirmed
for several cuprate superconductors (figure~\ref{Fig:Sommerfeld})
\cite{Mol94a,Che98a,Hus02a,Wen04a,Wan11a}.

In conclusion, the field dependence of the Sommerfeld coefficient cannot clearly
discriminate two-band from anisotropic single-band behavior. In some cases, the
experimental data seem to reveal a kink in $\gamma(H)$ at a field much lower
than the overall upper critical field, which would not be easy to explain
within the single-band theory. Further indications may be gained from comparing
the gaps and relative weights obtained from fits to $\gamma(H)$ with results
from analyzing other properties, such as the specific heat.

\subsection{Thermal conductivity}

Thermal conductivity is governed by several mechanisms in a superconductor,
which makes its interpretation often difficult. Nevertheless, at low
temperatures its field dependence might be useful for identifying two-band
superconductivity, as was demonstrated for MgB$_2$.

A temperature gradient in a material gives rise to heat flow, which is
characterized
by the thermal conductivity coefficient $\kappa$. In most cases, electrons
($\kappa_{\rm e}$) and phonons ($\kappa_{\rm p}$) contribute to the heat
transport and thus 
\begin{equation}
 	\label{eqn:tc:1}
	\kappa = \kappa_{\rm e} + \kappa_{\rm p}.
\end{equation}
The equilibrium distribution is reached by scattering processes, such as
electrons by impurities and phonons ($\kappa_{\rm e}^{-1}$ =
$\kappa_{\rm e,i}^{-1}$ + $\kappa_{\rm e,p}^{-1}$) and phonons by impurities
and electrons ($\kappa_{\rm p}^{-1}$ = $\kappa_{\rm p,i}^{-1}$ +
$\kappa_{\rm p,e}^{-1}$). Each of these processes leads to different temperature
dependencies, as for instance $\kappa_{\rm e,i} \propto T$ and
$\kappa_{\rm e,p} \propto T^{-2}$ at low temperatures, which may result in
different and complicated conductivity curves. At very low temperatures,
impurity or defect scattering often prevails. In normal-conducting metals the
low-temperature thermal flow is usually dominated by electrons, though
the situation may change in the superconducting state.  

The quasiparticle number and consequently, as Cooper-pairs do not contribute to the thermal flow, the electronic part of the thermal conductivity are reduced upon cooling in a superconductor. On the other hand, the smaller number of electrons reduces the number of scattering events by phonons and thus increases the corresponding phonon part $\kappa_{\rm p,e}$. We see that superconductivity changes not only the electron but also the phonon contribution, which makes interpreting the thermal conductivity often more difficult than for instance the specific heat. 

\begin{figure}
    \centering
    \includegraphics[clip, width = 8.5cm]{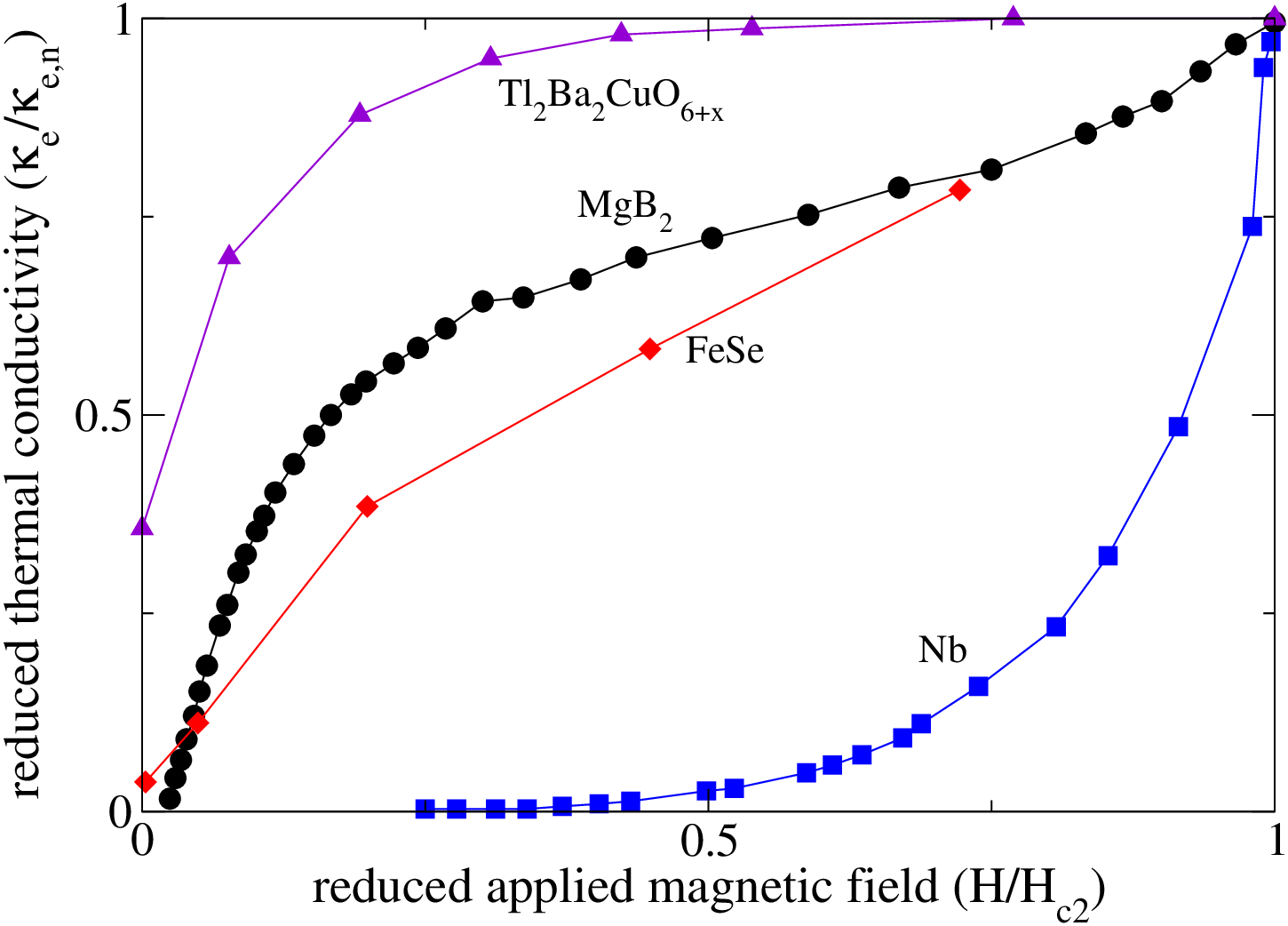}
   \caption{\label{Fig:ThermH} The low-temperature thermal
conductivity as a function of applied magnetic field for
Tl$_2$Ba$_2$CuO$_{6+x}$ (triangles), MgB$_2$ (circles), FeSe (diamonds), and Nb
(squares) \cite{Pro02a,Sol02b,Don09a,Low70a} in reduced units. The behavior
changes significantly from convex in the clean, near-isotropic single-band
s-wave material Nb, which matches the exponential trend predicted by BCS
theory, to concave in the d-wave material Tl$_2$Ba$_2$CuO$_{6+x}$. Two steep
slopes, one at low and one at high fields, interrupted by a flatter course at
intermediate fields were surmised for two-band (but also for anisotropic
single-band) materials, as is indeed backed by MgB$_2$ but barley visible in
FeSe.}
\end{figure}

The magnetic field dependence of the thermal conductivity may
vary considerably from material to material. At low temperatures and as the
magnetic field increases, we anticipate a drop of the conductivity
immediately above the lower critical field, which is eventually superseded by an increase at higher fields and a constant behavior above the upper critical
field. The drop follows from vortex formation above the lower critical field and
the corresponding proliferation of phonon scattering events at the vortex core
quasiparticles. The subsequent rise is carried by the delocalization of more
and more vortex core quasiparticles, while quasiparticles bound in a vortex
core do not participate in the thermal transport. The low-temperature thermal
conductivity of a conventional s-wave material in the clean
limit, such as Nb \cite{Low70a}, is expected to increase roughly exponentially
with field, i.e. it should slowly increase at low fields but rapidly just below the upper critical field, as a consequence of heavy vortex core overlapping, which is illustrated in figure~\ref{Fig:ThermH}. Increasing
the impurity density should produce
a larger slope at intermediate fields \cite{Wil76a}. In case of anisotropy, a
steeper increase at low field, followed by a flatter part at intermediate
fields, and again a steep slope below the upper critical field was reported
\cite{Kus02a}. This is basically similar to two-band effects. If
the gap has nodes, a significant amount of quasiparticles will be delocalized
even at low temperatures, which will render the thermal conductivity more linear
or even concave over the whole field range
(e.g. \cite{Sud97a,Pro02a} and figure~\ref{Fig:ThermH}). The curves often become
more complicated at higher temperatures. 

Let us discuss the thermal conductivity of MgB$_2$ \cite{Sol02b}. At low
temperatures the curve was observed to drop rapidly at low fields and then to
increase. This increasing part, shown in figure~\ref{Fig:ThermH}, was found to be
independent of the field orientation below about 0.5\,T and hence supposed to
reflect the isotropic $\pi$-band with an upper critical field of roughly 0.5\,T.
At higher fields the conductivity became flatter at first and then again steep
near the upper critical field and almost constant in the normal-conducting
state. The second part, i.e. that above 0.5\,T, was found to be strongly anisotropic
and hence assumed to reflect the conductivity of the anisotropic $\sigma$-band.
Thermal conductivity was measured in other materials such as NbSe$_2$, FeSe,
Lu$_2$Fe$_3$Si$_5$, etc. \cite{Boa03a,Don09a,Mac11a} and similarities to the
behavior of MgB$_2$ were considered to support the two-band scenario.

In conclusion, the effects of two-band superconductivity on the field dependence
of the thermal conductivity appear significant if the bands have different upper
critical fields. The expected curve is characterized by two steep slopes
just below each of the upper critical fields and a flatter slope in between.
Note, however, that a similar result was calculated for anisotropic
single-band superconductors \cite{Kus02a}. Finally, it should be borne in mind
that the behavior may significantly depend on the dominating scattering process,
and that less theoretical analysis than for some other properties, presented in
this paper, is available.

\subsection{Superfluid density and magnetic penetration depth}

The temperature dependencies of the superfluid density and the magnetic
penetration depth are well known for standard s-wave and d-wave superconductors.
In some materials, such as MgB$_2$, significant deviations from this standard
behavior, as for instance a faster drop of the superfluid density at low
temperatures, were observed, which could be interpreted in terms of a two-band
but in many cases, also of an anisotropic single-band model.

The London penetration depth is defined by
\begin{equation}
 	\label{eqn:ns:1}
	\Lambda_{\rm L} = \sqrt{\frac{m_{\rm s}}{\mu_0 n_{\rm s} q_{\rm s}^2}},
\end{equation}
with $m_{\rm s}$ the mass, $n_{\rm s}$ the density, and $q_{\rm s}$ the charge
of the superconducting charge carriers. It shows up in the second London
equation via $\vec{\nabla}^2 \vec{B}$ = $-\Lambda_{\rm L}^{-2} \vec{B}$ and via
$\vec{\nabla}^2 \vec{j}$ = $-\Lambda_{\rm L}^{-2} \vec{j}$, and
thus determines the penetration of the magnetic induction ($\vec{B}$) and of the
electrical current density ($\vec{j}$) at the surface of a superconductor in the
Meissner state. The London equation refers to the local limit and is hence valid
when the penetration depth is much larger than the coherence length. Corresponding relations
are acquired from Ginzburg Landau and BCS theory at low magnetic inductions.

\begin{figure}
    \centering
    \includegraphics[clip, width = 8.5cm]{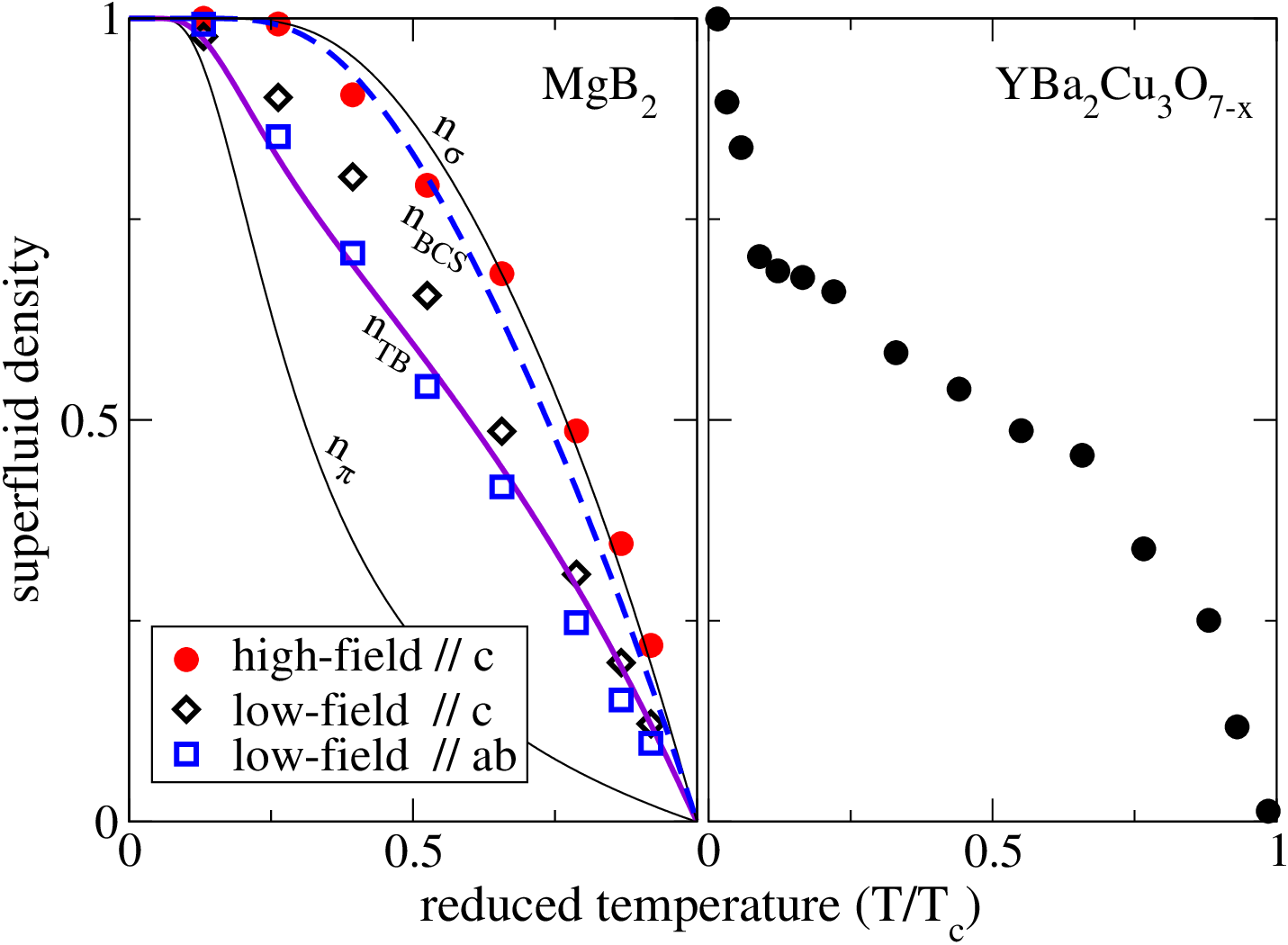}
   \caption{\label{SFLDensity} The superfluid density of MgB$_2$ 
(symbols \cite{Zeh04b}) compared with theoretical models (left panel) and of
YBa$_2$Cu$_3$O$_{7-x}$ \cite{Kha07a} (right panel) as a function of reduced
temperature. In the left panel, the experimental data were acquired from
evaluating reversible magnetization curves. The full circles refer to high
magnetic fields parallel to the uniaxial $c$ axis, while the
open symbols to low fields (parallel and perpendicular to $c$). The solid
lines
illustrate the two-band $\alpha$-model, from which the lower (thin) line refers
to $\Delta_1$ = 2\,meV ($n_\pi$), the upper (thin) line to $\Delta_2$ =
6.5\,meV ($n_\sigma$), and the (bold) one in the middle to the two-band
result $n_{\rm TB}$ = $0.43 n_\pi$ + $0.57 n_\sigma$. The selected energy gaps
and relative weights are reliable values for MgB$_2$. Finally, the dashed line
displays the BCS behavior. Note that at high fields the $\sigma$-band
properties dominate and thus single-band behavior is expected, which is indeed
confirmed by the closeness of  the high-field data (full
circles) to the single-band $n_{\rm BCS}$ and $n_\sigma$ curves in the
figure, while at low fields both bands influence the properties and the
experimental data (open symbols) are thus better described by the two-band
$\alpha$-model. In the right panel, the data were acquired from muon-spin
rotation measurements at an applied field of 0.1\,T. The low-temperature
behavior does not reflect the expectation for a d-wave superconductor;
instead, the significant rise below a reduced temperature of about 0.1 was
interpreted as evidence for an additional s-wave band in \cite{Kha07a}, but assumed to be induced by vortex pinning effects in \cite{Woj11a}. }
\end{figure}

It can be shown that  
\begin{equation}
 	\label{eqn:ns:3}
	\frac{\Lambda^2_{\rm L} (0)}{\Lambda^2_{\rm L} (T)} =
\frac{n_{\rm s} (T)}{n_{\rm s} (0)},  
\end{equation}
i.e. the temperature dependencies of the magnetic penetration depth and of the
Cooper-pair density are closely related. At low temperatures BCS predicts an
exponential behavior \cite{Fet03a}, similarly as for the specific heat
\begin{equation}
 	\label{eqn:ns:4}
	\frac{n_{\rm s} (T)}{n_{\rm s} (0)} \simeq 1 - D T^{-0.5} e^{-\Delta(0)
/ k_{\rm B} T},
\end{equation}
with $D$ a temperature independent constant. Reflecting the loss of
superconducting particles by thermal excitations, the curve decreases
continuously and reaches zero at the transition temperature. The weak-coupling BCS
behavior is universal for all materials (see dashed curve in
figure~\ref{SFLDensity}). At low temperatures the superfluid density of
fully-gapped materials is governed by the exponential function and thus almost
constant (and equal to 1) up to a particular temperature determined by $\Delta$.
Strengthening the coupling usually enhances the energy gap and hence enlarges
the range over which the superfluid density is nearly constant, while a smaller
gap reduces this range and makes the superfluid density decrease with
temperature more rapidly at low temperatures (see lines in
figure~\ref{SFLDensity}, where the gap of the $n_\pi$ curve is smaller and that
of the $n_\sigma$ larger than the BCS value). When the energy gap values diverge
in a material, the lower values govern the low-temperature behavior,
which is why the superfluid density drops faster than predicted by BCS theory,
while the larger gaps govern the high-temperature behavior. If parts of the gap
function are very small, the s-wave superconductor could even resemble the
d-wave behavior, for which we expect the superfluid density to decline linearly
at low temperatures.

Without interband interactions the superfluid density is just the sum of the contributions from the two bands, with the smaller gap-band dominating the low and the larger gap-band the high temperature region. The crossover at intermediate temperature is marked by a kink. As interband coupling and impurity scattering are turned on and then increased, the kink is smoothed and then fades away, so that the overall curve is shifted towards a linear and then towards a concave shape \cite{Gol02a,Nic05b}, as schematically presented in figure~\ref{Fig:EliashA}. Like the specific heat, the superfluid density of two-band materials has been fitted using the $\alpha$-model \cite{Man02a,Fle05a} (see figure~\ref{SFLDensity}). A more elaborate fit model including interband coupling was recently introduced by Kogan et al. \cite{Kog09a}. Note that a superconducting foreign phase with a different transition temperature may also produce a kink or a similar anomaly in the temperature dependence.

The superfluid density has been found to be almost linear or even convex in MgB$_2$
\cite{Man02a,Fle05a} (figure~\ref{SFLDensity}), NbSe$_2$ \cite{Fle07a} and other
materials. Khasanov et al. \cite{Kha07a,Kha07b} observed a very significant
change in the slope of $n_{\rm s}(T)$, measured by muon-spin rotation, in some
cuprates (e.g. YBa$_2$Cu$_3$O$_{7-\delta}$) at low temperatures, which is shown
in figure~\ref{SFLDensity} and was considered a confirmation of a second band
having s-wave symmetry in contrast to the established d-wave band. The s-wave
band contribution was found to be significant at low fields and to vanish smoothly
with increasing applied field, indicating a rather low upper critical field of
this band.

A lot of experimental data is available for iron-based superconductors
\cite{Mar09b,Has09a,Mal09a,Gor10a,Kim10a,Lua11a,Gug11a}. Though varying from
material to material and depending on doping, the superfluid density seems to
resemble the known two-band trends, i.e. it decreases rapidly at low
temperatures and becomes almost linear or even convex at elevated temperatures,
so that a conventional BCS fit does not match but a two-band one does.

Different energy gap values, from which the smaller ones dominate the low- and
the larger ones the high-temperature behavior, are also found in an anisotropic
single-band superconductor. Accordingly, the resulting curve could easily
resemble two-band behavior. Though it seems difficult to imitate the extreme
two-band case, in which the contributions of the two gaps can still be
distinguished (i.e. when interband effects are very weak), most experimentally
observed curves showing almost linear or slightly convex behavior over large
parts of the temperature range fit the anisotropic single-band model well as
was, for instance, demonstrated for MgB$_2$ \cite{Haa01a}.

In conclusion, in case of different energy gap values at the Fermi surface, the
low-temperature superfluid density behavior is governed by the smaller gaps and the high
temperature one by the larger gaps. Accordingly, the superfluid density of
both anisotropic and two-band s-wave materials, having a small and a large
gap, drops more rapidly than expected from BCS theory at low temperatures and
may become more linear or even convex at elevated temperatures. Consequently,
distinguishing anisotropy from two-band effects is again difficult.

\subsection{Upper critical field \label{Bc2}}

The upper critical field ($B_{\rm c2}$) was one of the first properties of
MgB$_2$, for which unconventional behavior was discerned, namely by a
pronounced upward (positive) curvature of its temperature dependence near the
transition temperature \cite{Zeh02a,Sol02a,Lya02a}. Since then, the effect has
been discovered in many materials and often considered as a confirmation of
two-band superconductivity. We shall point out, however, that the same effect
occurs in anisotropic single-band and even in d-wave superconductors.

The upper critical field can be easily determined in most superconductors, for
it marks the continuous phase transition from the super- to the
normal-conducting state, observable, for instance, by a jump in the electric resistivity or the specific heat, or by a kink in the reversible magnetization. Calculating the upper critical field within Eliashberg
theory allows taking into account different coupling strengths, impurity
scattering rates, order parameter symmetries, and arbitrary Fermi surfaces,
including multi-bands. 

The weak coupling BCS limit leads to the well-known WHH (Werthamer, Helfand, and
Hohenberg \cite{Hel66a,Wer66a}) behavior, predicting that the slope of $B_{\rm
c2} (T)$ is constant near the transition temperature and becomes gradually less
negative upon cooling. Universal behavior is reached by defining a reduced upper
critical field
\begin{equation}
 	\label{eqn:Bc2:1}
	b_{\rm c2}(t) = \frac{B_{\rm c2}(t)}{\partial_t B_{\rm c2}(t=1)}
\end{equation}
with $t = T / T_{\rm c}$. Ignoring Pauli spin paramagnetism, WHH found $b_{\rm
c2}(0) = 0.727$, which has been well confirmed not only for almost isotropic
conventional superconductors but also for many strongly anisotropic or two-band
materials along their uniaxial crystallographic axis (if available). For
instance, $b_{\rm c2}(0)$ equal to 0.75 was reported for MgB$_2$ in fields
perpendicular to the boron planes \cite{Zeh02a}. Those results are confirmed by
Eliashberg theory, which shows $b_{\rm c2}(0)$ to grow only slightly with
coupling strength \cite{Car90a}, e.g. to about 0.76 for $\lambda = 1.55$
(which corresponds to lead). Moreover, $b_{\rm c2}(0) = 0.69$, acquired in the
dirty limit of the weak coupling case, suggests that impurity scattering
effects are small.

\begin{figure}
    \centering
    \includegraphics[clip, width = 8.5cm]{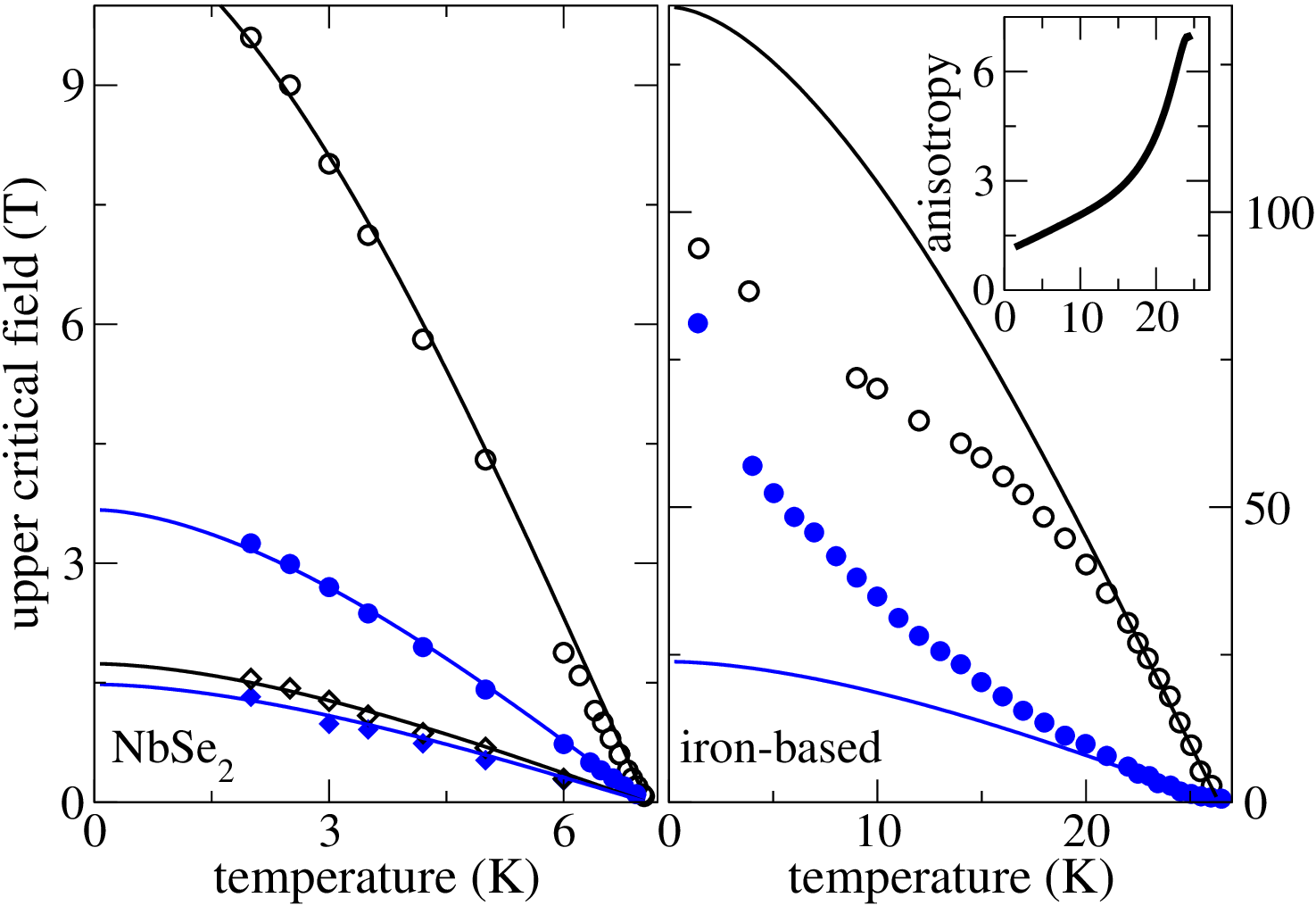}
   \caption{\label{Fig:BC2} The upper critical fields of NbSe$_2$ \cite{Zeh10a} (left panel) and of the iron-based superconductor Ca$_{10}$(Pt$_4$As$_8$)((Fe$_{1 - x}$Pt$_x$)$_2$As$_2$)$_5$ \cite{Mun12a} (right panel). The symbols indicate experimental data and the lines the clean-limit WHH model. In the left panel, the two upper curves show the (global) upper critical field of NbSe$_2$ at fields parallel (full circles) and perpendicular (open circles) to the uniaxial $c$-axis of the sample. For $H \parallel ab$ a slight upward curvature shows up near the transition temperature. The two bottom curves are the upper critical fields associated with the suppression of superconductivity in the band with the smaller gap. All curves roughly follow the clean-limit WHH trend. Note the quite different anisotropies, presented in figure~\ref{Anisotropy}, of the global and of the second-band upper critical fields, which should reflect the different anisotropies of the bands of NbSe$_2$. In the right panel, the curvature of the lower upper critical field (full symbols, $H \parallel c$) is positive over the whole temperature range, whereas the top curve (open symbols) is mostly concave. Accordingly, the upper critical field anisotropy, shown in the inset, increases upon warming. The WHH curves, chosen to match the symbols at high temperatures, deviate strongly from the experimental behavior of the iron-based superconductor.}
\end{figure}

Early measurements of the upper critical field in MgB$_2$ unveiled a remarkable
deviation from the WHH behavior, namely a pronounced upward curvature near the
transition temperature in fields parallel to the boron planes ($ab$ - direction)
but a conventional behavior along $c$ (the uniaxial axis), making the
anisotropy $\Gamma = B^{ab}_{\rm c2} / B^c_{\rm c2}$ decrease with temperature
\cite{Zeh02a,Sol02a,Lya02a}, as shown in figure~\ref{SepModel2}. These
peculiarities have also been reported for other materials, such as
Nb \cite{Wil70a,Web91a}, V \cite{Wil70a}, NbSe$_2$ \cite{Toy76a,San95a,Zeh10a}
(figure~\ref{Fig:BC2}), borocarbides and -nitrides \cite{Met97a,Shu98a,Man01a},
heavy fermion systems \cite{Mea04a}, iron-based
\cite{Hun08a,Jar08a,Qia09a,Kan09a,Sun09a} (figure~\ref{Fig:BC2}), and cuprate
\cite{Kor10a,Bos11a} superconductors, and have often been considered as an
indication or a confirmation of two-band superconductivity. The effect may
change from sample to sample and is sometimes quite small.

The upward curvature of the upper critical field of two-band superconductors has
also been confirmed by theory. Within Eliashberg theory, this curvature was shown
to appear when the Fermi velocities of the two bands are different and to
become more pronounced when the ratio of the velocities increases \cite{Man05a}.
This would explain why the feature is observed for $H \parallel ab$ but not for
$H \parallel c$ in MgB$_2$, for the mean Fermi velocities perpendicular to the
applied field are quite different in case of $H \parallel ab$ ($v_{\rm F,\sigma}
\ll v_{\rm F,\pi}$) but similar for $H \parallel c$ \cite{Bel01a}. Accordingly,
two-band superconductors do not necessarily display this feature for any field
direction. The curvature is affected  by the coupling and the impurity scattering
parameters in different manners. For instance, it was claimed that different
intraband scattering rates could produce the upward curvature even if the Fermi
velocities are similar \cite{Man05a}. A small positive $B_{\rm c2}(T)$ curvature
for the field along the uniaxial axis was indeed found after heavy neutron
irradiation in an MgB$_2$ single crystal \cite{Zeh03b}, but ascribed to inhomogeneities in the
transition temperature caused by the defect distribution.

In samples with uniaxial symmetry, the positive curvature usually occurs for measurements in fields perpendicular to this axis, which is mostly also the
direction of the maximal upper critical field values. A different behavior, namely an upward curvature for a field orientation not showing the highest upper critical field values, was
observed in some iron-based superconductors, as for instance in the 1111 and 122
compounds (cf. section \ref{sec:mat:Fe}) \cite{Hun08a,Jar08a,Qia09a,Kan09a,Sun09a}. Also in contrast to MgB$_2$, the
corresponding anisotropy was found to increase with temperature, which might be
an effect of Pauli paramagnetic pair breaking, occurring at low temperatures in
high magnetic fields and flattening the upper critical field behavior (cf.
figure~\ref{Fig:BC2}).

It should not be missed that an upward curvature of the upper critical field and
hence a temperature dependent anisotropy have been known for anisotropic s-wave superconductors for a long time. The theory based on the Eliashberg model was elaborated in
Refs.~\cite{Pro87a,Pit93a}. The effect was even observed in niobium and shown to
match theory well \cite{Web91a}. Similar results have been reported for other
materials. In many cases, the simple separable model, introduced in section
\ref{sec:sepmodel}, was found to fit the experiments quite well. Because this
model formally agrees with a spherical two-band model, similar effects by both
scenarios are anticipated. Indeed, the anisotropic single-band theory was shown
to reproduce the upper critical field of MgB$_2$ \cite{Zeh03a} (see figure
\ref{SepModel2}) and other potential two-band materials \cite{Man01a} nicely.
The upward curvature of $B_{\rm c2}(T)$ in single-band superconductors was also
derived from the Eilenberger equations \cite{Mir03a}. In principle, theory shows the
upper critical field to be governed by an integral over the Fermi surface which
includes the Fermi velocities perpendicular to the field orientation. Thus,
details of the anisotropy do not show up in the upper critical field and hence quite different Fermi surfaces with a similar mean anisotropy
may lead to a similar $B_{\rm c2}(T)$ behavior.

The upward curvature was also predicted for d-wave symmetry
\cite{Pro90a}, which was recently confirmed by measurements of a temperature
dependent anisotropy in SmBa$_2$Cu$_3$O$_x$ \cite{Kor10a} and
YBa$_2$Cu$_3$O$_7$ \cite{Bos11a}.

Differences between the single and the two-band model might become visible when
introducing impurities. In the single-band scenario, scattering by non-magnetic
point defects smears out the anisotropy of the Fermi surface and thus reduces
the transition temperature until saturation is reached. At the same time $B_{\rm
c2}(T)$ becomes steeper near the transition temperature, which may lead to an increase in $B_{\rm c2}(0)$. In case of the two-band scenario, the different
channels for impurity scattering affect the upper critical field and the
transition temperature in different ways \cite{Man05a,Bro10a}. For instance,
if the bands are spherical, intraband impurity scattering increases the upper
critical field, but does not change the transition temperature (as known from
the Anderson theorem), while interband scattering modifies both quantities.
Consequently, in contrast to anisotropic single-band materials, the two-band scenario
would allow us to increase the upper critical field without lowering the
transition temperature, if we succeeded in changing impurity scattering only for
a selected channel.

To conclude, identifying two-band materials from measurements of the upper
critical field appears hardly feasible, for the characteristic upward
curvature of the upper critical field near the transition temperature and the
corresponding temperature dependence of the anisotropy may also 
emerge in anisotropic single-band and d-wave superconductors.

\subsection{Torque \label{sec:torque}}

The anisotropies of the magnetic penetration depth and of the coherence length have
often been claimed to be different in two-band superconductors, which can be
verified by studying the angular dependence of the magnetic torque. We will see
that the experiments have not confirmed this statement thus far. Moreover,
torque experiments are an efficient tool for acquiring several superconducting
properties from a single measurement.

The single-band Ginzburg Landau model for uniaxial superconductors needs but
one anisotropy parameter $\Gamma = \sqrt{m_c / m_{ab}}$, where $m_c$ and
$m_{ab}$ are the effective masses  of the principal crystallographic axes. The
same quantities determine the ratio of the penetration depths ($\Lambda$) and
of the upper critical fields ($B_{\rm c2}$), i.e.,
\begin{equation}
 	\label{eqn:Tor1}
	\Gamma = B^{ab}_{\rm c2} / B^c_{\rm c2} =  \Lambda_{c} / \Lambda_{ab}
\end{equation}
Here, uniaxial anisotropy, with $c$ the uniaxial direction and $ab$
perpendicular to it, is assumed. In equation~(\ref{eqn:Tor1}), the indices of
the lengths indicate the flow direction (e.g. of the currents) and the
superscripts the field orientation. The anisotropy of the
coherence length is usually supposed to match that of the upper critical
field: $\Gamma = \xi_{ab} / \xi_{c}$.

The magnetic torque is defined by $\vec{\tau} = \vec{m} \times \vec{B}$,
with $\vec{m}$ the magnetic moment of a sample with volume $V$. The reversible
torque of a superconductor can be derived within London
theory \cite{Kog88a,Far88a}. If we assume the magnetic induction to be
equal to the applied field (which usually holds well for not too low fields),
we obtain
\begin{equation}
 	\label{eqn:Tor2}
	\tau(\vartheta) = -\frac{V H_{\rm a} \Phi_0}{16 \pi \Lambda_{ab}^2}
\left( 1 - \Gamma^{-2} \right) \frac{\sin 2 \vartheta}{\epsilon} \ln
\left(\frac{\eta B^c_{\rm c2} }{\epsilon \mu_0 H_{\rm a} } \right)
\end{equation}
with
\begin{equation}
 	\label{eqn:Tor3}
	\epsilon = \epsilon(\vartheta, \Gamma) = \sqrt{\Gamma^{-2} \sin^2
\vartheta + \cos^2 \vartheta}
\end{equation}
Here, $\vartheta$ denotes the angle between the field orientation and the
$c$-axis of the sample, $H_{\rm a}$ the applied magnetic field, and $\eta \sim
1$ is a parameter that depends on the vortex configuration. Usually, different
background signal terms are to be added.

Equation~(\ref{eqn:Tor2}) can be adjusted to experimental data by varying the
parameters $\Lambda_{ab}$, $B^c_{\rm c2}$, and $\Gamma$ freely. The magnetic
penetration depth is simply a proportionality factor, the upper critical field
is highly sensitive to the curvature of the torque, and the anisotropy is mainly
determined by the slope near the $ab$ direction. Thus there is a good chance of
determining each variable quite independently from the others. Nevertheless,
reducing the number of fit-parameters by taking results from other experiments,
as for instance the upper critical field from SQUID measurements, should improve the
quality of the fit procedure. Usually, the reversible torque has to be acquired
from the irreversible branches, obtained by measuring at opposite rotation
directions. Evaluation errors, in particular those for the anisotropy, will become substantial with increasing hysteresis width between the irreversible branches.

\begin{figure}
    \centering
    \includegraphics[clip, width = 8.5cm]{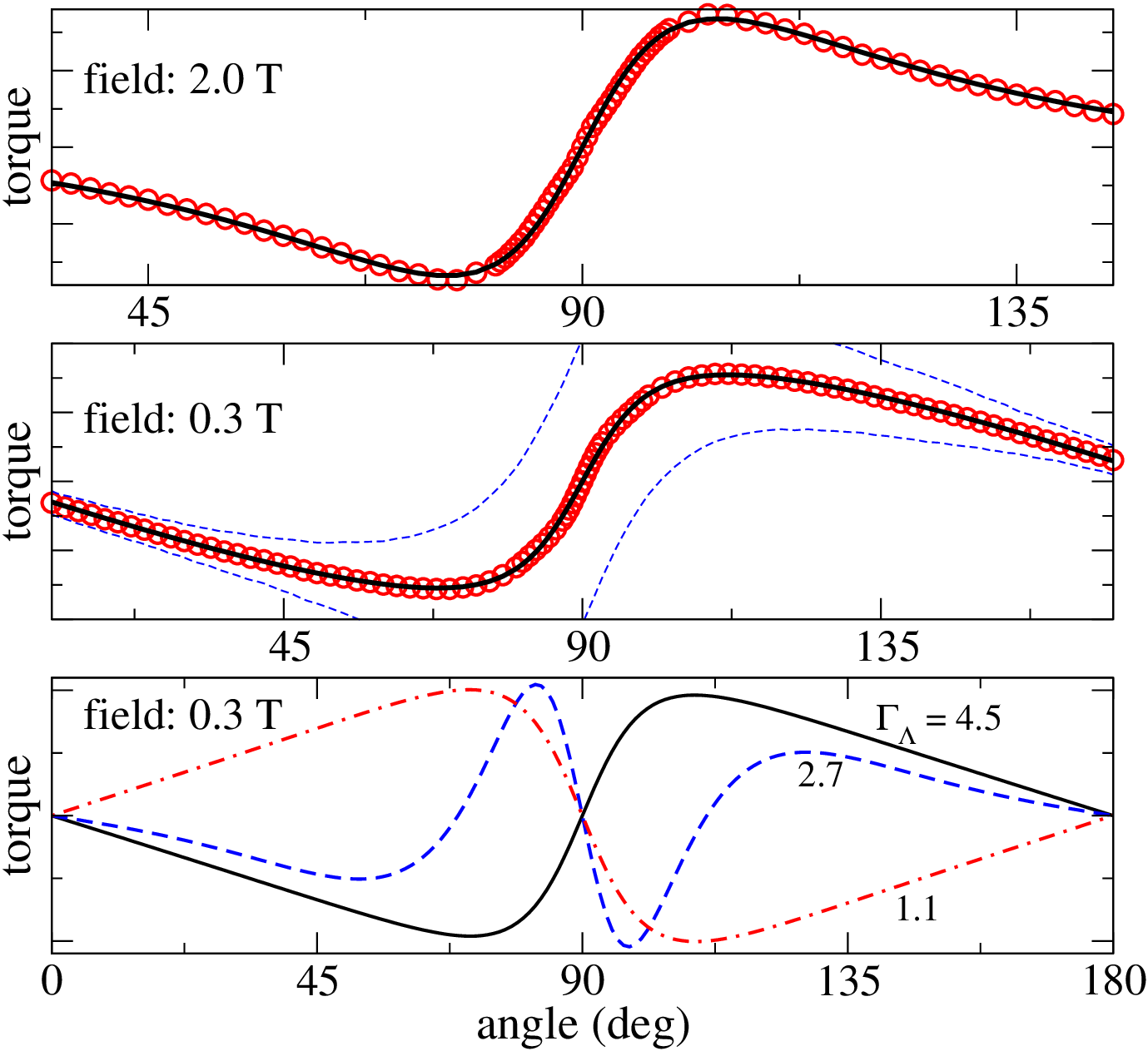}
   \caption{\label{Torque} The torque as a function of angle. The upper panels
show experimental data of MgB$_2$ \cite{Zeh04b} compared with theory. The open
circles display the reversible torque obtained from the irreversible branches,
indicated by dashed lines, as discussed in the text. At 2.0\,T (upper panel) the
reversible and irreversible curves coincide. The solid lines show the
single-band fits according to equation~\ref{eqn:Tor2}. Notice that the data at
2\,T, displayed in the upper panel, refers to the high-field single-band regime,
while the data at 0.3\,T, displayed in the middle panel, to the low-field
two-band regime of MgB$_2$, and yet, both curves excellently match the
single-band model, in which the anisotropies of the penetration depth and the
coherence length are assumed to be equal. The bottom panel illustrates the
effect of diverging anisotropies according to equation~\ref{eqn:Tor6} at a field
of 0.3\,T. The upper critical field (or coherence length) anisotropy is held
constant at 4.5, while the penetration depth anisotropy changes from 4.5 (solid
line) to 2.7 (dashed line) to 1.1 (dashed-dotted), which modifies the curves in
a profound way. All data refer to MgB$_2$ at 5\,K and a $B^c_{\rm c2}$ of
2.8\,T.}
\end{figure}

Quite early, very pure single crystals of MgB$_2$, whose magnetic properties
were almost fully reversible and which were thus excellently suited for torque
experiments, have been made available \cite{Kar03a}. Describing the
torque data by equation (\ref{eqn:Tor2}) was shown to work well, as
illustrated in figure~\ref{Torque}, and to reveal
parameters in nice agreement with results from other experiments, for instance
with the upper critical field and its anisotropy determined from SQUID
magnetometry \cite{Zeh04d}.

As already mentioned, the anisotropies of the upper critical field
\begin{equation}
 	\label{eqn:Tor4}
	\Gamma_{\rm H} = B^{ab}_{\rm c2} / B^c_{\rm c2}
\end{equation}
and of the magnetic penetration depth
\begin{equation}
 	\label{eqn:Tor5}
	\Gamma_{\Lambda} = \Lambda_{c} / \Lambda_{ab}
\end{equation}
were proposed to be unequal, i.e. $\Gamma_{\rm H} \neq
\Gamma_{\Lambda}$, in
two-band superconductors. 
Equation (\ref{eqn:Tor2}) holds only when both anisotropies are equal; for
$\Gamma_{\rm H} \neq \Gamma_{\Lambda}$ Kogan \cite{Kog02a} derived:   
\begin{eqnarray}
 	\label{eqn:Tor6}
	\tau(\vartheta) &=& -\frac{V H_{\rm a} \Phi_0}{16 \pi \Lambda_{ab}^2}
\left(
1 -
\Gamma^{-2}_\Lambda \right) \frac{\sin 2 \vartheta}{\epsilon_\Lambda} \nonumber
\\
\nonumber &\times& \bigg[ \ln \left(\frac{\eta B^c_{\rm c2} }{\mu_0 H_{\rm a}}
\frac{4 e^2 \epsilon_\Lambda}{(\epsilon_\Lambda + \epsilon_{\rm H})^2} \right)
\\&-&
 \frac{2 \epsilon_\Lambda}{\epsilon_\Lambda + \epsilon_{\rm H}}
  \left( 1 + \frac{(1 - \Gamma^2_{\rm H})\epsilon_\Lambda}{(1 -
\Gamma^2_\Lambda) \epsilon_{\rm H}} \right)
 \bigg],
\end{eqnarray}
with $\epsilon_\Lambda = \epsilon(\vartheta, \Gamma_\Lambda)$ and
$\epsilon_{\rm H} = \epsilon(\vartheta, \Gamma_{\rm H})$ (cf.
equation \ref{eqn:Tor3}).

An upper critical field anisotropy ($\Gamma_{\rm H}$) of about 4-6 and a
penetration depth anisotropy ($\Gamma_{\Lambda}$) of about 1-2 were proposed for
MgB$_2$  at low temperatures \cite{Cub03a,Fle05a}; yet, applying expression
(\ref{eqn:Tor6}) to torque data of MgB$_2$ suggested $\Gamma_{\rm H} \simeq
\Gamma_{\Lambda}$, and that both anisotropies were in close agreement with
$\Gamma$ evaluated via the single-band expression (\ref{eqn:Tor2})
\cite{Zeh04b,Zeh04d}. A closer examination \cite{Zeh04d} of equation
(\ref{eqn:Tor6}) revealed that small variations in $\Gamma_{\Lambda}$ increase
the fit error strongly, whereas variations in $\Gamma_{\rm H}$ are less
significant for the case of MgB$_2$; in other words, uncertainties in $\Gamma_{\Lambda}$ are
much smaller than in $\Gamma_{\rm H}$, yet $\Gamma_{\rm H} \simeq
\Gamma_{\Lambda}$ still holds. Moreover, $\Gamma_{\rm H}$ was directly
determined from the upper critical fields, obtained from magnetization
measurements in a SQUID, and found to excellently agree with $\Gamma_{\Lambda}$ from torque
data. Equation (\ref{eqn:Tor6}) predicts qualitative changes, as for
instance additional zeros, in the angular dependence of the torque when
$\Gamma_{\rm H}$ and $\Gamma_{\Lambda}$ differ significantly, as shown in the
bottom panel of figure~\ref{Torque}. If $\Gamma_{\rm H} \sim 4-6$ and
$\Gamma_{\Lambda} \sim 1-2$, as suggested for MgB$_2$, equation (\ref{eqn:Tor6})
makes the torque change its sign over the whole angular range. As far as I know,
such effects have never been reported and thus no indication of such different
anisotropies in MgB$_2$ and similar materials seem to be available. Note that
the experiments were also carried out at low magnetic fields, e.g. from 0.1 -
0.5\,T at 5\,K, where both bands should affect the properties of MgB$_2$
\cite{Zeh04b,Zeh03c} (cf. figure~\ref{Torque}). We will see in
section~\ref{sec:aniso}, that a pronounced field dependence of the anisotropy
might mainly be responsible for the reports of different values for different
quantities.

Torque measurements were also carried out on iron-based superconductors and
analyzed via equations (\ref{eqn:Tor2}) and (\ref{eqn:Tor6}), but the results
suffer from large hysteresis widths between the two irreversible branches. In
\cite{Wey09a}, XFeAsO$_{0.8}$F$_{0.2}$ (X = Nd or Sm) single crystals with
transition temperatures between 44 and 48\,K were analyzed at a field of 1.4\,T
and at temperatures from about 20 to 44\,K. Employing expression
(\ref{eqn:Tor2}) led to excellent agreement with experiment, and the anisotropy
was found to decrease from about 15-20 at 20\,K to 7 near the transition
temperature. Since the results were in striking contrast to resistivity
measurements, from which $\Gamma_{\rm H} \simeq 5$ at 34\,K was obtained
\cite{Jar08a}, equation (\ref{eqn:Tor6}), with $\Gamma_{\rm H}$ fixed by the
resistivity data, was applied. Again, good agreement with experiment was reached
and $\Gamma_\Lambda$ found to be equal to $\Gamma$ from the single-band evaluation
(\ref{eqn:Tor2}), i.e. 15-20 at 20\,K to 7 near the transition. This
demonstrates that $\Gamma_{\rm H}$ does virtually not affect the quality of the
fit and that the relation between the anisotropies could thus not be determined
in this case. Note that $\Gamma_{\rm H}$ from the resistivity measurement
referred to much higher fields than the results from the torque experiments.
Further experiments on iron-based materials have been carried out
\cite{Li11a,Ben10a} and could be well described by the single-band expression
(\ref{eqn:Tor2}). Applying the method to LaFeAsO$_{0.9}$F$_{0.1}$ ($T_{\rm c}
\sim 15$\,K) revealed a temperature and field dependent anisotropy
\cite{Li11a}. 

In conclusion, the magnetic torque is not able to unveil two-band behavior
of a superconductor directly, for the data are usually well described by the
conventional expression, valid for single-band materials. The
temperature and, in particular, the field dependence of the evaluated
properties, such as the anisotropy and the magnetic penetration depth, may,
however, help to identify a two-band material, as we will see later.

\subsection{Reversible magnetization \label{sec:RevMag}}

The reversible magnetization, $M_{\rm r}$, of a superconductor is another
property that may be sensitive to two-band effects. Indeed, its field
dependence should allow us to discriminate the two-band from the
anisotropic single-band scenario. Unfortunately, acquiring samples in which
the reversible magnetization can be determined often proves difficult due to
the interfering effects caused by flux-line pinning.

Measurements of the magnetic moment are routinely performed in SQUIDs and
vibrating-sample-magnetometers. Direct access to the
reversible part is often blocked by the irreversible properties, coming along
with flux-line pinning and showing up in a hysteresis of the
magnetization loop. The volume averaged reversible magnetization can then be
calculated via 
\begin{equation}
 	\label{eqn:Rev1}
	M_{\rm r}(H) = \frac{m(H_+) + m(H_-)}{2 V}
\end{equation}
with $m$ the measured magnetic moment, $V$ the sample volume, and $H_+$ and
$H_-$ the applied field $H$ for the increasing ($H_+$) and for the decreasing
field ($H_-$) sweep branch; $H_+$ and $H_-$ should roughly refer to the same
magnetic induction $B$ in equation (\ref{eqn:Rev1}). When the irreversible
contributions are significantly larger than the reversible part, the result of
the above equation will usually become highly unreliable. The hysteresis width
can, however, be reduced by employing the so-called vortex shaking technique
\cite{Wil98a,Bra02a}.

For some materials, such as MgB$_2$, NbSe$_2$, V$_3$Si, Nb, etc., pure single
crystals are available, in which the hysteresis is small or completely absent over a large
field and temperature range. If additionally the upper critical
field of that material is not too high, the reversible magnetization can be
determined over a large part of the superconducting phase diagram. Calculating
$B = \mu_0 (H - D M_{\rm r} + M_{\rm r})$, where $D$ is the demagnetization
factor of the sample, gives $M_{\rm r}(B)$, i.e. the reversible magnetization as
a function of the magnetic induction, which we can compare with theory. 

The theoretical $M_{\rm r}(B)$ curve can be taken from any model; for instance,
Ginzburg Landau theory appears quite convenient - not only because simple
interpolation formulas, provided by Brandt \cite{Bra03a,Bra04a}, are available.
The Ginzburg Landau model depends on two parameters, namely on the upper
critical field, at which $M_{\rm r}(B)$ vanishes, and on the Ginzburg Landau
parameter $\kappa$. Both quantities can be acquired by fitting theory to the
experimental $M_{\rm r}(B)$ data. In Ginzburg Landau theory, anisotropy is
specified by the appropriate effective masses at the Fermi surface (cf. equation
\ref{eqn:GL_fi1}), but it was shown that the anisotropic can be mapped onto the
isotropic single-band model, when the magnetic field points along a principal
axis of the sample \cite{Kle80a,Hao91a}. Thus, in contrast to two-band effects,
anisotropy should not significantly affect the shape of $M_{\rm r}(B)$.

\begin{figure}
    \centering
    \includegraphics[clip, width = 8.5cm]{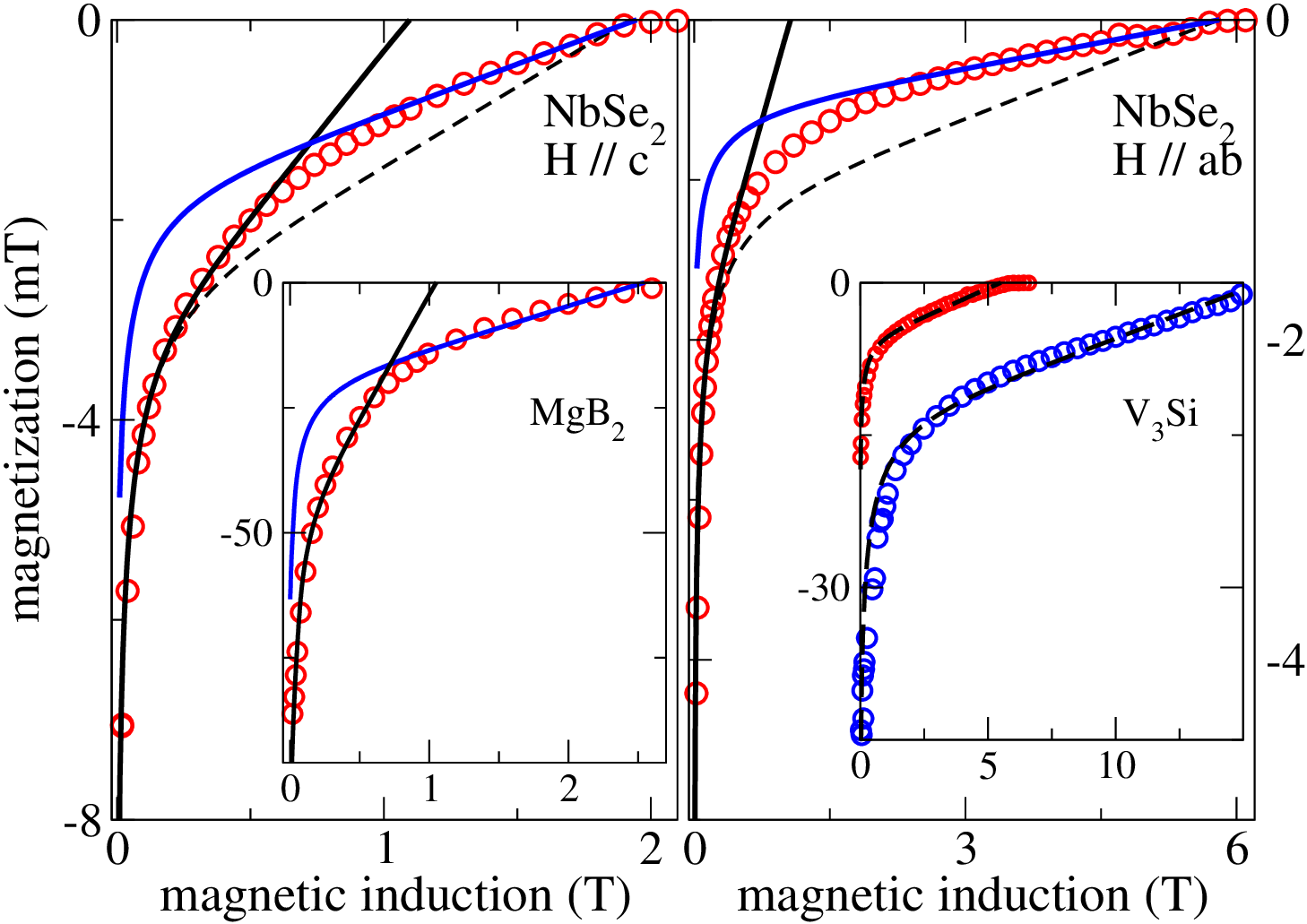}
   \caption{\label{Mrev} The reversible magnetization of several
single-crystalline samples as a function of magnetic induction (open circles)
compared with single-band Ginzburg Landau theory (lines) \cite{Zeh10a,Zeh04b}.
The two main panels show results on NbSe$_2$ for fields parallel ($c$, left
panel) and perpendicular ($ab$, right panel) to the uniaxial axis at a
temperature of 4.2\,K ($T/T_{\rm c}$ $\simeq$ 0.6). The left inset displays data
of MgB$_2$ for $H \parallel c$ at 5\,K ($T/T_{\rm c}$ $\simeq$ 0.13) and the
right one of V$_3$Si at 13.5\,K ($T/T_{\rm c}$ $\simeq$ 0.8, upper curve) and
7\,K ($T/T_{\rm c}$ $\simeq$ 0.4). The dashed lines
illustrate fits to the whole experimental field range, while the solid lines
either to the low- or to the high-field regime alone. We find qualitative
disagreement between single-band Ginzburg Landau theory and NbSe$_2$ or MgB$_2$
and that two different theoretical curves, one for the
low and one for the high-field regime, can cover most part of the experimental
data. The deviation from single-band theory does not essentially change at
other temperatures in these materials. In contrast, reliable agreement between
single-band theory and reversible magnetization over the whole field range is
observed in V$_3$Si at high and, though slightly worse, at low
temperatures \cite{Zeh12a}.}
\end{figure}

Fig.~\ref{Mrev} presents reversible magnetization curves of NbSe$_2$ and MgB$_2$
as a function of field \cite{Zeh04b}. The dashed lines display the best
single-band Ginzburg Landau fits when the upper critical field is fixed by the
field where the experimental data become zero. The striking differences between
theory and experiment provide evidence for a non-single-band behavior in these
materials. 

Ginzburg Landau theory is known to hold strictly only sufficiently close to the
transition temperature and the upper critical field. As the temperature
decreases, larger deviations between the Ginzburg Landau and the experimental
behavior are likely, although we do not expect considerable qualitative
effects if adjusting the parameters of the model in a convenient way. This was
verified for V$_3$Si (right inset of figure~\ref{Mrev}), for which reliable
agreement was not only found at high ($T/T_{\rm c}$ $\simeq$ 0.8), but also at
relatively low temperatures ($T/T_{\rm c}$ $\simeq$ 0.3 - 0.4) \cite{Zeh12a},
and for Nb. Nice matching was also reported for the anisotropic superconductors
YBa$_2$Cu$_4$O$_8$, Nd$_{1.85}$Ce$_{0.15}$CuO$_{4-\delta}$, and
YBa$_2$Cu$_3$O$_{7-\delta}$ \cite{Pog01a,Ach05a,Kar06a}. It should be noted,
however, that for all those samples two-band superconductivity has been
suspected in some publications. Nevertheless, I point out that for MgB$_2$ and
NbSe$_2$ the disagreement between experiment and single-band theory is not
essentially changed at higher temperatures.

To evaluate two-band properties, we can analyze the experimental data by
applying two independent single-band fits, one for the low and one for the high
field regime, as illustrated in Fig.~\ref{Mrev}. It should be kept in mind that
this procedure does not yield a two-band model, for interband effects are not
taken into account, but rather resembles the $\alpha$-model used for fitting the
specific heat (cf. section~\ref{sec:SH}) and other properties. The point is that
in two-band superconductors, such as MgB$_2$ and NbSe$_2$, the properties of one
of the bands are apparently suppressed above a particular magnetic field,
associated with the upper critical field of that band, and thus a nearly
single-band behavior, matching the single-band Ginzburg Landau model, appears
at high fields, while both bands significantly contribute to the superconducting
state at low fields. The high-field fit reveals $\kappa$, $B_{\rm c2}$, and, by
applying well-known Ginzburg Landau relations \cite{Tin75a}, further critical
fields and lengths as well as the anisotropy of the band with the higher upper
critical field. Even at low fields, a single-band model was found to describe
MgB$_2$ and NbSe$_2$ well, but the corresponding fit leads to effective
quantities representing both bands, though the properties of the second band
might dominate  in this field region in the cases of MgB$_2$ and NbSe$_2$. Note
that, although the global upper critical fields are quite different for the $c$ and the
$ab$ direction in these materials, the upper critical fields from the low-field
fits are almost equal, as shown in figure~\ref{Fig:BC2} and \ref{Anisotropy}. This reflects the shape of
the two bands, from which one is strongly anisotropic and the second almost
isotropic.

The procedure  not only probes two-band superconductivity, but can also
reveal the field dependence of several quantities. A more elaborated model of
the two-band Ginzburg Landau magnetization curve, including interband effects,
was applied in \cite{Eis05a}, and let to more detailed results of the
temperature and field dependence of several superconducting properties (e.g.
top right panel of figure~\ref{Anisotropy}).

The above method was successfully applied to MgB$_2$ and NbSe$_2$ 
\cite{Zeh04b,Zeh10a}. Unfortunately, iron-based superconductors usually have a
large hysteresis, so that the reversible part has not reliably been determined
so far, which is similar to the situation in many cuprates. Moreover,
the upper critical fields of these materials are often much larger than the
maximum field provided by the experimental equipment.

To conclude, comparing the measured field dependence of the reversible
magnetization with the theoretical single-band behavior appears to be a useful
tool for probing two-band superconductivity. This holds at least when the
superconducting properties of one of the bands are significantly suppressed at
higher magnetic fields and, as a consequence, a single-band fit of $M_{\rm r}(B)$ does not match over the whole field range. Using the Ginzburg Landau model
seems sensible, though some uncertainties remain due to the restrictions of the
model. The single-band fits allow us to extract the field and temperature
dependence of several superconducting properties.

\subsection{Anisotropy and field dependence of the characteristic lengths
\label{sec:aniso}}

Peculiarities in the anisotropy have frequently been considered a strong
confirmation of two-band superconductivity. Indeed, while a temperature
dependence of the anisotropy is rather common, a strong field dependence is
difficult to explain by non-two-band effects. Those field dependencies are
mirrored by different anisotropies of, for instance, the magnetic penetration depth and
the upper critical field when these properties are measured at different fields.

Anisotropy shows up in many superconducting properties, such as the magnetic
penetration depth ($\Gamma_\Lambda$), the coherence length ($\Gamma_\xi$), the
Ginzburg Landau parameter ($\Gamma_\kappa$), the upper ($\Gamma_{\rm Bc2}$ =
$\Gamma_{\rm H}$), and the lower critical field ($\Gamma_{\rm Bc1}$), etc., when
different crystal directions are probed. In the following, I will concentrate on
uniaxial anisotropy, since most samples discussed in this text have uniaxial or
near-uniaxial symmetry. In simple single-band materials, all anisotropies are equal, in accordance with Ginzburg Landau theory, i.e. $\Gamma_\Lambda$ =
$\Gamma_\xi$ = $\Gamma_\kappa$ = $\Gamma_{{\rm Bc2}}$ = $\sqrt{m_{c} / m_{ab}}$.

In case of two- or multi-band superconductivity, a more complicated situation
arises, as different bands may have different Fermi surface shapes. In fact,
when people determined the anisotropy of MgB$_2$ they reported quite different
results for different quantities and techniques. Basically, the anisotropy of
the upper critical field was found to be large ($\sim 5$) at low temperature and to
decrease upon warming, while that of the penetration depth small ($\sim 1$) and
to increase, so that about the same value was reached at the transition
temperature \cite{Cub03b,Fle05a}. This could be explained by theory. Within the
Eilenberger model, the ratio of the penetration depths at 0\,K \cite{Kog02b} was
found to depend only on the ratio of the Fermi velocities, which gives roughly
one in MgB$_2$, while the larger values at higher temperature were explained by
the additional influence of the gap anisotropy. Concerning the anisotropy of the
upper critical field \cite{Mir03a}, the same expression as for the penetration
depths was derived at the transition temperature, which explains the
experimentally observed merging of the two anisotropies at this point, while a
rather large anisotropy, corresponding to the shape of the band dominating at
high fields, was theoretically derived and experimentally acquired at low
temperatures.

It should be noted that in the above experiments and theoretical calculations,
the anisotropies of the upper critical fields and penetration depths refer to
different magnetic fields, namely to low or zero field in the case of the
penetration depth but to high fields in the case of the upper critical field.
Some methods allow us to determine the penetration depth and thus
$\Gamma_\Lambda$ at different magnetic fields. A simple method providing two
values, one for the high and one for the low-field region, is fitting two
single-band curves to the reversible magnetization, as presented in the previous
subsection (\ref{sec:RevMag}), which was reported to yield rather small values 
- $\Lambda_{ab} \simeq 50$\,nm and $\Lambda_{c} \simeq 60$\,nm - at low magnetic
fields, and larger values - $\Lambda_{ab} \simeq 80$\,nm and $\Lambda_{c} \simeq
360$\,nm at high fields in MgB$_2$ \cite{Zeh04b}; accordingly, also this
anisotropy increased from about 1 to 4.5 with increasing field at 0\,K. A more
elaborated evaluation of the reversible magnetization within two-band Ginzburg
Landau theory \cite{Eis05a,Eis07a} revealed a smoother behavior of the
quantities over the whole field range and basically confirmed the above
mentioned trend. A quite different technique, neutron scattering by flux lines
of MgB$_2$ \cite{Cub03a}, suggested $ 30 - 60 \%$ larger penetration depths,
but the same trend for the field dependence of its anisotropy. 

\begin{figure}
    \centering
    \includegraphics[clip, width = 8.5cm]{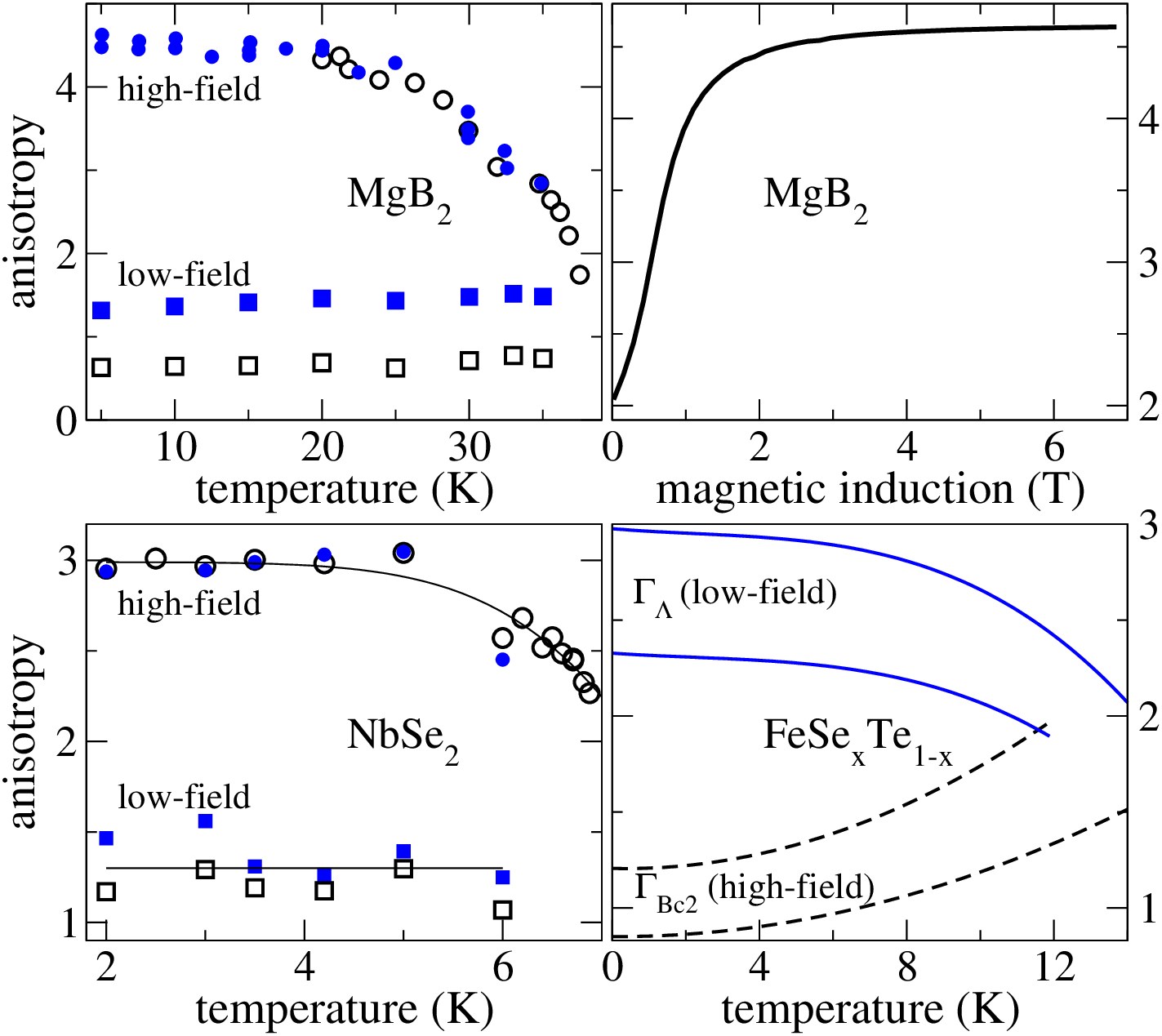}
   \caption{\label{Anisotropy} The anisotropies of several
single-crystalline samples at different temperatures and magnetic fields. The
left panels show anisotropy versus temperature in MgB$_2$ (top) and NbSe$_2$
(bottom), obtained from single-band fits to the low- or to the high-field
region, as illustrated in Fig.~\ref{Mrev}. The circles (upper curves)
refer to evaluations at high and the squares (lower curves) to evaluations at
low fields. Open symbols indicate upper critical field
or coherence length anisotropies. For MgB$_2$ the full symbols show the
penetration depth anisotropies obtained from the reversible magnetization in
the low-field and from the torque in the high-field case. For
NbSe$_2$ the full symbols show the anisotropies of the Ginzburg Landau
parameter. Notice that for both samples, any quantity measured at high fields
has high, and any quantity measured at low fields has low anisotropy. This is
illustrated in the upper right panel, where the field dependence of the
anisotropy of MgB$_2$ at 20\,K, evaluated by applying a real two-band Ginzburg
Landau model, is depicted \cite{Eis05a}. The bottom right panel refers to the
iron-based superconductor FeSe$_x$Te$_{1-x}$, with $x \sim 0.4 - 0.5$. The two
dashed curves roughly encircle the area in which the upper critical field
anisotropies have been found, while the two solid lines indicate the
low-field penetration depth anisotropies \cite{Ben10a,Fan10a,Lei10a}. Note
that, opposite to MgB$_2$ and NbSe$_2$, the high-field measurements led to low,
and the low-field measurements to high anisotropies in the iron-based
materials.}
\end{figure}

Because talking about the field dependence of the upper critical field
anisotropy does not make sense, we will turn to the closely related coherence length. The coherence length, acquired from the reversible magnetization of MgB$_2$ \cite{Zeh04b}, was reported to be  about 10.7 and 2.3\,nm at high and 17.4 and 34\,nm at low
fields, which gives an anisotropy comparable with that of the penetration depth
in the same field regime. Again, similar but more reliable results were obtained
from the two-band Ginzburg Landau model \cite{Eis05a,Eis07a}. Klein et
al.~\cite{Kle06a} determined the field dependence of the coherence length from
the Sommerfeld coefficient by assuming $\gamma (B)$ to be proportional to the
core size ($\sim \xi^2(B)$) and to the number of vortices ($\sim N \propto B$),
i.e. $\xi^2(B) \propto \gamma(B) / B$; their experimental finding that
$\gamma(B)$ of MgB$_2$ roughly behaves like $\sqrt{B}$ resulted in
$\xi(B) \propto B^{-0.25}$. A strong field dependence of the coherence length
was also suggested from scanning tunneling spectroscopy measurements of the
vortex core, indicating that the low-field coherence length might be much larger
(factor $\sim 5$) than that expected from Ginzburg Landau theory at the
upper critical field \cite{Esk02a} ($B_{\rm c2} = \Phi_0 / 2 \pi \xi^2$). In
NbSe$_2$, muon spin rotation measurements \cite{Cal05a} showed the coherence
length to decrease strongly with increasing field at low fields and to become
almost constant above a certain field, associated with the upper critical field
of the first band, which confirmed the results from the reversible magnetization
in this material \cite{Zeh10a}. The same held for the magnetic penetration
depth. 

It should be noted that a field dependent coherence length is not a specific
property of multi-band materials but shows up even in s-wave single-band
materials by a roughly $1/\sqrt{B}$ behavior \cite{Ich99a,Kog05a}, which is a consequence of vortex overlapping and the corresponding
delocalization of core quasiparticles \cite{Son04a,Son07a}. The field dependence
turned out to be particularly large in clean samples and at low temperatures
and might even become comparable with the effect in MgB$_2$ or
NbSe$_2$ for $H \parallel c$, where $\xi(0) / \xi(B_{\rm c2}) \sim 2$, but for
$H \parallel ab$, these ratios may be larger  in the two-band materials.
Furthermore, an abrupt change, as seen in NbSe$_2$ \cite{Cal05a} due to the
suppression of one of the bands, is not expected in conventional materials. 

We see that the anisotropies of the coherence length and of the magnetic
penetration depth display similar behavior in several two-band superconductors,
such as MgB$_2$ and NbSe$_2$. They are not only temperature but also
significantly field dependent, as can be seen in figure~\ref{Anisotropy}. For
instance, the anisotropies of MgB$_2$ and NbSe$_2$  are small at low fields, increase strongly with field, and are rather constant
at high fields. Obviously that reflects the dominant influence of the more
isotropic band at low and that of the more anisotropic band a high fields. The
correspondence of the coherence length and the penetration depth anisotropy at
all fields and temperatures cannot unambiguously be proved experimentally but has not been
refuted by the known experiments thus far. At least, the anisotropies were shown
to differ not significantly, as was also confirmed by the torque results of section
(\ref{sec:torque}). Moreover, the anisotropy of the lower critical field of
MgB$_2$ \cite{Lya04a}, which naturally corresponds to the low-field regime, was
found to be in reasonable agreement with the low-field $\Gamma_\Lambda$. Also, the
anisotropy of the Sommerfeld coefficient, acquired from specific heat
measurements at different magnetic fields, agreed with the other anisotropies
\cite{Pri07a,Bou02a}. Finally, direct tunneling spectroscopy measurements
indicated a small vortex lattice anisotropy, i.e. an almost undistorted
hexagonal lattice, and rather circular vortex cores at low fields perpendicular
to $c$ \cite{Esk03a}.

The low-field penetration depth and the high-field upper critical field
anisotropies are significantly different in the iron-based superconductors as
well, though their functionalities are contrary to MgB$_2$ and NbSe$_2$, as
illustrated in figure~\ref{Anisotropy}. The
question of different anisotropies at the same field and temperature has not
been addressed thus far. Torque data could be well described by the single-band
model with equal anisotropies and at the same time by a model including
different anisotropies \cite{Wey09a}.

In conclusion, considering a significant amount of experiments on MgB$_2$ and
similar materials leads us to the suggestion that the anisotropies of the
superconducting properties, in particular of the penetration depth and the
coherence length, are equal or almost equal in these two-band samples. These
anisotropies change with temperature and magnetic field. The field dependence
of the characteristic lengths and of their anisotropies reflects the properties and
anisotropies of the different bands. The situation is currently
unsettled in  the iron-based materials.

\subsection{Superconducting energy gap}

The most direct evidence for two-band superconductivity should be gained from a
measurement of the energy gap structure on the Fermi surface, which appears to
be a straightforward task. However, things are again not that simple as we will
see in the following. For instance, not all energy gaps necessarily show up in
tunneling or point-contact spectroscopy measurements; moreover, anisotropy would
also lead to varying gap values. Finally, angle-resolved photoemission
spectroscopy, which allows a direct assignment of gaps to certain bands, does
not work for all materials.

Studying two-band materials with distinct gaps reveals two independent gaps with
BCS-like temperature behavior and two different transitions when interband
coupling is ignored. According to calculations \cite{Nic05b}, slightly turning on
interband coupling hardly changes the larger gap, while the lower gap
curve should become flat near its single-band transition, so that it eventually
vanishes at the same temperature as the larger gap, as indicated in
figure~\ref{Fig:EliashB}. As interband coupling grows,
the lower gap becomes larger and smoother but still deviates from a BCS-like
shape, particularly at high temperature. The larger gap is less affected, though
some deviations from BCS behavior are to be expected. The ratio of the gaps is
reduced. Almost the same effects were predicted from increasing the interband
impurity scattering rate. Notice that both parameters also affect the transition
temperature. 

Scanning tunneling spectroscopy (STS) provides a direct way of measuring the
superconducting gap structure. In principle, a metallic tip is brought very
close ($\sim 1$\,nm) to the superconducting surface, so that charge carriers can
tunnel between them when a bias voltage $V$ is applied. At $T = 0$\,K tunneling
is only possible for $|V| > |\Delta| / e_0$, with $e_0$ the positive elementary
charge, and the gap magnitude, $|\Delta|$, thus becomes apparent when we record the tunneling current as a function of voltage (e.g. \cite{Fis07a}). The expression (e.g.
\cite{Che08a})
\begin{equation}
 	\label{eqn:gap:1}
	\frac{{\rm d} I}{{\rm d} V} \propto \rho_{\rm s}(E_{\rm F} + e_0V)
\end{equation}
provides an approximate relation between the tunneling current $I$ and the local quasiparticle
density of states of a superconductor $\rho_{\rm s}$, where $E_{\rm F}$ is the
Fermi energy, which will be set to zero from now on; the density of states of
the tip is assumed constant near the Fermi level. The density of states of a conventional s-wave superconductor, acquired from tunneling spectroscopy, may be described by
\cite{Dyn78a}
\begin{equation}
 	\label{eqn:gap:2}
	\rho_{\rm s}(E, \Upsilon) = {\rm Re} \frac{|E| + i \Upsilon}{\sqrt{(|E|
+ i \Upsilon)^2 - \Delta^2}}.
\end{equation}
Setting $\Upsilon = 0$ gives the standard BCS behavior of a
superconductor. The parameter $\Upsilon$ - usually denoted by $\Gamma$
in literature - takes the broadening of the curve features into account, which is mainly 
a consequence of impurity scattering.

The spectra of two-band superconductors have usually been modeled by $\rho_{\rm
s}$ = $n_1 \rho_{1}$ + $n_2 \rho_{2}$, with $\rho_{\rm i}$ the densities of the
different bands and $n_{\rm i}$ their relative contributions to the tunneling
current ($n_1 + n_2$ = 1). Dependent on the five parameters $\Delta_1$,
$\Delta_2$, $\Upsilon_1$, $\Upsilon_2$, and $n_1$, the density of states may show two
peaks on each side of the Fermi level or one peak and one shoulder at the
positions of the gap energies, as illustrated in figure~\ref{Fig:BCS}.
Unfortunately, similar structures can be acquired from anisotropic single-band 
(e.g. \cite{Haa01a}), while d-wave materials are indicated by a v-shaped density
of states around the Fermi level.

\begin{figure}
    \centering
    \includegraphics[clip, width = 8.5cm]{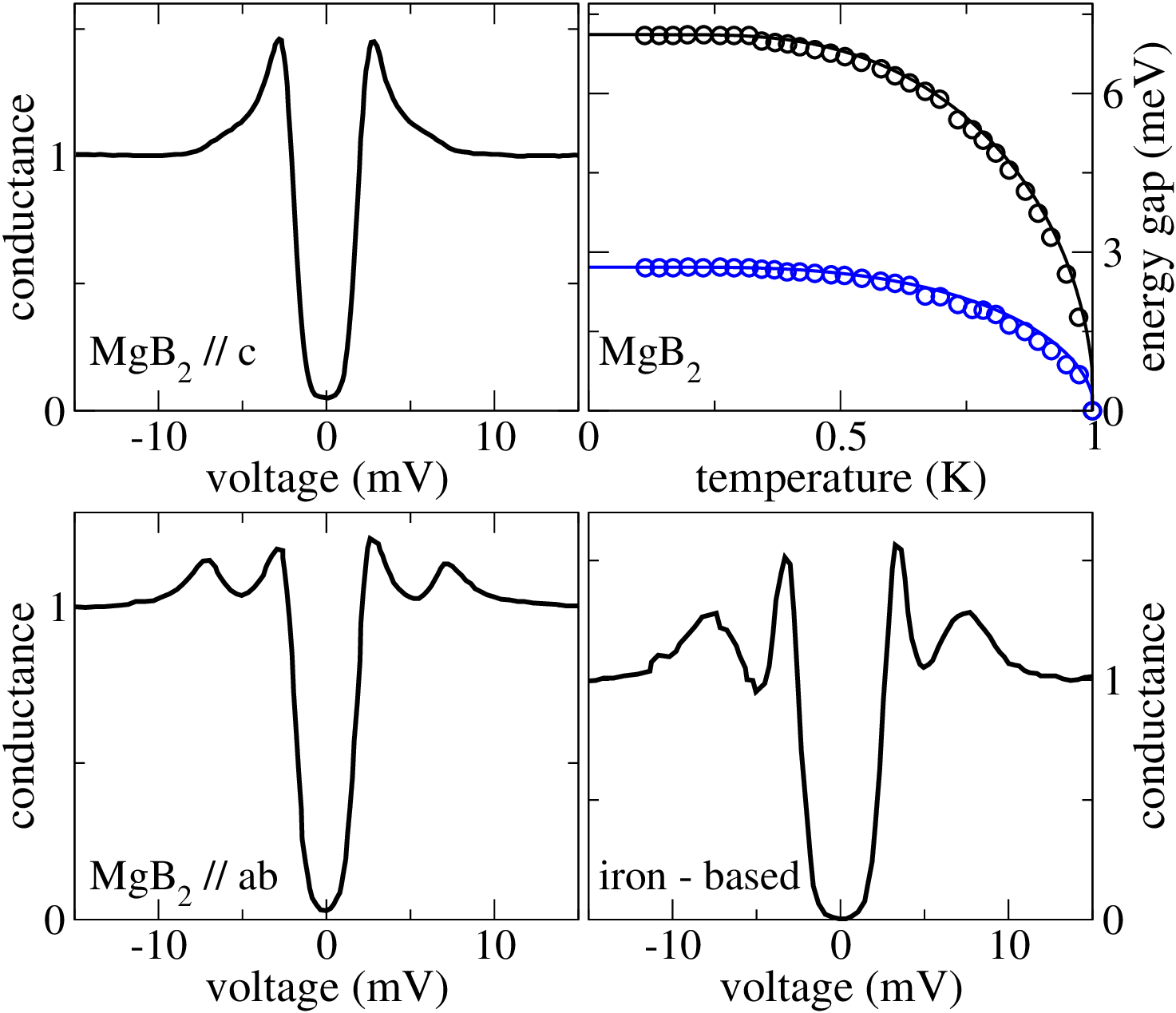}
   \caption{\label{Gap} The tunneling conductance (in arbitrary units) of
MgB$_2$ and of the iron-based material Ba$_{0.6}$K$_{0.4}$Fe$_{2}$As$_{2}$ as a
function of bias voltage and the energy gaps of MgB$_2$ as a function of
temperature \cite{Esk02a,Esk03a,Sha11a,Gon03a}. The left panels show
tunneling
spectroscopy results on MgB$_2$ for the tunneling current parallel (upper
panel) and perpendicular (lower panel) to the uniaxial $c$-axis of the
single-crystalline material at very low temperatures in zero magnetic field.
Notice that, when the current flows along $c$, only one peak, namely that
associated with the $\pi$-band, appears on each side of the Fermi level, while
only a vague shoulder at a higher voltage might indicate the second gap. In
contrast, two peaks are clearly visible for currents along $ab$. The circles in
the upper right panel display the energy gap values corresponding to the peak
positions in the spectroscopy measurements, though the displayed data were
extracted from point-contact spectroscopy curves at different temperatures.
Finally, the lower right data depicts the conductance of a
Ba$_{0.6}$K$_{0.4}$Fe$_{2}$As$_{2}$ single crystal at 3\,K, in which a two-peak
structure was found.} \end{figure}

Tunneling spectroscopy of MgB$_2$ revealed only one gap for currents parallel to
the uniaxial $c$-axis \cite{Esk02a}. This result was actually expected, for the
contribution of the $\sigma$-band to the conductivity was estimated to be not
more than about 1\% of that of the $\pi$-band. \cite{Bri02a} (cf. also
figure~\ref{Fig:BCS}). For currents perpendicular to $c$ the ratio of $n_\sigma$
to $n_\pi$ is about 1 : 2, and thus a double-peak structure could be detected
\cite{Esk03a}. In NbSe$_2$ and NbS$_2$, experiment revealed one peak and one
shoulder at a lower energy value for the tunneling current parallel to $c$
\cite{Gui08a,Gui08b}.

In many materials surfaces appropriate for tunneling spectroscopy are difficult
to prepare. Point-contact spectroscopy often places less demands on the sample
preparation; for a detailed review of point-contact spectroscopy see
\cite{Deu05a}.  In principle, a metallic tip is brought into contact with a
superconducting surface. To obtain reliable spectroscopy results, the size of
the contact should be smaller than the mean free scattering length of an
electron. Usually, an additional barrier, whose height is characterized by the
dimensionless parameter $Z$, emerges between the metal and the superconductor.
For $Z = 0$ a direct contact between the metal and the superconductor is
established. When we apply a voltage $V$ smaller than the gap, i.e. $e_0 |V| <
|\Delta|$, at $T = 0$\,K, only Andreev reflection takes place, which doubles the electrical
conductance at the interface (with respect to a metal - normal conducting
interface); if the voltage becomes larger ($e_0 |V| > |\Delta|$), electrons may
also directly propagate from the normal- to the superconducting part and the
conductance drops. Usually, a finite barrier is present ($Z > 0$), at which electrons can be simply reflected, which leads to a peak at the position of the gap and a reduction of the
conductance at lower values of $|V|$; eventually, at about $Z > 10$ the behavior
matches that of the tunneling spectroscopy (equation \ref{eqn:gap:2}). A
theoretical description of the conductance curves is given by the so-called BTK
model \cite{Blo82a}. 

A second band was shown to entail an additional peak or shoulder in the
density of states, as recently reviewed in \cite{Dag10a}. Similar to the tunneling
results, the conductance can be described by the sum of two weighted
BTK single-band curves. Though this model includes seven independent fit
parameters, $\Delta_1$,
$\Delta_2$, $\Upsilon_1$, $\Upsilon_2$, $Z_1$, $Z_2$, and $n_1$, the values
obtained for the two gaps are often reliable. Again, single-band anisotropy
provides similar modifications \cite{Dag10a}.

Several point-contact spectroscopy investigations have been carried out on
MgB$_2$. Applying the two-band fits led to good matching with the experimental
data and revealed gaps in good agreement with results from other methods. Also
the weights, $n_\pi$ and $n_\sigma$, were found in fair agreement with the
theoretical prediction \cite{Bri02a}, namely, for instance, $n_\sigma$ very
small and the corresponding gap thus hardly detectable for the $c$ direction.
Applying a magnetic field of 1\,T or more suppressed the $\pi$- and unveiled the
larger $\sigma$-gap \cite{Gon03a}.

Point-contact and tunneling spectroscopy experiments on the iron-based
materials often suffer from difficulties in surface preparation.
Nevertheless, two peaks or a peak and a shoulder in the
density of states have been observed by both methods
\cite{Sha11a,Hof11a,Tea11a,Dag11a} (see also figure~\ref{Gap}). 

Angle-resolved photoemission spectroscopy (ARPES) should provide the
most direct way of detecting superconducting gaps on different Fermi surface
sheets \cite{Dam03a}. In principle, the kinetic energy and the momentum vector (i.e. the
angular dependence) of electrons, that are activated by photons of known energy
and thus emitted from the sample by the photoelectric effect, are determined. As
a result, the Fermi surface and, if present, the corresponding gap amplitude can
be acquired. It should be noted that only surface states can be probed by this
methods, and that the surface preparation and a high stability are thus critical.
Moreover, the energy resolution may be an issue in some cases. Nevertheless, results that might be representative for the superconducting bulk state were obtained in many samples and compared with numerical calculations of the Fermi surface.

ARPES on NbSe$_2$ \cite{Yok01a} revealed three Fermi surface sheets, from which two
exhibited a superconducting gap at 5.3\,K. In MgB$_2$ the $\sigma$- and
$\pi$-band gaps were determined \cite{Tsu03a} in nice agreement with results
from other methods. Finally, several ARPES studies, showing up to 5 gaps on 5
different Fermi surface sheets, have been performed on the iron-based materials
\cite{Din08a,Ume12a,Zha12a}. 

To conclude, finding two peaks or a peak and a shoulder in tunneling or
point-contact spectroscopy measurements provides confirmation of a two-band
system, though the anisotropic single-band scenario cannot be completely
excluded. On the other hand, the absence of such structures does certainly not
disprove two-band superconductivity. Cogent evidence for the two- or
multi-band scenario should be gained from angle-resolved photoemission
spectroscopy by directly observing superconducting gaps on different bands.

\section{Materials \label{Sec:Materials}}
In this section, I will describe several materials for which attributes of
two- or multi-band superconductivity have been reported. I will mainly take
those into account that have been investigated after the discovery of
MgB$_2$, as this was the origin of a broader comprehension of the phenomenon,
while earlier reports on two-band effects in materials, such as
Nb \cite{She65a,Haf70a}, Ta \cite{She65a}, V \cite{She65a,Rad66a}, Nb-doped
SrTiO$_3$ \cite{Bin80a}, Mo$_{0.4}$Tc$_{0.6}$ \cite{Ste78a}, etc., are not
further discussed here. In the following, the more prominent materials, namely
MgB$_2$, NbSe$_2$ (with NbS$_2$), iron-based superconductors, cuprates, and
borocarbides, will be introduced in some detail. Then several other potential
two-band superconductors are presented more briefly. The list is not exhaustive, but I hope that not too many important materials were missed. In
some of these materials, multi-band superconductivity is quite well
established, in others the situation has yet to be clarified by further
experiments, and in some the reports are rather questionable. I will mainly
report on experimental data suggesting support or confirmation of the two-band
state, and I will not point out possible different interpretations, for instance in
terms of anisotropic single-band behavior, in each case.

\subsection{MgB$_2$} 

Since the discovery of its superconductivity in 2001 \cite{Nag01a}, MgB$_2$ has
become one of the most intensively studied superconducting materials, not only
because of its potential for applications, but also because it was the first
material in which two-band effects have been established. Today, it can be
considered as the prototype of a two-band superconductor, with which other
two-band candidates are compared, and to which most theoretical
calculations refer. 

MgB$_2$ has a hexagonal crystal structure, built up by a hexagonal magnesium
cell
and a ring of 6 boron atoms in the interior. The lattice parameters are
about 0.31 and 0.35\,nm. The Fermi surface of MgB$_2$ was calculated by several
groups and found to be not very complicated, though quite anisotropic
\cite{Kor01a,Cho02a}. Four bands crossing the Fermi surface were identified,
two $\sigma$-bands that have cylindrical shape and are strongly anisotropic,
and two $\pi$-bands that are rather isotropic. The $\sigma$-bands are usually
treated as one ($\sigma$) band and likewise the $\pi$-bands, making a two-band
description feasible. The energy gap has s-wave symmetry and opens on all
bands; the corresponding absolute values are around 7\,meV in the $\sigma$ and
2\,meV in the $\pi$-band. Two-band effects are experimentally observable
because both bands have similar densities of states and thus
contribute to superconductivity significantly but at the same time
diverge considerably in their properties.

A multitude of experiments and in particular their quantitative agreement with
theoretical results established that electron-phonon interaction drives
superconductivity in MgB$_2$. For instance, measurements of the transition
temperature showed the isotope effect, namely $T_{\rm c} \propto
M^{-\alpha}$, where $M$ is the mass of the element, with exponents $\alpha
\simeq 0.26 - 0.30$ for boron \cite{Hin01a,Bud01a} and $\alpha \simeq 0.02$ for
magnesium \cite{Hin01a}, which could be well reproduced by Eliashberg
theory \cite{Cho02a}. Thus, apart from the two-band effects, MgB$_2$ is
classified as a conventional superconductor.

The two-band nature of MgB$_2$ was suggested quite early and confirmed by
many experiments, yet most of them can also be interpreted by other models as
discussed in the previous sections. Some of these experiments have been
described above (cf., for instance, figures~\ref{SepModel1}, \ref{SepModel2},
\ref{Fig:Sommerfeld},  \ref{Fig:ThermH}, \ref{SFLDensity}, \ref{Torque}, 
\ref{Mrev},  \ref{Anisotropy}, and  \ref{Gap}) and shall not again be discussed here;
more can be found in review papers, such as Refs.~\cite{Eis07a,Xi08a}, and
references therein. Unambiguous evidence that energy gaps exist on two different
bands was given by  ARPES experiments \cite{Tsu03a,Tsu05a}, showing $\Delta
\simeq 5.5 - 7$ on the $\sigma$ and about 2\,meV on the $\pi$-band at low
temperatures.

Interesting effects are revealed when MgB$_2$ is probed in different magnetic
fields. Due to interband interactions, both bands have the same upper critical
fields, but as the field increases the relative contribution of the $\pi$-band
to the overall superconducting quantities apparently diminishes, so that
eventually the $\sigma$-band dominates. The field above which the $\pi$-band
becomes negligible, which is around 1\,T at low temperatures, is usually denoted
the second upper critical field of this band (yet, due to interband coupling,
traces of superconductivity are still expected at higher fields). The
suppression of the $\pi$-band makes the field dependence of diverse properties
different from that of an anisotropic single-band material. For instance, the
anisotropic single-band model was shown to fail in describing the reversible
magnetization of MgB$_2$ over the whole field range from 0\,T to $B_{\rm c2}$
\cite{Zeh04b}. Instead, a single-band fit worked well only for fields above the
$\pi$-band upper critical field, thus indicating a near single-band ($\sigma$)
behavior at high fields in agreement with the above mentioned suppression of the
$\pi$-band. The corresponding $\sigma$-band properties were reported to be about
3\,T for the upper, 0.07\,T for the lower, and 0.3\,T for the thermodynamic
critical field, as well as 77\,nm for the magnetic penetration depth and 11\,nm
for the coherence length for fields parallel to $c$ at 0\,K. The low-field
region could be described by a different single-band fit, but the corresponding
results include both $\pi$ and $\sigma$-band contributions and are thus
effective parameters only, though more indicative of the $\pi$-band. For fields
parallel to $c$ and low temperatures, an upper critical field of about 1\,T, a
lower critical field of 0.11\,T, a magnetic penetration depth of about 51\,nm
and a coherence length of 17\,nm were determined for this region. Those absolute
values are in good agreement with a more sophisticated two-band evaluation of
the reversible magnetization that gives a smooth field dependence of the
properties \cite{Eis05a,Eis07a}. To conclude, most quantities depend on the
magnetic field in a different way than expected from single-band theory.

Like the characteristic fields and lengths, also the corresponding anisotropies
change with magnetic field. At the upper critical field, the anisotropy is
mainly determined by the $\sigma$-band and thus about 4.5. As the field
decreases, the anisotropy remains almost constant at the beginning and is then,
starting at about the upper critical field of the $\pi$-band, continuously
reduced to about 1 at 0\,T due to the emerging influence of the $\pi$-band.
Close to the transition temperature, the field dependence of the anisotropy
becomes considerably weaker. As for the temperature dependence, the high-field
anisotropy is about 4.5 over a large range and slightly decreases close to the
transition temperature, while at low fields, the quantity is about 1 over a
large range and slightly increases close to the transition temperature, as
shown in figure~\ref{Anisotropy}. The experimental differences between the
anisotropies of different quantities, such as between the coherence length and
the penetration depth, at the same field and the same temperature were found to be
small, and are usually within the expected experimental uncertainties
\cite{Zeh04d,Zeh04b,Eis07a}. Assuming the anisotropy of all relevant quantities
to be equal in MgB$_2$, we can nicely explain most of the diverging reports on
this property in the literature by considering the field dependence of the
anisotropy. 

\subsection{NbSe$_2$ and NbS$_2$}

NbSe$_2$ is one of the most interesting and most studied superconductors. 
Although its properties have been analyzed for many years, its two-band
character was realized only recently in the wake of MgB$_2$. Today, there is
little doubt that NbSe$_2$ is a multi-band superconductor. The two-band effects
that show up in this material are quite similar to those of MgB$_2$.

NbSe$_2$ has a layered structure with a hexagonal unit cell. The layers consist
of two parallel Se planes and a Nb plane in between. Along the uniaxial
$c$-axis, these layers are only weakly bound by van der Waals forces. The
transition temperature of NbSe$_2$ is about 7.2\,K and its upper critical field
about 4 in $c$ and 12\,T in $ab$ (perpendicular to $c$) direction \cite{San95a},
giving an anisotropy of 3. Accordingly, most parts of the superconducting phase
diagram are available for experiments. Moreover, large pure single crystals,
whose superconducting magnetic properties are reversible over most of
the phase diagram, can be grown.

Thanks to the properties mentioned above, NbSe$_2$ is well suited for a wide
range of investigations. In particular, its layered structure makes preparing an
atomically flat surface by cleaving a crystal along the $ab$ planes easy, and,
in contrast to most other materials, this surface is very stable even in
air. Therefore, NbSe$_2$ was the first material to which scanning tunneling
microscopy was successfully applied for imaging vortex cores \cite{Hes90a} and
distributions \cite{Hes89a} (at arbitrary magnetic fields), and it is still
widely used for such investigations. Below about 33\,K we can study the charge
density wave state \cite{Mon75a} and at even lower temperatures its possible
competition with the superconducting gap \cite{Kis07a,Bor09a}. The existence of
high quality single crystals, showing almost no vortex pinning effects, allows
us to investigate the reversible magnetic properties and by introducing small
defects suitable for vortex pinning the second magnetization peak \cite{Bha93a}
(fishtail).

Following the discovery of MgB$_2$, new experiments were carried out on NbSe$_2$
and interpreted in terms of a two-band scenario, although most could as well be
described by the anisotropic single-band model. Some of these experiments were
presented in the previous sections and partly illustrated in
figures~\ref{Fig:BC2}, \ref{Mrev}, and \ref{Anisotropy}. The Fermi surface of
NbSe$_2$ was found to consist of three bands, two rather anisotropic (Nb 4d
derived) bands and a more three dimensional (Se 4p) band \cite{Ros01a}. ARPES
\cite{Yok01a} revealed gaps of about 0.9 - 1.0 meV at 5.3\,K on the two
cylindrical bands, while no gap was detected on the Se-band. Later, a gap
variation from 0.3 to 1.2 meV was measured, though again on the two Nb
bands \cite{Kis07a}, which was in better agreement with tunneling spectroscopy
showing a gap range from about 0.4 to 1.4 meV close to 0\,K \cite{Rod04a,Gui08a}. Two-band fits resulted in gaps of 0.73 and 1.26 meV from the specific heat \cite{Hua07a} and in comparable values from penetration depth measurements \cite{Fle07a}.

Recently, it was shown that the  field dependence of the reversible
magnetization of a NbSe$_2$ single crystal \cite{Zeh10a} cannot be properly
described by a single-band but by a simplified two-band model. Separating the
reversible magnetization into a high- and a low-field region allowed assessing
some two-band properties by fitting single-band Ginzburg Landau theory to both
regions separately, in the same way as done for MgB$_2$. As in MgB$_2$,
single-band behavior was found in the high-field and two-band behavior in the
low-field region. The field below which effects of the second band emerge is
usually proclaimed the upper critical field of this band and was found at
about 2\,T for both field orientations at 0\,K, as shown in
figure~\ref{Fig:BC2}. It was concluded that at least two bands contribute to the
superconducting state of NbSe$_2$, one is rather isotropic and suppressed above
roughly 2\,T (at 0\,K), and the second is strongly anisotropic ($\Gamma \simeq
3$). It appears natural to identify the isotropic part with the more three
dimensional Se 4p band, but it should be recalled that ARPES did not indicate a
gap on this band (at 5.3\,K), hence this point remains to be clarified. As in
MgB$_2$, the anisotropy of NbSe$_2$ was reported to change strongly with
applied field, namely from a large value of about 3 at high fields
to almost isotropic behavior at low fields and to depend only slightly on the
temperature. No significant differences in the anisotropy of the characteristic lengths were
detected when acquired at the same field and temperature (cf.
figure~\ref{Anisotropy}).

NbS$_2$ belongs to the same family, NbX$_2$, with X a chalcogen element, as
NbSe$_2$ and exhibits similar behavior. Its transition temperature is slightly
lower, about 6\,K, while the upper critical field anisotropy rises to
about 7-8 \cite{Ona78a,Kac10a}. Tunneling spectroscopy revealed one peak at about 1 and one
shoulder at about 0.5 meV in the density of states   \cite{Gui08b}; similar
gap values resulted from a two-band fit to specific heat data \cite{Kac10a} and
to the superfluid density \cite{Die11a}. Although these results do not prove
NbS$_2$ to be a two-band superconductor, its close relation to NbSe$_2$ makes it
probable.

\subsection{Iron-based superconductors \label{sec:mat:Fe}}

Though the first of the new iron-based superconductors was discovered in 2006
(LaOFeP \cite{Kam06a}; $T_{\rm c} \sim 4$\,K), the current boom broke out only
after a much higher transition temperature, namely 26\,K \cite{Kam08a}, was reported for LaO$_{1-x}$F$_x$FeAs. Since
then, a lot of materials with similar iron structures, reaching transition
temperatures of up to 56\,K, have been discovered. The main components of this
kind of superconductors are the parallel FeX layers with X a pnictogen (usually
P or As) or, less frequently, a chalcogen (Se,Te, and S) atom. The pure layers
form the so-called '11' family with transition temperatures of up to 14\,K
(FeTe$_{x}$Se$_{1-x}$). Putting an additional layer between the FeX blocks
yields further families. Most investigated are '122' materials with alkaline
earth metals, such as Ba, Sr, and Ca, in between (e.g. BaFe$_2$As$_2$), '111'
with alkali metals (e.g. LiFeAs), and '1111' with LnO layers, where Ln is a
metal of the lanthanide series (La, Ce, Sm, Nd, etc.), e.g. LaFeAsO. Almost all
elements can be partly replaced by different dopants, which may have great
effects on the superconducting properties. The highest transition temperatures
of about 56\,K were observed in the '1111' family. The unit cells are
usually orthorhombic, though the in-plane anisotropy of the lattice parameters
is almost negligible. For recent reviews see \cite{Ste11a,Oh11a,Asw10a}.

Many properties of the iron-based superconductors resemble those of the
cuprates. For instance, antiferromagnetism appears in both materials in the
strongly underdoped regime but is successively suppressed by doping. Eventually,
antiferromagnetism vanishes and superconductivity, with a dome-like shaped
transition temperature as a function of doping, shows up. In
both materials the superconducting pairing mechanism is yet to be found,
though electron-phonon coupling is excluded by most people, while spin
fluctuations are a good candidate. Contrary to the cuprates, the
iron-based materials are metallic in the underdoped regime. Moreover, the order
parameters of the materials are apparently different, namely d-wave
symmetry in the cuprates but most likely s-wave in most iron-based.

Though many aspects of the iron-based superconductors are not settled, there is
little doubt that they are multi-band materials. It was theoretically and
experimentally shown that most materials have two anisotropic electron Fermi sheet
pockets around the 'M' symmetry point of the Brillouin zone and at least two
anisotropic hole pockets around the '$\Gamma$' point; additional hole pockets
around '$\Gamma$' that may be more isotropic have occasionally been reported.
The superconducting order parameter is commonly believed to have s-wave symmetry
and to appear on the hole and the electron Fermi surface sheets and is supposed
to change its sign, in other words the phases might differ by an angle $\pi$ on
different sheets, which defines the so-called $s_\pm$ scenario. The question of
whether the electron band-gap has (accidental) nodes or not has been
discussed more controversially. Both viewpoints have been supported by a couple
of experiments. At present, there are indications that different scenarios may
prevail in different samples. In particular, it was reported \cite{Gof11a} that
optimally doped Ba211 samples might have a fully gapped order parameter, while
under- and overdoping might induce nodes. A review of the gap symmetry and
structure was recently presented \cite{Hir11a}. 

A large number of experiments has been published on the multi-band nature of
the iron-based superconductors. Of particular interest are angle-resolved photoemission spectroscopy
(ARPES) measurements, as such experiments are capable of revealing the energy
gap at different positions of the Fermi surface (restricted to the sample
surface, however). Results on different samples suggest gaps on several Fermi
sheets with different values. For instance, in Ba$_{0.6}$K$_{0.4}$Fe$_2$As$_2$,
gaps of about 12\,meV were assigned to the anisotropic hole and the electron
sheets, and a smaller 6\,meV gap to a larger hole sheet also surrounding the
$\Gamma$ point \cite{Din08a}. A more recent study on LiFeAs revealed (nodeless)
gaps on all four Fermi sheets, with larger gap values of around 4 - 5 meV on one
of the two hole and one of the two electron-like sheets and smaller values of
around 2.5-2.8 meV on the remaining sheets \cite{Ume12a}. Gaps on five sheets
were identified in BaFe$_2$(As$_{0.7}$P$_{0.3}$)$_2$ \cite{Zha12a}. For a
review, see \cite{Ric11a}. The results indicate up to five superconducting bands
in the iron-based materials, but since the gap magnitudes seem to cluster around
two different values, one larger and one smaller than the BCS gap, a two-band description might  be sufficient for describing most properties.

Further experimental results have partly been mentioned in the previous
sections. For instance, specific heat measurements have been carried out on
different samples. In contrast to the isotropic single-band model, the two-band
model usually worked well, though a shoulder at intermediate temperatures, like
that in MgB$_2$, was observed only in some iron-based samples. The two-band
evaluations typically resulted in gaps in the ratio of two to three, one gap
larger and one gap lower than the BCS prediction. The corresponding weights of
the density of states were about 0.3 : 0.7 to 0.5 : 0.5 (or even 0.1 : 0.9 in
one case) \cite{Har10a,Gof11a,Mu11a,Lin11a,Wei10a,Sto11a,Pra11a,Pop10a}. Probing
the superfluid density by two-band fits resulted in similar gap ratios.
\cite{Mar09b,Has09a,Mal09a,Gor10a,Kim10a,Lua11a,Gug11a}; see \cite{Pro11a} for a
review. As in MgB$_2$, the anisotropy of the penetration depths was found to be
temperature dependent, but contrary to MgB$_2$ decreasing upon warming
\cite{Mar09a,Kha09a,Wey09a,Ben10a}.

The upper critical fields of many iron-based materials were reported to match
those of the cuprates in absolute values ($\sim 100$\,T) for $H \parallel c$, yet their
shapes and anisotropies may be quite different. In most materials, an upward
curvature of the upper critical field emerges for magnetic fields perpendicular
to a uniaxial or near-uniaxial crystal axis, but in the iron-based materials, this feature
was observed for fields parallel to this axis and, contrary to other materials,
to appear over the whole or a large part of the temperature range in some samples
(e.g. '1111' and '122')
\cite{Hun08a,Jar08a,Qia09a,Kan09a,Sun09a,Ben10a,Khi10a,Fan10a,Mun12a} (see
figure~\ref{Fig:BC2}). For fields parallel to the iron layers ($H \parallel ab$),
higher upper critical fields, displaying no or only a slight upward curvature
near the transition temperature, were measured. At low temperatures, the upper
critical fields of many iron-based samples rise beyond the
Pauli paramagnetic limit, which is about 1.84 $T_{\rm c}$ in units of tesla in
the BCS limit, and the curves thus become noticeably flatter. The
corresponding Zeeman energy is rather isotropic, which could partly explain why
the upper critical field anisotropy of iron-based superconductors was found to
grow with increasing temperature
\cite{Jar08a,Kan09a,Sun09a,Ben10a,Fan10a,Put10a,Zha11a,Mun12a}, which is
contrary to the behavior of most other known two-band superconductors, as
illustrated in figure~\ref{Anisotropy}, and to anisotropic single-band
superconductors. The maximum anisotropy near the transition temperature was
reported between about 2 and 7. For a review of the upper critical fields in the
iron-based materials, see \cite{Gur11a}.

As for the anisotropy of the iron-based superconductors, we can summarize that
the low-field penetration depth anisotropy was reported to decrease while the
upper critical field anisotropy to increase upon warming \cite{Wey09a,Ben10a}
towards a common value at the transition temperature ($\sim$ 2-7). This
resembles the behavior of MgB$_2$ when the upper critical field and the
penetration depth anisotropy behaviors are swapped, as depicted in
figure~\ref{Anisotropy}. The temperature dependence of the penetration depth
anisotropy is usually assumed to be a consequence of the multi-band character,
while that of the upper critical field of the Pauli paramagnetism limit, but
microscopic confirmations and detailed multi-band model calculations are yet to
be provided for this material. In MgB$_2$ the differences seem to disappear when
the anisotropies, e.g. of the penetration depth and the coherence length, are
determined at the same field and temperature. This issue has not been addressed
in the iron-based samples thus far, though torque measurements
\cite{Wey09a,Wey10a,Li11a,Bos12a} did not indicate an unconventional behavior (e.g.
additional nodes in the reversible torque vs. angle curve) as it may be the case
for different anisotropies.

Tunneling spectroscopy has been carried out on several samples, though
uncertainties due to sample surface quality issues are to be expected. Moreover,
it was predicted that tunneling mainly probes the bands around the '$\Gamma$'
point, whereas the electron pockets around '$M$' provide only negligible
contributions to the tunneling current \cite{Lee09a}, which might explain the
single-band behavior found in many tunneling spectroscopy studies. Nevertheless,
two gaps were resolved in \cite{Tea11a} Ba(Fe$_{1-x}$Co$_x$)$_2$As$_2$ and
\cite{Lei11a,Sha11a} Ba$_{0.6}$K$_0.4$Fe$_2$As$_2$, in both cases with a gap
ratio of about 1 : 2. For a review, see \cite{Hof11a}. 

Point-contact spectroscopy on iron-based samples also struggles with
several problems that led to different results and quite different
interpretations of the gap structure \cite{Dag10a}, but yet, clear
two band spectra with either two peaks or a peak and a shoulder were found and
fitted using the two-band models \cite{Yat09a,Sam09a,Sza09a,Dag09a}. A
review was recently given in \cite{Dag11a}.

\subsection{Cuprates}

The cuprate superconductors, discovered in 1986 by Bednorz and M\"uller
\cite{Bed86a}, reach the highest known transition temperatures. Though
 still among the most intensively studied materials, they are far from
being fully understood.

In the following, I will point out some cornerstones of the cuprates, though for
details the reader is referred to one of the numerous books or reviews on this
subject (e.g. \cite{Poo07a,Hack11a,Lee06a,Bas05a}). The superconducting cuprates
have a layered structure with a strongly anisotropic orthorhombic or tetragonal
unit cell. The copper-oxygen (CuO$_2$) planes are common to all varieties and
are the places where superconductivity is induced. They are sandwiched between
other non-metallic layers supplying charge carriers. Among the most prominent
representatives are YBa$_2$Cu$_3$O$_{7-x}$ and
Bi$_2$Sr$_2$Ca$_{n-1}$Cu$_n$O$_{2n+4-x}$, having transition temperatures of up
to 92 and of up to 110\,K; the highest known transition temperature of
135\,K is reached in HgBa$_2$Ca$_2$Cu$_3$O$_8$. In the undoped state (e.g.
YBa$_2$Cu$_3$O$_6$) the materials are antiferromagnetic Mott insulators. Adding
dopants (e.g. oxygen atoms) gradually reduces their Neel temperature to zero and
eventually induces the superconducting state, displaying a dome-like shape in
the transition temperature versus doping diagram. Optimal doping refers to the
maximum transition temperature. The cuprates are classified as extreme type II
superconductors, for their Ginzburg Landau parameter is often in the range of
100, they have the highest-known transition temperatures thus far, and their
anisotropy reaches values from about 5 to more than 100. A lot of questions
concerning these materials are still not resolved including that on the
mechanism responsible for Cooper-pairing. The most frequently mentioned
candidate is the coupling by antiferromagnetic spin fluctuations, which should
lead to d-wave symmetry of the order parameter.

There is wide consensus that d-wave symmetry predominates in the cuprates
\cite{Tsu00a}, yet not all experiments are in full agreement with this
assumption. For instance, Khasanov et al. \cite{Kha07b} determined the in-plane
magnetic penetration depth of La$_{1.83}$Sr$_{0.17}$CuO$_4$ by muon-spin
rotation and found the temperature dependence of the superfluid density to rise
abruptly below about 10\,K at low magnetic fields, which was no longer
observed at higher fields ($> 0.64$\,T). This anomaly was assumed to indicate a
second band. Fitting the curves with the two-band $\alpha$-model revealed gap
values of 8.2\,meV for the dominant d-wave band and 1.6\,meV for the second
band, for which s-wave character was supposed. Similar results were reported for
YBa$_2$Cu$_3$O$_{7-x}$, presented in figure~\ref{SFLDensity}, and
YBa$_2$Cu$_4$O$_8$ \cite{Kha07a,Kha08a}. Conclusions on a possible s-wave
admixture to the d-wave symmetry were also drawn from other techniques
\cite{Smi05a,Fur08a,Bus12a,Kir06a}; a second d-wave component was considered in
\cite{Val10a}. A temperature dependent anisotropy, resembling that of MgB$_2$,
was detected in Sm and Y based cuprates \cite{Kor10a,Bos11a}, which, however, is
not exclusively explicable by two-band effects, as mentioned in section
\ref{Bc2}. Application of ARPES in samples such as YBa$_2$Cu$_4$O$_8$ suffers
from sample surface preparation difficulties. Nevertheless, deviations from a
pure d-wave symmetry were suggested, and a $d+is$ symmetry was proposed
\cite{Oka09a}.

Not all unexpected results in the cuprates have been interpreted in terms of a
second band. For instance, Wojek et. al \cite{Woj11a} reproduced the
low-temperature low-field anomaly of the superfluid density in
La$_{1.83}$Sr$_{0.17}$CuO$_4$ but attributed it to vortex pinning effects and
were thus able to describe the whole temperature dependence with a single d-wave
gap-band. Clearly, many experiments have pointed to a single d-wave band
\cite{Tsu00a}, yet it was suggested that the suppression of the second-band
contribution may be a consequence of surface sensitive measurements
\cite{Mul07a}.

To conclude, most experiments on superconducting cuprates are well explicable by a single d-wave band. Nevertheless, the discrepancies found in some experiments, which were partly supposed to indicate a second band, should be clarified.

\subsection{Borocarbides}

Borocarbides were discovered in 1994 \cite{Cav94a,Nag94a}. They have
transition temperatures of more than 20\,K and exhibit unconventional
properties. Most investigated are the rare-earth nickel borocarbides
RNi$_2$B$_2$C with R = Y, Lu, Tm, Er, etc. They have a strongly anisotropic
tetragonal unit cell with lattice parameters of about 0.35 in $a$ and 1\,nm in
$c$ direction and consist of alternate stacks of Ni$_2$B$_2$ and RC layers.
Several properties make the borocarbides an interesting object for
investigations, such as for instance the competition and partial coexistence of
superconducting and magnetic ordering. The borocarbides were originally thought
to be conventional phonon-mediated s-wave superconductors. Later, a strong
anisotropy of the superconducting gap, which is not mirrored by the
rather isotropic normal conducting electronic properties, was noticed, and nodes
in the gap were indicated by several experiments. This led to the suggestion
that $s + g$ - wave symmetry \cite{Mak02a}, which means a gap with point nodes,
exists in these materials. Alternatively, experiments have also been described
by a two-band model. For reviews see \cite{Mul01a,Nie11a,Gup06a}. In the
following I will concentrate on the non-magnetic borocarbides, such as
YNi$_2$B$_2$C and LuNi$_2$B$_2$C. They have transition temperatures of about 15
- 16\,K and Ginzburg Landau parameters of 10 - 20, and their upper critical
fields may slightly surpass 10\,T. The electron - phonon coupling strength was
shown to be close to 1. 

The Fermi surface of the borocarbides was experimentally and numerically
determined by several groups
\cite{Dug99a,Sta08a,Ber08a,Dug09a,Bab10a,Mat94a,Kim95a,Yam04a} and found to be
multi-sheeted. ARPES of RNi$_2$B$_2$C showed that a superconducting gap opens on
at least two Fermi sheets \cite{Bab10a}, making multi-band superconductivity
possible. The two-band model for describing borocarbides was first reported by
Shulga et al. \cite{Shu98a} even before the advent of MgB$_2$. They reported a
pronounced upward curvature in the upper critical field of YNi$_2$B$_2$C and
LuNi$_2$B$_2$C, which they could bring into agreement with the two-band
Eliashberg theory. Later, a similarly good agreement was achieved by applying
the single-band separable model \cite{Man01a}. Four different scenarios (line
nodes, point nodes, s+g, and the two-band model) were adjusted to specific heat
data of YNi$_2$B$_2$C in \cite{Hua06a}, from which the two-band model worked
best, while the s+g was rather ruled out by the authors. The resulting gap
energies were about 2.7 and 1.2 meV. Additionally, they found the Sommerfeld
coefficient to increase with magnetic field $H$ as $H^{0.47}$, which is actually
close to the $d$ - wave forecast, though likewise explicable by the two-band
model. Multi-band superconductivity was also suspected from point-contact
spectroscopy measurements \cite{Bas07a},  which revealed gap values of ca. 2.3
and 1.5 meV in the same material. Finally, the field dependencies of the two
gaps were studied by point-contact spectroscopy \cite{Muk05a}. The
larger gap was reported to vanish at the upper critical field but the smaller
one at a much lower field, which was considered as evidence for the two-band
scenario. Other experiments such as the point-contact spectroscopy presented in
\cite{Lu11a} did not reveal any signs of a second band. Two gaps were also claimed
to exist in the magnetic borocarbide TmNi$_2$B$_2$C as a result of analyzing
point-contact spectroscopy curves \cite{Nai11a}. 

To conclude, the borocarbides are serious candidates for two-band
superconductivity, as seen from their multi-sheeted Fermi surface and several
experiments such as upper critical field, specific heat, and point-contact
spectroscopy measurements. For an unambiguous decision more
experimental and theoretical studies are necessary.

\subsection{Further potential multi-band superconductors}

After presenting the most prominent potential two-band superconductors, I will
attach shorter descriptions of further materials in which possible two-band
effects were suspected. The status of verification differs from material to
material, but to unambiguously clarify whether or not multi-band superconductivity
occurs calls for further experiments in most cases.

\textit{Nb$_3$Sn.} Nb$_3$Sn is one of the best-known superconducting materials
since many years. After niobium-titanium, it is the most important commercial
superconductor, mainly used for high-field magnets. Its transition temperature
is about 18\,K, and upper critical fields of up to 30\,T can be reached; see
\cite{God06a} for a recent review. A specific heat anomaly at low temperatures
has long been known and led to early speculations about a two-band scenario in
this material \cite{Vie68a}. This was supported by the recent finding of a
shoulder in the specific heat of a polycrystalline sample at about 0.25 $T_{\rm
c}$ \cite{Gur04a}. Applying the two-band $\alpha$-model revealed two gaps with
0.6 and 3.7\,meV and corresponding band weights of 7.5 : 92.5. The anomaly
vanished at magnetic fields above about 5\,T. Several other explanations for the
shoulder, including the possible existence of a different superconducting phase,
were rejected. Point-contact spectroscopy on samples from the same source
confirmed the specific heat results on the gaps \cite{Mar10a}. In contrast,
later specific heat measurements on several Nb$_3$Sn single crystals did not
show any trace of the low-temperature anomaly, and the authors \cite{Esc09a}
thus concluded that Nb$_3$Sn were a single-band material. 

\textit{V$_3$Si.} Several authors quoted V$_3$Si as an example of two-band
superconductivity, and several authors quoted V$_3$Si as an example of
single-band superconductivity. The typical transition temperature of this
material is about 17\,K and the low-temperature upper critical field roughly
20\,T. In \cite{Nef05a}, the microwave response of V$_3$Si single crystals
was shown to deviate significantly from BCS behavior. The corresponding temperature dependence
of the superfluid density was fitted applying a two-band model with two gaps of about 2.5 and 1.3\,meV; the interband coupling was
claimed to be very weak in this material. These results were confirmed in
\cite{Kog09a}. A two-band state was also proposed from infrared spectroscopy
data of oriented films \cite{Per10a}; in the same work, the Fermi surface was
calculated and shown to be crossed by four bands. 

Other studies led to contrary conclusions. For instance, the field dependence of
the reversible magnetization was shown to match the predicted single-band
behavior reliably well over the whole field range \cite{Zeh10a}, as shown in
figure~\ref{Mrev}, which is in stark contrast to the two-band materials NbSe$_2$ and
MgB$_2$ \cite{Zeh04b}. This was confirmed by thermal conductivity and specific
heat experiments, in which V$_3$Si displayed the expected single-band field
dependence \cite{Boa03a}. In \cite{Son04b}, the magnetic field effects on
the vortex core size and on the magnetic penetration depth, measured via
muon-spin rotation, could be explained within the single-band picture (see also
\cite{Son07a} for a comparison with two-band materials). In \cite{Gur04a}
the temperature dependence of the specific heat of V$_3$Si poly- and single
crystals was determined (to compare it with Nb$_3$Sn) and no low-temperature
anomaly was reported. Such a low-temperature anomaly is not an inevitable
consequence of two-band superconductivity, as mentioned in a previous section,
but rather expected when interband coupling is weak, and interband coupling was
claimed to be weak in this material \cite{Nef05a,Kog09a}. To conclude,
even for such a long-known superconductor as V$_3$Si, further experiments are
necessary to clarify the possible two-band state. 

\textit{CeCoIn$_5$.} The heavy fermion superconductor CeCoIn$_5$ has a
transition temperature of 2.3\,K and displays d-wave gap symmetry \cite{Pet01a}.
Point-contact spectroscopy revealed a multiple-structured curve \cite{Rou05a},
interpreted as a reflection of two superconducting d-wave gaps of about 0.95 and
2.4\,meV. This interpretation was questioned \cite{She06a,Par06a}, and it was
shown that the Andreev reflection data may also be described within a single-band
model \cite{Par05a}. Thermal conductivity measurements seemed
to support the two-band scenario \cite{Sey08a}, though with much smaller gaps
than claimed in \cite{Rou05a}. A two-band scenario was also proposed from the
temperature and field dependence of the anisotropy determined by torque
measurement \cite{Xia06a}, though the effects were not very significant.
Finally, band structure calculations revealed a multi-sheeted Fermi surface
\cite{Mae03a}. 

\textit{PrX$_4$Sb$_{12}$ (X = Os, Ru), LaRu$_4$As$_{12}$, and
LaOs$_4$Sb$_{12}$.} For PrOs$_4$Sb$_{12}$ ($T_{\rm c} \simeq
1.85$\,K; $B_{\rm c2}$(0\,K)$\simeq$ 2.2\,T) two-band superconductivity was
assumed due to the unconventional field dependence of the thermal conductivity
\cite{Sey05a}. The low-temperature thermal conductivity was reported to
increase
strongly at very low fields, less steeply  at intermediate fields, and again
strongly near the upper critical field, which is similar to the results on
MgB$_2$, while in conventional single-band materials a more gradual
increase is expected. As in MgB$_2$, a band with a smaller upper critical field
($\sim 0.07 B_{\rm c2}$) was held responsible for the steep rise at low fields
and a second band with a higher upper critical field for the rise at high
fields. In \cite{Sey06a} two s-wave gaps with a ratio of about 3 were proposed,
while in \cite{Hil08a} only one of the bands was assumed to be fully gapped and
the other one to have nodes. Two bands with s-wave symmetry were also reported
for PrRu$_4$Sb$_{12}$, based again on thermal conductivity measurements
\cite{Hil08a}. LaRu$_4$As$_{12}$ has a transition temperature of about 10\,K
and an upper critical field of 12\,T. Observations of an upward curvature in the
upper critical field, a non-BCS specific heat temperature dependence, though
without a shoulder at intermediate temperatures, and a non-linear field
dependence of the Sommerfeld coefficient led to the suggestion of a two-band
scenario \cite{Nam08a,Boc12a}. In LaOs$_4$Sb$_{12}$ ($T_{\rm c} \simeq
0.74$\,K) such a suggestion came from a convex temperature dependence of the
superfluid density \cite{Tee12a}.

\textit{Ba$_8$Si$_{46}$.} Ba$_8$Si$_{46}$ has a transition temperature of about
8\,K, an upper critical field of roughly 6\,T at 0\,K, and might be mediated by
electron - phonon coupling. The gap function appears to have s-wave symmetry,
and the Fermi band was found to be crossed by several bands \cite{Tse05a}.
Specific heat curves were published in \cite{Lor08} and claimed to be best
described by a two-band model with one small gap of about 0.35\,meV, which
contributes only 10\% to the data, and one dominating second gap, which is about
four times larger. A possible two-band state, though with a smaller gap ratio of
about 1.4, was also deduced from tunneling spectroscopy results
\cite{Noa10a}.

\textit{Na$_{0.3}$CoO$_2 \cdot$ 1.3H$_2$O.} Samples of Na$_{0.3}$CoO$_2 \cdot$
1.3H$_2$O have a maximum transition temperature of about 4.5\,K, but were
shown to change their superconducting properties with time when kept at room
temperature. Specific heat measurements revealed resemblance to MgB$_2$ and were
therefore analyzed by applying the two-band s-wave $\alpha$-model \cite{Oes08a}.
Two different samples were investigated in which the gap ratio, obtained from a
two-band fit, changed from about 3 in the first sample to 2 in the second and
the corresponding density of states ratio, from about 1 to 4. The upper critical
fields of the first sample were roughly estimated to be about 2 for the first
and 8-9\,T for the second band.

\textit{OsB$_2$.} The layered boride material OsB$_2$ becomes superconducting
below about 2.1\,K. Its upper critical field is roughly 20\,mT and its Ginzburg
Landau parameter rather small though still in the type-II regime. A two-band
state was concluded \cite{Sin10a} from measurements of the superfluid density,
which could not be well described by a single-band s-wave model but by a
two-band model with two s-wave gaps, from which one was larger and one smaller
than the BCS value and of which the ratio was about 0.66. The assumption was
supported by the specific heat jump at the transition temperature, which was
found to be smaller than expected from BCS theory (as is also the case for
MgB$_2$), namely $\Delta C(T_{\rm c}) / (\gamma_{\rm n} T_{\rm c}) \simeq 1.3$. 

\textit{XMo$_6$S$_8$ (X = Sn, Pb).} The superconducting Chevrel phases are known
for their rather high transition temperatures of up to 15\,K and very high upper
critical fields, about 40\,T in SnMo$_6$S$_8$ and 80-90\,T in
PbMo$_6$S$_8$. In a recent paper, Petrovic et al. \cite{Pet11a} reported
possible two-band effects in these materials from tunneling spectroscopy and
specific heat measurements. The density of states, obtained from tunneling
spectroscopy, partly showed a shoulder in addition to the main coherence peak. A
simple fit model led to energy gaps of about 2.95 and 1.05\,meV with relative
density of states of 0.62 and 0.38 for SnMo$_6$S$_8$ and about 3.1 and 1.4\,meV
with relative density of states of 0.66 - 0.9 and 0.34 - 0.1 for PbMo$_6$S$_8$. Analyzing specific heat measurements, which
probe the bulk state, using the two-band $\alpha$-model confirmed the gap
values but resulted in somewhat different weights, namely about 0.96 : 0.04
in SnMo$_6$S$_8$ and in 0.9 : 0.1 in PbMo$_6$S$_8$. Based on a small kink in the
field dependence of the Sommerfeld coefficient, the authors estimated the
smaller band upper critical fields to be about 2.8 and 4.3\,T in the two
samples.

\textit{ZrB$_{12}$.} ZrB$_{12}$ becomes superconducting below about 6\,K. It has
a rather low Ginzburg Landau parameter (even a change from type I to type II
behavior with decreasing temperature was assumed \cite{Wan05a}), upper critical
fields of roughly 100\,mT, and electron - phonon coupling seems to be
responsible for Cooper-pairing. Two-band effects were suggested from an
unconventional temperature dependence of the superfluid density and of the upper
critical field in a single crystal \cite{Gas06a}. A two-band fit of the
superfluid density, showing a shoulder similar to MgB$_2$ at about 4\,K, led to
gap values of 2.1 and 0.73\,meV. The upper critical field was found to increase almost
linearly even close to 0\,K, which was explained by the dirty
limit two-band model, as discussed in \cite{Gur03a}. The conclusions were backed
by showing the Fermi surface to consist of several distinct sheets
\cite{She03a}. In \cite{Slu11a} the anomaly at about 4-5\,K, observed in this
case in the specific heat, was supposed to be caused by a possible structural
phase transition similar to that in LuB$_{12}$. Other experimental data,
namely from tunneling \cite{Tsi04a}, point-contact spectroscopy \cite{Dag04a},
and specific heat measurements \cite{Lor05a} also pointed to single-band s-wave
superconductivity in this material.

\textit{X$_2$C$_3$ (X = La, Y).} The sesquicarbides La$_2$C$_3$ and Y$_2$C$_3$
reach transition temperatures of up to 18\,K and upper critical fields of
some 10\,T. Recent nuclear-magnetic-resonance measurements \cite{Har07a}
suggested that Y$_2$C$_3$ were a noncentrosymmetric, yet a spin-singlet and
s-wave multi-band superconductor. The temperature dependence of the nuclear
spin-lattice relaxation rate showed an atypical kink at about 5\,K, which was
taken as a sign of two-band effects. Fitting these data with an
$\alpha$-model-like curve revealed two gaps with sizes of 3.4 and 1.4\,meV and
relative weights of 0.75 : 0.25. Confirmation came from muon spin relaxation
measurements in Y$_2$C$_3$ and La$_2$C$_3$ \cite{Kur08a}. In La$_2$C$_3$
polycrystals the temperature dependence of the superfluid density showed an
abrupt kink or shoulder, indicating two gaps of 2.7 and 0.6\,meV and relative
weights of 0.38 : 0.62. In Y$_2$C$_3$ the two-band effects were less obvious,
but still a single-band BCS fit did not work well. The two-band fit resulted in
slightly larger gaps than in La$_2$C$_3$ but in smaller gaps than those from the
nuclear-magnetic-resonance method, namely 3.1 and 0.7\,meV and weights
of 0.86 : 0.14.  For both materials s-wave symmetry was assumed, and different
interband coupling strengths were held responsible for parts of the differences.
On the other hand, based on tunneling spectroscopy results, the authors of
\cite{Eki12a} rejected multiband superconductivity in Y$_2$C$_3$ and ascribed
its seeming appearance in diverse experiments to local phase differences and
inhomogeneities in the investigated samples.

\textit{X$_2$Fe$_3$Si$_5$ (X = Lu, Sr).} Lu$_2$Fe$_3$Si$_5$ is a member of the
ternary-iron silicide superconductors X$_2$Fe$_3$Si$_5$ (X = Lu, Y, Sc, Tm, Er). They have some interesting and peculiar properties from which some were
assumed to originate from two-band effects already in 1983 \cite{Vin83a}.
Lu$_2$Fe$_3$Si$_5$ has a transition temperature of about 6\,K and upper critical
fields in the range of 10\,T with a moderate anisotropy of roughly 2. Band
structure calculations showed the Fermi surface to be crossed by three bands
having different anisotropies \cite{Nak08a}. As in MgB$_2$, the specific heat
of single crystals \cite{Nak08a,Bis11a} displayed a distinct shoulder at about
0.2\,T$_{\rm c}$, and the normalized jump $\Delta C(T_{\rm c}) / (\gamma_{\rm n}
T_{\rm c}) \simeq 1.1$ was found to be smaller than the BCS value of 1.43, as
presented in figure~\ref{Fig:Spec}.
Two-band superconductivity was thus proposed and probed by the $\alpha$-model
fit, resulting in two s-wave gaps with the ratio of 1 to 4 but almost
equal Sommerfeld constants. Penetration depth measurements \cite{Gor08a,Bis11a}
confirmed two, though slightly smaller, gaps.
The field dependence of the thermal conductivity \cite{Mac11a} indicated
two bands with upper critical fields of approximately 6.4 and 0.26\,T. This was
confirmed by measurements of the Sommerfeld coefficient, revealing an almost
isotropic upper critical field of circa 0.33\,T on the smaller gap-band and a
more anisotropic behavior ($\sim 2$) on the larger gap-band. Sr$_2$Fe$_3$Si$_5$
has a slightly smaller transition temperature of about $5$\,K and smaller upper
critical fields though with a similar anisotropy. The temperature dependence of
the specific heat was reported to have a shoulder as well; the corresponding
two-band fit led to a gap ratio of about 2 and to relative weights of the
two bands of 0.36 : 0.64 \cite{Tam08a}. The specific heat jump at the transition
temperature was observed to be only about half of the BCS height.

\textit{MgCNi$_3$.} The cubic antiperovskite compound MgCNi$_3$ is usually
described as a fully gapped, rather strong coupling phonon-mediated
superconductor with a transition temperature of 7\,K and an upper critical field
in the order of 10\,T. Although specific heat and upper critical field data of a
polycrystal could be reasonably well described by a single-band model, the
authors of \cite{Wal04a} suggested a two-band model, which, they claimed,
would be necessary for explaining literature data of the Hall-conductivity and
thermopower and reports on different gap values on the Fermi surface.
Data on single crystals, including the specific heat, point-contact
spectroscopy, and the penetration depth \cite{Pri11a,Die09a}, did not indicate
multi-band effects. 

\textit{Sr$_2$RuO$_4$.} Sr$_2$RuO$_4$ is certainly one of the most
unconventional and hence interesting superconducting materials. Its tetragonal
crystal structure is highly anisotropic and similar to that of the cuprates. The
transition temperature is not more than about 1.5\,K; the upper critical field
is strongly anisotropic ($\sim$ 20) and reaches a maximum of roughly 1.5\,T
\cite{Aki99a}. Experiments indicated triplet pairing and broken
time-reversal symmetry, suggesting a chiral p-wave order. Moreover, topological
superconductivity and Majorana fermions have been studied in this material.
For details, I refer to recent review papers \cite{Kal12a,Mae12a} and
references therein. To make things even more complicated, multiband effects were
suggested in order to resolve some of the discrepancies between theory and
experiment. The cylindrical-like Fermi surface is crossed by three bands
\cite{Ber00a} named $\alpha$, $\beta$, and $\gamma$, from which the
$\gamma$-band, pocketing about 60\% of the density of states, has usually been
assumed to dominate the superconducting properties. Measurements of the specific
heat showed that the jump at the transition temperature is much smaller than the
BCS prediction and that a significant kink in the field dependence at about
0.15\,T \cite{Nis00a,Deg04a} appears, which was assumed to mark the low-field influence
and upper critical fields of the $\alpha$ and the $\beta$ band. While
the $\gamma$-band was assumed to be fully gapped, nodes were supposed for the
gaps of the other bands. Other studies rejected the two-band scenario for
Sr$_2$RuO$_4$. For instance, the anomalous field dependence of the specific heat
was attributed to strong Pauli-paramagnetic effects, which was confirmed by
calculations within the Eilenberger model \cite{Mac08a}. Tunneling spectroscopy
\cite{Sud09a} uncovered but one (fully open) gap with a size of about 0.3 meV,
which is close to the predicted BCS value. 

\textit{URu$_2$Si$_2$.} URu$_2$Si$_2$ is a heavy fermion compound that becomes
superconducting below about 1.5\,K and has an anisotropic upper critical
field between ca. 3 and 12\,T. The material has attracted interest due to
its phase transition to a still unknown ordered state, called a hidden order
state, at about 17\,K; for a review see \cite{Myd11a}. In
\cite{Kas07a} the field dependence of the thermal conductivity was regarded as a
confirmation of two-band superconductivity. The authors observed a
steep increase below about 0.4\,T for different field orientations, as expected
from an isotropic band with a low upper critical field, and a flatter, more
anisotropic behavior at higher fields, assumed to mirror the second anisotropic
band. Both bands were supposed to have nodes. In \cite{Oka10a} parts of the
lower critical field could be described by a two-band model. Some of the
unconventional properties of the thermal conductivity were found to be
reflected in the field dependence of the Sommerfeld coefficient, namely the
rather isotropic behavior below about 0.5\,T and the strongly anisotropic
behavior above \cite{Yan08a}. These authors, however, held nodes in the gap and
the influence of the Paul paramagnetism at higher fields responsible for the
effects. Low-temperature tunneling spectroscopy detected just one gap of about
0.2\,meV with nodes \cite{Mal12a}.

\textit{UPt$_3$.} The heavy fermion compound UPt$_3$ becomes antiferromagnetic
below about 5\,K and superconducting below 0.5\,K and has an upper critical
field of roughly 2 - 3\,T at 0\,K. Its field versus temperature diagram
decomposes into three regions of different superconducting phases with different
order parameter symmetries, from which two exist at zero field. Multi-band
superconductivity was suggested in view of the several bands that cross the
Fermi surface \cite{Mcm08a}. Confirmation by the methods presented in this paper
is difficult since the expected effects seem to be covered by other
unconventional properties of this compound. For instance, specific heat
measurements \cite{Bri94a} showed two distinct transitions at different
temperatures, indicating the different phases, and a considerable upturn below
about 0.1\,K. For reviews see \cite{Joy02a,Pfl09a}. 

\textit{UCoGe.} UCoGe orders ferro-magnetically below ca. 2.5\,K and
condensates into the superconducting state at ca. 0.6\,K. The upper critical
field is strongly anisotropic and reaches fields of up to 30\,T. Possible
two-band effects were indicated by an upward curvature of the upper critical
field near the transition temperature \cite{Huy08a}, but alternative
explanations related to the interplay with the ferromagnetic phase were also provided
\cite{Aok09a}.

\section{Summary}

Many superconducting properties and potential two-band effects as well as several two-band candidate materials have been reviewed. The two-band effects were compared with those expected from anisotropy in the single-band model and, in some cases, with those from different energy gap structures. Basically, the temperature dependence of many properties was found to be similar in the two-band and in the anisotropic single-band model, making a distinction between the two scenarios difficult in these cases. On the other hand, the field dependencies may be quite different in the two models because in two-band materials, such as MgB$_2$ and NbSe$_2$, superconductivity is apparently suppressed in one of the bands at sufficiently high fields, making the field dependent behavior of some properties a better candidate for successfully discriminating the two-band from other scenarios.

In section~\ref{Sec:Models}, the theoretical results predicted by the two-band
and by other models within Ginzburg Landau, separable model, BCS,
quasi-classical model, and Eliashberg theory have been reviewed. The
main results are:

\begin{itemize}
  \item Many experimentally observed two-band effects can be well reproduced by the theoretical models, even when we assume simple spherical Fermi surfaces. This indicates that basic properties of two-band superconductors do not strongly rely on details but rather on the averaged values of the anisotropy and the coupling strength.
  \item Ignoring coupling between the two bands, which are assumed to have distinctly different energy gap magnitudes and may differ also in other properties, leads to two transitions in the temperature
dependence of diverse properties, such
as the specific heat, superfluid density, etc. As interband coupling or
interband impurity scattering is enhanced, the low-temperature transition,
usually referring to the band with the smaller gap, is washed out and eventually
fades away, making the curves similar to
single-band behavior. Any small interband interaction leads to a common
transition temperature and upper critical field of both bands, though
superconductivity of one band may be strongly suppressed above a certain
magnetic field, which is then usually denoted the upper critical field of that
band.
  \item Increasing the gap ratio or the coupling strength often has opposite
effects. For instance, the specific heat jump becomes lower as the gap
ratio increases, but it becomes higher as the coupling or
interband interaction increases. Similarly, the low-temperature specific heat
and the superfluid density change more rapidly with temperature as
the gap ratio increases but more slowly as coupling increases. 
  \item The upward curvature of the upper critical field is sensitive to
several properties but basically appears and becomes more prominent as the
Fermi velocity ratio becomes larger.
  \item The separable model can be interpreted as an anisotropic
single-band as well as a two-band model with two spherical Fermi surfaces.
Anisotropy effects in a single-band material and two-band effects would thus
lead to the same unconventional behavior in this model, making a 
distinction between the two scenarios by comparing the temperature dependence of
experimental data with theoretical models often problematic.
\end{itemize}

In section~\ref{Sec:Exp}, several experimental techniques were briefly
described and some results obtained for possible two-band and other
materials reviewed. Two-band effects, in most cases investigated
in MgB$_2$, have been found to result in the following modifications:
\begin{itemize}
  \item The \textit{specific heat} increases more rapidly upon warming at low
temperatures, and its jump-height at the transition temperature is reduced. A
kink may show up at elevated temperatures if not suppressed by large interband
coupling or impurity scattering. Basically, the same effects
are predicted for the anisotropic single-band model.
  \item As the applied magnetic field increases, the \textit{Sommerfeld
coefficient} was found to grow rapidly at low fields and, when superconductivity is
suppressed in one of the bands, to grow at a lower rate at high fields.
Experiments indicate that distinguishing this behavior from that in non-two-band
materials might prove difficult.
  \item The field dependence of the low-temperature \textit{thermal
conductivity} was reported to be rather flat at intermediate, and steep  at
low and high fields, which again is similarly predicted for anisotropy effects.
  \item As the temperature increases, the \textit{superfluid density} decreases
faster at low and may become linear or even convex at intermediate temperatures.
The
same holds for the anisotropic single-band model.
  \item The curvature of the \textit{upper critical field} may become positive near
the transition temperature as is also well-known for non-two-band materials
having a non-spherical Fermi surface. In some iron-based superconductors, the
upward curvature stretches over the whole temperature range.
  \item The angular dependence of the \textit{magnetic torque} usually fits the
conventional single-band model quite well, assuming the same anisotropy for the coherence
length and the magnetic penetration depth, even in two-band samples.
  \item In contrast to the anisotropic single-band model, the \textit{reversible
magnetization} of two-band superconductors cannot be well described by the
single-band Ginzburg Landau theory over the whole field range. Instead, two
fits, one for the low and one for the high-field regime, are needed, which
reflects the different upper critical fields of the two bands.
  \item The \textit{anisotropy} may change significantly with magnetic field. In
MgB$_2$ and NbSe$_2$, the anisotropy is high at high and low at low magnetic
fields,
explained by a strongly anisotropic band dominating at high fields and an
almost
isotropic band dominating at low fields in these materials. The high-field
anisotropy decreases, while the low-field anisotropy increases upon warming,
so that both seem to merge at the transition temperature. No significant
differences of the anisotropies of different properties have been observed when
the results refer to the same field and temperature. The situation appears
different in the iron-based superconductors, where the high-field upper
critical field anisotropy was found to be small and that of the low-field magnetic
penetration depth rather large.
  \item Measurements of the \textit{superconducting gaps} by
tunneling or point-contact spectroscopy may reveal a single peak, two peaks, or
a peak and a shoulder in the energy dependence of the electrical
conductance. Anisotropy in single-band materials may lead to similar structures.
A clear proof of gaps on
different bands can be provided by angle-resolved photoemission spectroscopy.  
\end{itemize}

In the final section~\ref{Sec:Materials}, I have listed several potential
two-band materials, from which the more prominent ones were described in
more detail.
\begin{itemize}
  \item \textit{MgB$_2$}, being unquestionably a two-band superconductor,
consists of a quite isotropic band having a rather small s-wave energy gap and
a quite
anisotropic band having a large s-wave energy gap. In the isotropic band,
superconductivity is obviously suppressed at sufficiently high magnetic
fields. Most results and explanations of this review refer to this material.
  \item \textit{NbSe$_2$} has unconventional properties that are often
amazingly similar to those of MgB$_2$. As in MgB$_2$ the experiments indicate an
isotropic band with a small gap and an anisotropic band with a large gap as
well as suppression of superconductivity in the isotropic band at high fields.
Possible two-band effects have also been found in \textit{NbS$_2$}. 
  \item Concerning the \textit{iron-based superconductors}, the whole family is widely
accepted to show two- or multi-band superconductivity, though many of their properties
deviate significantly from those of MgB$_2$. The Fermi surface is crossed by
several bands on which a superconducting gap opens. These gaps might have
s-wave symmetry though different signs on different bands and 
nodes may exist in some cases.
  \item The \textit{cuprates} are actually known for their d-wave character,
but some experiments could not be explained within this model. For
instance, a significant boost of the superfluid density at low temperatures
measured by muon-spin rotation was regarded as evidence for a second band by one
group of authors but as an effect coming just from vortex pinning by
another group. 
  \item The \textit{borocarbides} were analyzed within different models.
A possible two-band scenario has been confirmed by many experimental and
theoretical results showing similar properties as in MgB$_2$, though a clear
distinction to all other models is currently not fully accepted.
  \item A lot of other materials were considered as two-band superconductors
on the basis of various experiments. In some of these materials, the results
 strongly point to the two-band scenario, while in others the data are not
convincing, but in most cases a clear interpretation demands more  experiments.
The materials  listed in this paper are Nb$_3$Sn, V$_3$Si, CeCoIn$_5$,
PrX$_4$Sb$_{12}$ (X = Os, Ru), LaRu$_4$As$_{12}$,
LaOs$_4$Sb$_{12}$, Ba$_8$Si$_{46}$, Na$_{0.3}$CoO$_2 \cdot$
1.3H$_2$O, OsB$_2$, XMo$_6$S$_8$ (X = Sn, Pb), ZrB$_{12}$, X$_2$C$_3$ (X = La,
Y), X$_2$Fe$_3$Si$_5$ (X = Lu, Sr), MgCNi$_3$, Sr$_2$RuO$_4$, URu$_2$Si$_2$,
UPt$_3$, and UCoGe.
\end{itemize}

\begin{acknowledgments}
  I would like to thank Prof. H. W. Weber, M. Eisterer, and F. Sauerzopf for their
support and numerous discussions, and the first two for proofreading the manuscript.
This work was supported by the Austrian Science Fund under Contract No. 21194
and 23996. 
\end{acknowledgments}

%

\end{document}